%% file: paper.tex
\documentclass[manuscript]{acmart}

\AtBeginDocument{%
  \providecommand\BibTeX{{
    \normalfont B\kern-0.5em{\scshape i\kern-0.25em b}\kern-0.8em\TeX}}}

\renewcommand\footnotetextcopyrightpermission[1]{}

\usepackage[inline]{enumitem}
\usepackage{xspace}

\usepackage{graphicx}
\usepackage{subfigure}
\usepackage{listings}
\usepackage{multirow}
\usepackage{gensymb}
\usepackage{marvosym}

\usepackage{float}
\usepackage{color}
\usepackage{listings}
\usepackage{caption}
\newfloat{lstfloat}{htbp}{lop}
\floatname{lstfloat}{Listing}

\newcounter{nalg}
\DeclareCaptionLabelFormat{algocaption}{Algorithm \thenalg}

\newcommand\figref[1]{Fig.\,\ref{#1}}
\newcommand\secref[1]{Sec.\,\ref{#1}}

\renewcommand\eqref[1]{Eq.\,(\ref{#1})}

\newcommand{\fakepar}[1]{\vspace{1mm}\noindent\textbf{#1.}}

\newlist{mylist}{enumerate*}{1}
\setlist[mylist]{label=(\roman*)}

\title{Dynamic Voltage and Frequency Scaling for Intermittent Computing}

\newcommand\sceptic{\textsc{ScEpTIC}\xspace}
\newcommand\dvfs{\textsc{D$^2$VFS}\xspace}
\newcommand\fbtc{\textsc{FBTC}\xspace}

\newcommand\default{\texttt{Default}\xspace}
\newcommand\noadcoff{\texttt{NOADCOFF}\xspace}
\newcommand\twovmin{\texttt{ADCMINV}\xspace}

\author{Andrea Maioli}
\email{andrea1.maioli@polimi.it}
\affiliation{%
  \institution{Politecnico di Milano}
  \country{Italy}
}

\author{Kevin A. Quinones}
\email{kevinalessandro.quinones@mail.polimi.it}
\affiliation{%
  \institution{Politecnico di Milano}
  \country{Italy}
}

\author{Saad Ahmed}
\email{sahmed@gatech.edu}
\affiliation{%
  \institution{Georgia Institute of Technology}
  \country{U.S.}
}

\author{Muhammad H. Alizai}
\email{hamad.alizai@lums.edu.pk}
\affiliation{%
  \institution{Lahore University of Management Sciences}
  \country{Pakistan}
}

\author{Luca Mottola}
\email{luca.mottola@polimi.it}
\affiliation{%
  \institution{Politecnico di Milano, Italy and Uppsala University}
  \country{Sweden}
}

\begin{document}

\input{abstract}

\maketitle

\input{introduction}

\input{background}

\input{technique}

\input{implementation}

\input{evaluation}

\input{conclusion}

\bibliographystyle{ACM-Reference-Format}
\bibliography{paper}

\end{document}

%% file: abstract.tex
\begin{abstract}
    We present hardware/software techniques to intelligently regulate supply voltage and clock frequency of intermittently-computing devices.
    These devices rely on ambient energy harvesting to power their operation and small capacitors as energy buffers.
    Statically setting their clock frequency fails to capture the unique relations these devices expose between capacitor voltage, energy efficiency at a given operating frequency, and the corresponding operating range.
    Existing dynamic voltage and frequency scaling techniques are also largely inapplicable due to extreme energy scarcity and peculiar hardware features.
    We introduce two hardware/software co-designs that accommodate the distinct hardware features and function within a constrained energy envelope, offering varied trade-offs and functionalities.
   Our experimental evaluation combines tests on custom-manufactured hardware and detailed emulation experiments. 
   The data gathered indicate that our approaches result in up to  $3.75 \times$ reduced energy consumption and $12 \times$ swifter execution times compared to the considered baselines, all while utilizing smaller capacitors to accomplish identical workloads.
\end{abstract}

%% file: introduction.tex
\section{Introduction}
\label{sec:intro}

Ambient energy harvesting enables battery-less embedded sensing~\cite{batteryless-future, soil-termoelectric, bridge-sensor, traffic-flow-sensor, sensys20deployment, flicker, permadaq}.
However, energy from the environment is generally erratic, causing frequent and unanticipated energy failures.
Executions thus become \textit{intermittent}, as they consist of intervals of active operation interleaved by periods of recharging energy buffers~\cite{harvesting-survey}.

\begin{figure}[h]
    \centering
    \resizebox{0.75\columnwidth}{!}{\includegraphics{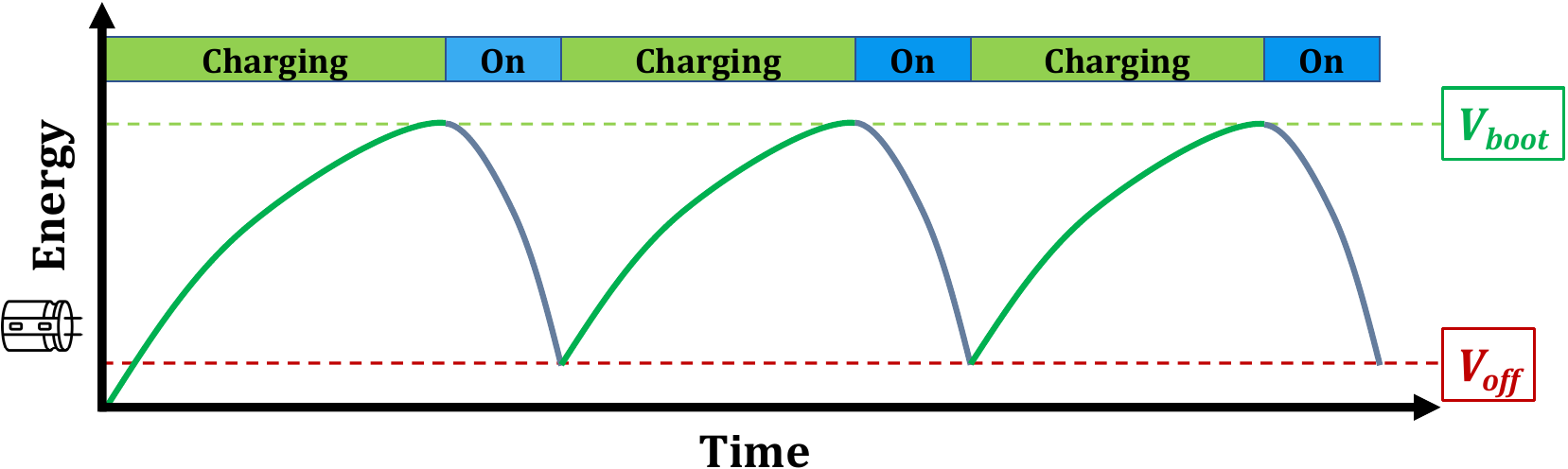}}
    \caption{Example of intermittent execution.}
    \label{fig:intermittent}
\end{figure}

Battery-less devices typically employ capacitors as energy buffers.
As intuitively shown in \figref{fig:intermittent}, as long as the capacitor voltage is below a predetermined \emph{boot threshold}, the device rests dormant until the buffered energy is sufficient to boot.
An \emph{energy cycle} then starts when the device actively operates.
The energy consumption during this cycle typically exceeds the ambient energy intake, leading to a net negative energy balance.
Consequently, the capacitor voltage drops below the \emph{operating voltage}, causing the device to shut down, at which point a new charging phase begins.

Due to extreme resource constraints of the target platforms, applications run on bare hardware without proper operating system support.
Energy failures thus normally cause devices to lose computational and peripheral states.
To ensure forward progress across energy failures, techniques~\cite{ALFRED, lucia-peripherals, maioli21ewsn, maioli19lctes, karma, sytare, restop, Mementos, Hibernus, Hibernus++, DINO, alpaca, chinchilla, HarvOS, clank, chain} exist that, at the cost of significant overhead, allow the system to save the computational and peripheral state onto non-volatile memory (NVM) locations, which persist across energy failures.
Once the boot threshold is attained again, the state is restored from NVM, and execution picks up near the point where the energy failure occurred.

\begin{figure}[t]
    \centering
    \parbox{0.47\columnwidth}{%
        \resizebox{0.47\columnwidth}{!}{\includegraphics{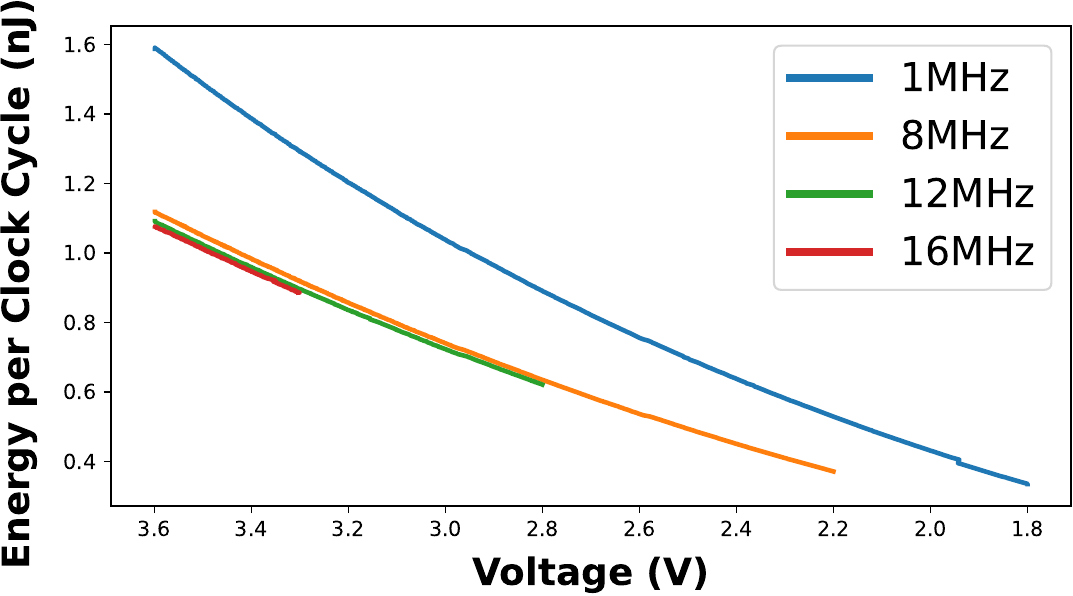}}
        \caption{Energy consumption per clock cycle at various voltage and frequency ranges for the MSP430-G2553~\cite{epic,msp430g2553}.}
        \label{fig:energy_per_cc}
    }%
    \qquad
    \parbox{0.47\columnwidth}{%
        \vspace{5.8mm}
        \resizebox{0.47\columnwidth}{!}{\includegraphics{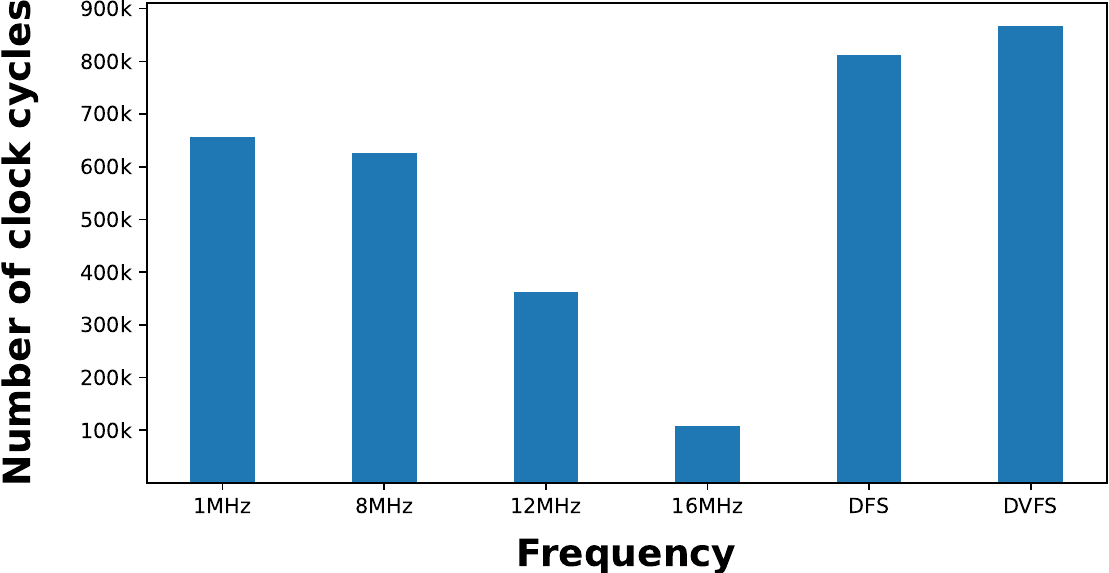}}
        \caption{Clock cycles executed in a single discharge from $3.6V$ of a $100 \mu F$ capacitor for various frequency configurations for the MSP430-G2553~\cite{msp430g2553}.}
        \label{fig:clock_cycles_power_cycle}
    }%
\end{figure}


\fakepar{Frequency, voltage, and the rest} With low-power microcontrollers, system efficiency is typically dictated by the rate of energy consumption and execution speed.
These parameters are influenced by the running frequency, supply voltage, and operating range~\cite{epic}.

Consider the MSP430-G2553~\cite{msp430g2553} microcontroller unit (MCU) of the TI MSP430 series, that is, arguably the most used MCU platform in battery-less devices.
\figref{fig:energy_per_cc} shows the energy consumption per clock cycle at the four factory-calibrated operating frequencies.
The higher the frequency, the faster the computation and the lower the energy consumption per clock cycle.
For example, running the MCU at $16MHz$ is on average $47\%$ more energy efficient per clock cycle and $16x$ faster than the $1MHz$ setting.
However, compared to the latter, running the MCU at $16MHz$ limits the operating voltage range: as soon as the supply voltage falls below $3.3V$, the MCU shuts down.
Differently, if the MCU is set to run at $1MHz$, it can continue operating until the supply voltage reaches $1.8V$.

\figref{fig:clock_cycles_power_cycle} demonstrates the impact of these trade-offs on the number of clock cycles the MCU can execute, given a fixed energy budget.
Although the $16MHz$ setting offers faster execution and superior energy efficiency per clock cycle, its narrowed operating voltage range results in $3.75x$ fewer clock cycles compared to the slower, yet less energy-efficient $1MHz$ setting.
This latter configuration enables the MCU to compute for an extended duration, specifically as long as the supply voltage remains above~$1.8V$.
Fundamentally, the $1MHz$ setting allows the system to harness more energy—and consequently more useful work—from an identical initial capacitor charge.

\fakepar{Challenge} Similar trade-offs are seen also in regular processors and routinely exploited to improve execution speed and/or energy consumption~\cite{shirvani2020survey}.
In mobile platforms, for instance, the dynamic adjustment of operating frequencies and supply voltage enables systems to respond to sudden surges in system load, while conserving energy during periods of lighter loads~\cite{kim2013stabilizing}.
To achieve this, dedicated hardware and software components are employed, collectively referred to as Dynamic Voltage and Frequency Scaling (DVFS)~\cite{eyerman2011fine}.

DVFS techniques used in mainstream platforms are not applicable to battery-less devices.
Resource constraints and different performance metrics demand a different design rationale.
As an example, employing hardware support for DVFS from mainstream platforms in battery-less devices would be impractical due to the excessive energy consumption~\cite{eyerman2011fine}.
Conversely, the lack of a proper operating system renders existing software drivers outright unusable.

Crucially, the application and system requirements of battery-less embedded computing diverge significantly from those in mainstream computing.
Energy consumption is the primary, and often \emph{only} metric of interest.
To conserve energy~\cite{aic}, application developers often prioritize energy savings over other metrics of interest, such as execution speed or data processing accuracy.
Conserving energy extends the duration of energy cycles, consequently reducing the overhead associated with NVM operations.

Further, charge-discharge cycles are frequent in battery-less devices, as the push for miniaturization prompts energy storage facilities to be minimized as well.
For example, harvesting energy from RF transmissions to compute a simple CRC may lead to $16$ energy failures over a $6$ seconds period~\cite{harvesting-survey}.
The improvements in energy consumption, leading to prolonged energy cycles and lower overhead, are going to have a magnifying effect on other metrics of interest, including data throughput.

\fakepar{Contribution and road-map}
As we discuss \secref{sec:background}, only a few efforts exist to apply DVFS to battery-less devices~\cite{power-neutral-dvfs, power-neutral-dfs}.
Research most similar to ours primarily targets multi-core processors equipped with DVFS hardware support, which are distinctly different from MSP430-class microcontrollers.
While their focus is on achieving power neutrality by adjusting power consumption to match harvested energy, they do not account for the implications of NVM operations.

We demonstrate that it is possible to achieve DVFS functionality in a much more limited energy envelope, throughout intermittent operations, and consequently unlock significant performance gains.
\secref{sec:design} illustrates the design rationale, whereas \secref{sec:impl} provides concrete evidence based on two hardware/software co-designs that expose different trade-offs and functionality.
The two distinct implementations, \dvfs and \fbtc, were developed to balance simplicity, efficiency, and configurability in achieving DVFS in batteryless embedded devices.
\dvfs serves as a reference design, straightforward but occasionally less efficient, emphasizing the gains in performance even with the energy costs of its DVFS circuitry.
On the other hand, \fbtc improves upon \dvfs by reducing energy overhead and introducing a configurable startup voltage threshold, offering developers a means to tailor energy dynamics to specific deployment scenarios.
This design choice underscores a pragmatic approach: providing a baseline system that demonstrates the benefits of DVFS while also offering a more advanced alternative that optimizes for energy efficiency and provides greater flexibility for real-world applications.
Both implementations use the same MCU and voltage regulator, but their different architectures highlight the balance between energy efficiency, system responsiveness, and hardware complexity, addressing distinct use cases and optimization priorities in the domain of energy-harvesting systems.

\secref{sec:eval} presents an extensive evaluation of both designs.
We compare their performance against a stock MSP430 microcontroller that is statically set to one of the four factory-calibrated frequencies.
This configuration fails to capture the trade-offs illustrated in \figref{fig:clock_cycles_power_cycle}.
Our results demonstrate that both \dvfs and \fbtc can achieve up to $3.75x$ lower energy consumption and $12x$ faster execution time than the considered baselines, while requiring a smaller energy buffer and thus reducing recharging times and mitigated energy waste due to leakage.

%% file: background.tex
\section{Background and Related Work}
\label{sec:background}

We offer a primer on intermittent computing and delve into the challenges and prevailing solutions related to DVFS for both mainstream computing platforms and battery-less devices.

\subsection{Intermittent Computing}

The pattern of intermittent computing necessitates specialized system support to bridge periods of energy scarcity.
Numerous techniques have been developed to ensure forward progress in battery-less devices despite energy disruptions. Some strategies implement checkpoints at compile-time based on execution patterns~\cite{Mementos, chinchilla} or program structures~\cite{DICE, HarvOS, Mementos}, while others utilize supplementary hardware to initiate proactive checkpointing~\cite{Hibernus, Hibernus++, QuickRecall}.
There are also approaches that offer developers task-based programming abstractions with transactional semantics~\cite{chain, alpaca, coala}.
Specialized solutions have been designed to preserve peripheral states through energy disruptions~\cite{karma, restop, sytare}.

However, the majority of techniques in intermittent computing primarily aim to minimize the energy overhead associated with maintaining application progress.
They often overlook the dynamics of supply voltages and MCU frequency adjustments.
Thus, the application of DVFS presents a distinct challenge, influencing system performance \emph{within} an energy cycle—by enhancing energy efficiency, for instance—rather than spanning multiple energy cycles.

\subsection{DVFS}

DVFS includes two key mechanisms: voltage and frequency scaling.
Each processor possesses distinct operational ranges, with each range characterized by a frequency and voltage tuple $(f,V)$.
Mainstream computing platforms utilize advanced software and hardware mechanisms that allow for precise control over voltage and frequency configurations~\cite{dvfs1, dvfs2}.

In the following, we will focus our discussion on related works pertaining to embedded systems, as they closely align with battery-less devices.

\fakepar{Real-time embedded systems}
Salehi~et~al.~\cite{dvfse} present an adaptive voltage and frequency scaling technique that rapidly tracks the workload changes to meet soft real-time deadlines.
Their work demonstrates considerable energy savings and fewer frequency updates compared to DVFS systems based on fixed update intervals.
HyPowMan~\cite{hypowman} considers the problem of minimizing energy consumption for periodic real-time tasks scheduled over multiprocessor platforms.
The technique takes a set of well-known existing DVFS policies, each performing well for given conditions, and adapts at runtime to the best-performing policy for a given workload.

Huang~et~al.~\cite{mixc} apply DVFS to mixed-criticality systems and show that DVFS helps critical tasks meet deadlines by speeding up the processor when it is bound to miss a deadline.
Liu~et~al.~\cite{liu} employ DVFS to optimize system thermal profiles to prevent run-time thermal emergencies and to minimize cooling costs.
RT-DVFS~\cite{rtdvfs} modifies the OS's real-time scheduler and task management service to provide energy savings while maintaining real-time deadline guarantees.
Generalized Shared Recovery (GSHR)~\cite{gshr} efficiently uses DVFS techniques to achieve a given reliability goal for real-time embedded applications.

While these works offer essential insights into the application of DVFS in embedded systems, their design objectives diverge significantly, rendering their techniques less suited for direct application to battery-less devices.
The latter rarely deal with real-time deadlines, whereas reducing energy consumption for a fixed workload is key.

\fakepar{Wireless sensor networks}
Kulau~et~al.~\cite{tp,idealvolting,uvdcoss} analyze the effects of undervolting a wireless sensor node.
They show that such a device can still work reliably, even if the voltage recommendations are violated, because a correlation exists between temperature and probability of error at a given voltage level.
Powell~et~al.~\cite{bsn} design DVFS hardware to meet battery life and form factor expectations of body area sensor networks.
Similar to these works are also the efforts on developing DVFS techniques in distributed microsensor networks~\cite{dmsn} and in sensor networks with deadlines~\cite{deadlines}.

As most of these works aim to conserve energy, many of them are similar to ours in spirit, yet the authors consider battery-powered devices with \emph{finite} energy supplies and tend to accept performance penalties to increase lifetime.
On the contrary, we deal with intermittent but unbounded energy supplies, with the goal of increasing the amount of work achieved in an energy cycle.

\fakepar{Battery-less devices} EA-DVFS~\cite{eadvfs} presents a high-level simulation study on the advantages of DVFS for real-time operation in battery-less devices. Due to the lack of a corresponding hardware implementation, it does not serve as a suitable baseline for our investigation.
Lin~et~al.~\cite{allmath} model a framework for concurrent task scheduling and dynamic voltage and frequency scaling in real-time embedded systems with energy harvesting.
Li~et~al.~\cite{blabla} also provide early insights into jointly scaling workload, voltage, and frequency in multi-core sensor networks using energy harvesting.

These studies offer valuable preliminary perspectives on the application of DVFS in energy harvesting devices.
However, our work is the first concrete implementation of any such technique, complemented by a comprehensive evaluation that distinctly underscores the advantages of applying DVFS in battery-less environments.

\fakepar{Summary}
Numerous efforts exists to enhance energy efficiency, particularly in environments with stringent energy constraints.
The primary focus of these works is on devices with \emph{finite} energy sources.
These works, although foundational, often diverge in design goals and cannot be applied ``as-is'' to battery-less devices.

Our research pivots from these traditional paradigms.
Instead of finite energy reserves, we consider intermittent, yet potentially perpetual energy supplies.
Our primary objective is not merely to conserve energy but to maximize the amount of usefule work accomplished within each active cycle.

%% file: technique.tex
\section{Design Rationale}
\label{sec:design}

The fundamental element enabling DVFS for a target MCU is the identification of the available \emph{performance windows}, which consist in a plaform-specific combination of voltage and frequency settings.

Indeed, most low-power MCUs feature dozens of possible frequency settings.
We concentrate on a subset of them, usually the factory-calibrated ones, where the datasheet also explicitly reports the corresponding minimum supply voltage.
At a given frequency setting, the minimum supply voltage yields the lowest energy consumption~\cite{epic}. 
For instance, with the MSP430-G2553~\cite{msp430g2553} MCU, we examine the four factory-calibrated frequency settings with the corresponding minimum supply voltages, thereby determining \emph{four (ordered) performance windows}:
\begin{mylist}
    \item $16MHz$ at $3.3V$,
    \item $12MHz$ at $2.8V$,
    \item $8MHz$ at $2.2V$, and
    \item $1MHz$ at $1.8V$
\end{mylist}.

\begin{figure}[t]
    \centering
    \parbox{0.47\columnwidth}{%
        \resizebox{0.47\columnwidth}{!}{\includegraphics{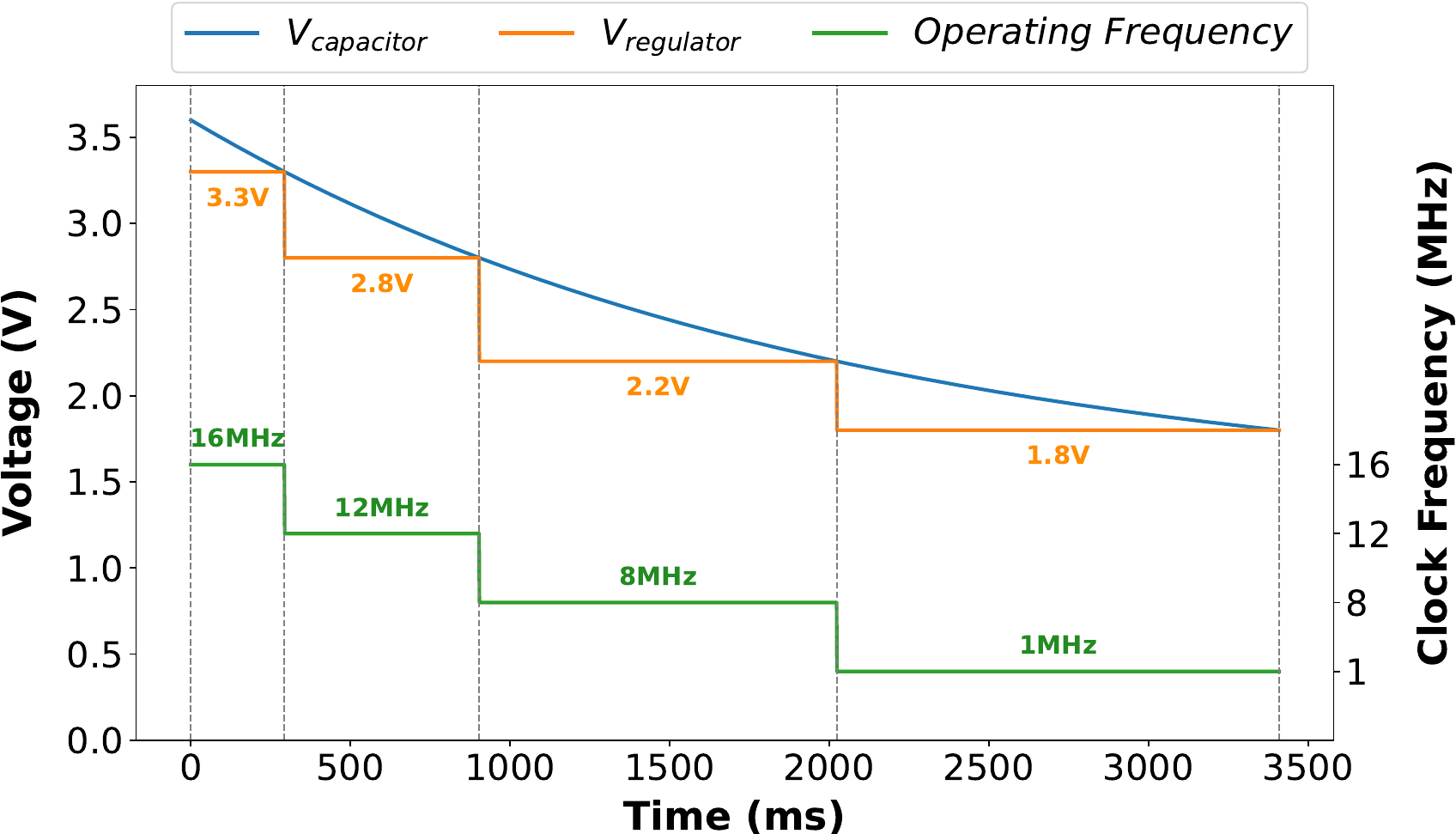}}
        \caption{System behavior when capacitor discharges.}
        \label{fig:abstract_discharge}
    }%
    \qquad
    \parbox{0.47\columnwidth}{%
        \resizebox{0.47\columnwidth}{!}{\includegraphics{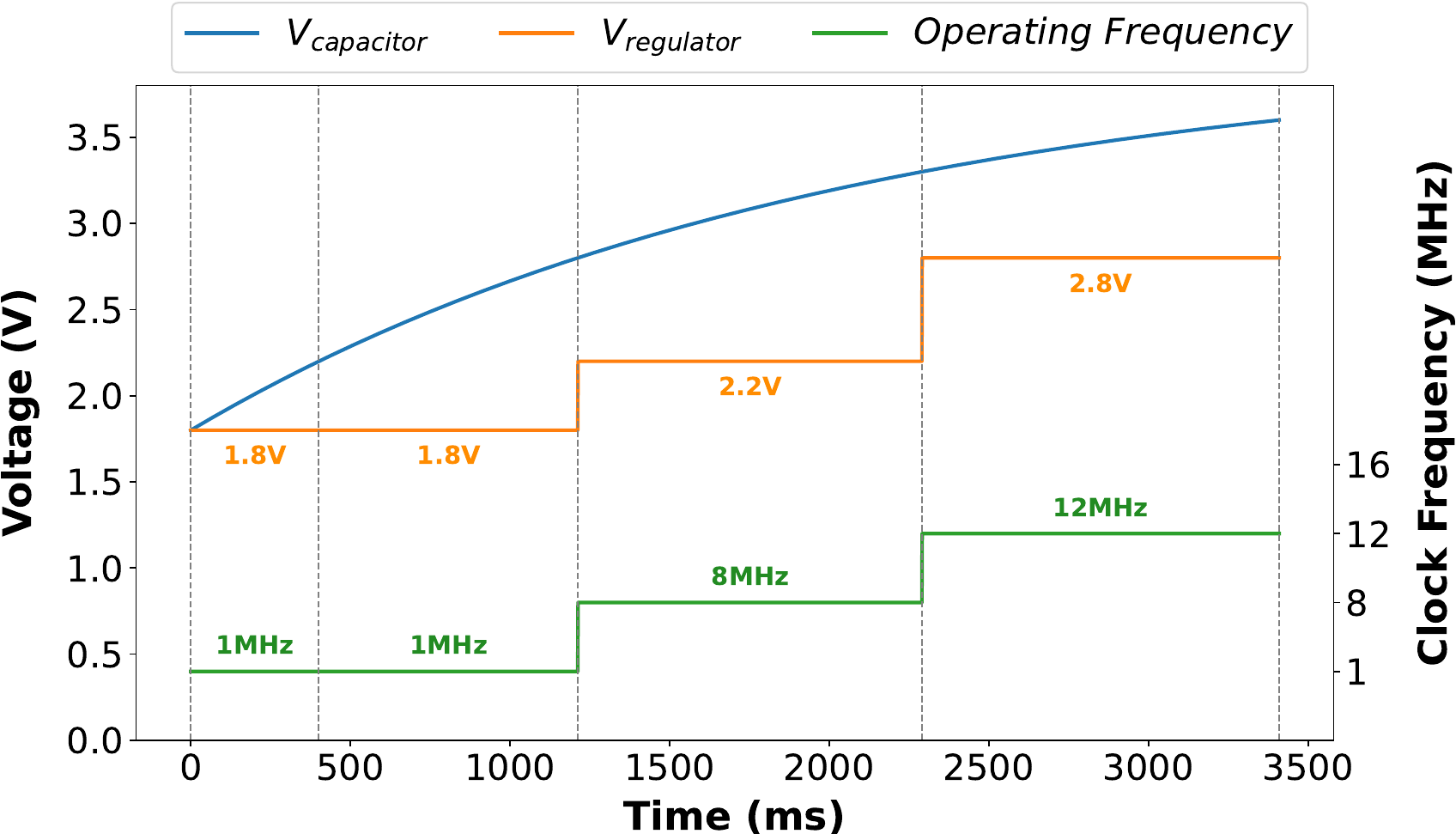}}
        \caption{System behavior when capacitor charges.}
        \label{fig:abstract_charge}
    }%
\end{figure}

\fakepar{Scaling down}
The blue and orange curves depicted in \figref{fig:abstract_discharge} illustrate the expected performance across the four performance windows of the MSP430-G2553 during capacitor discharge.

As long as the capacitor voltage is above the minimum supply voltage of a certain performance window, the supply voltage is regulated to \emph{exactly} this minimum, which provides the best energy efficiency at the corresponding frequency.
As soon as the capacitor voltage crosses the lower bound of the \emph{current} performance window, frequency and voltage settings are scaled to enter the \emph{lower} performance window.
For example, when the capacitor discharges from $3.6V$ to $3.3V$, frequency changes to $12MHz$  and supply voltage is scaled to $2.8V$, thus moving from window~(i) to window~(ii).
Transitioning to a lower performance window necessitates altering the frequency settings prior to adjusting the supply voltage; reversing this sequence would result in device shutdown due to the supply voltage dipping below the minimum threshold for the given frequency setting.

\fakepar{Scaling up}
This rationale is also applicable when the capacitor voltage rises, albeit with a nuance as depicted in \figref{fig:abstract_charge}.

Energy consumption per clock cycle increases when moving from a lower to a higher frequency setting.
Should the device fail to harvest sufficient energy, the heightened energy consumption per clock cycle could precipitate an immediate reduction in capacitor voltage, thereby compelling the system to revert promptly to a lower performance window.
Following the adjustment, as the energy consumption per clock cycle decreases, the net energy balance may shift to positive, leading to a subsequent rise in capacitor voltage. 
This increase can trigger a transition back to the higher performance window.
This behavior may repeat indefinitely, entering a sort of livelock.
To avoid this, we cautiously wait until the capacitor voltage reaches the \emph{upper bound} of the upper performance window before changing frequency and voltage settings accordingly.
Symmetrically, to avoid shutting down the system when transitioning to the upper performance window, we change supply voltage first, then frequency.

\fakepar{Towards implementation}
Realizing this behavior concretely hinges on a careful consideration of trade-offs between the energy overhead attributed to supplementary hardware components and the resulting gain in flexibility.

For example, to change supply voltage, an external voltage regulator may be required, as regular low-power MCUs are usually not equipped with it.
Detection of the capacitor voltage reaching a threshold that necessitates a change in performance window can be accomplished either by periodic polling or by employing specialized circuitry that asynchronously alerts the MCU of particular conditions occurring at the capacitor.
Conversely, existing low-power MCUs are capable of altering frequency settings via software:
using MSP430-class MCUs~\cite{msp430g2553}, frequency settings are programmatically set by changing the values of specific registers.

%% file: implementation.tex
\section{Implementation}
\label{sec:impl}

The design rationale is materialized in two distinct implementations, each elucidating different trade-offs and functionalities.
The first implementation we present is called \dvfs (Discrete Dynamic Voltage and Frequency Scaling) and may be regarded as a reference implementation of sorts.
It achieves DVFS functionality in the simplest, but not necessarily the most efficient or flexible way.
As illustrated in \secref{sec:eval}, despite the energy overhead due to the circuitry realizing DVFS functionality, \dvfs already provides great performance advantages compared to a static setting.

The second implementation is called \fbtc (Fixed Boot Threshold Controller) and improves over \dvfs \emph{in three ways}.
The circuitry realizing DVFS functionality imposes a much lower energy overhead compared to \dvfs.
Further, \fbtc avoids the fluctuation problem mentioned in \secref{sec:impl} by design, without requiring a delay in the changes to upper performance windows during the capacitor charge.
This results in a faster and more efficient change of operating setting compared to \dvfs.
The corresponding energy savings are spent in useful application processing, boosting the overall energy efficiency.
Finally, \fbtc allows developers to configure the voltage threshold to boot the system, providing a knob that may be useful to capture deployment-specific energy dynamics~\cite{sensys20deployment}.

Both implementations are centered around the MSP430-G2553~\cite{msp430g2553} MCU and use the TPS62740~\cite{tps62740} voltage regulator.
The performance windows are those in \secref{sec:design}.

\subsection{\dvfs}
\begin{figure}[t]
    \subfigure[\dvfs logic.]{
        \centering
        \resizebox{0.49\columnwidth}{!}{\includegraphics{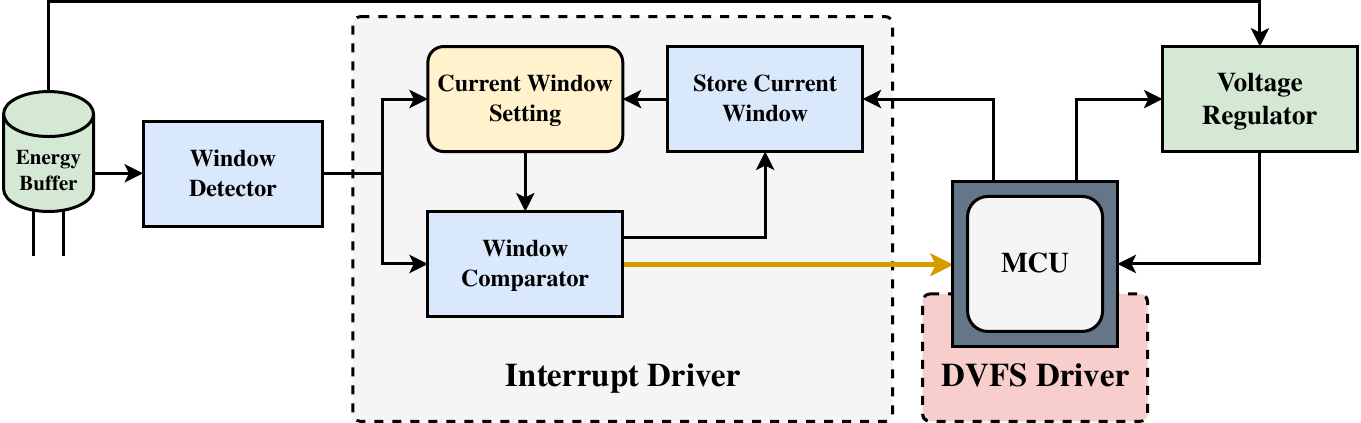}}
        \label{fig:dvfs_logic}
    }
    \subfigure[\dvfs schematics.]{
        \centering
        \resizebox{0.48\columnwidth}{!}{\includegraphics{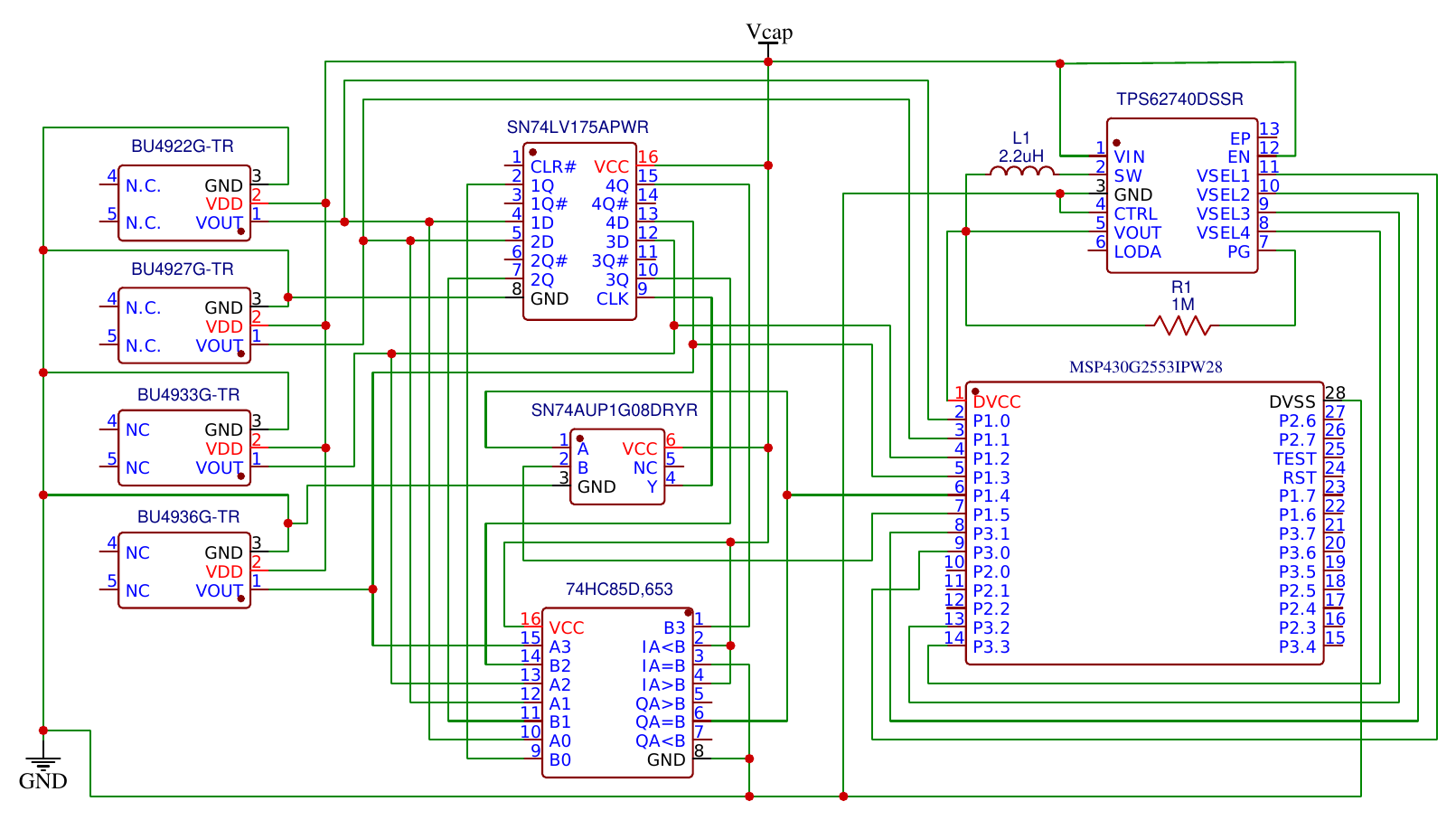}}
        \label{fig:dvfs_schematics}
    }
    \caption{\dvfs design.}
    \label{fig:dvfs}
\end{figure}

\figref{fig:dvfs} illustrates the design of \dvfs; \figref{fig:dvfs_logic} describes the logical components and \figref{fig:dvfs_schematics} shows the schematics.

\fakepar{Logical components}
The \emph{Window Detector} in \figref{fig:dvfs_logic} determines the valid performance window based on capacitor voltage.
To circumvent the energy-intensive process of periodic polling by the MCU's ADC, we employ four TI BU49XXG~\cite{bu49xxg} voltage comparators, as illustrated in \figref{fig:dvfs_schematics}; one for each performance window.
Each comparator takes as input the capacitor voltage $V_{cap}$ and outputs a signal that indicates if $V_{cap}$ is higher than the threshold.

The MCU is required to ascertain shifts in the current performance window to adjust its operating frequency and supply voltage appropriately.
One approach could involve periodic software polling of the \emph{Window Detector}'s output.
However, this method is fraught with several drawbacks: it imposes extra latency dependent on the polling interval, risks interrupting the flow of application processing, and leads to superfluous energy expenditure, as each non-revealing check essentially constitutes wasted effort.
We anticipate that such unproductive checks would predominate.

We opt for a design that employs a hardware interrupt mechanism to notify the MCU of a change in the performance window.
This functionality is shown as \emph{Interrupt Driver} in \figref{fig:dvfs_logic}.
The key is to maintain a small dedicated memory that reflects the active performance window—specifically, the current configuration of the MCU's frequency and supply voltage—as depicted in \emph{Current Window Setting} in \figref{fig:dvfs_logic}.
A dedicated \emph{Window Comparator} monitors both the output of the \emph{Window Detector} and the \emph{Current Window Setting}; whenever the two differ, it signals an interrupt to the MCU.
This informs the MCU that the capacitor voltage entered a new performance window.
As a result, the \emph{Store Current Window} function updates the \emph{Current Window Setting} to reflect the new information accurately.

The \emph{Interrupt Driver} is implemented using three hardware components, each chosen for its energy-saving potential, as depicted in \figref{fig:dvfs_schematics}:
\begin{mylist}
    \item a SN74LV175A~\cite{sn74lv175a} D-type flip-flop that stores the \emph{Current Window Setting},
    \item a 74HC85~\cite{74hc85} 4-bit comparator working as the \emph{Window Comparator}, and
    \item a SN74AUP1G08~\cite{sn74aup1g08} \texttt{AND} gate operating as the \emph{Store Current Window}  block.
\end{mylist}

\begin{figure}[t]
    \centering
    \hspace{30mm}
    \resizebox{0.30\columnwidth}{!}{\includegraphics{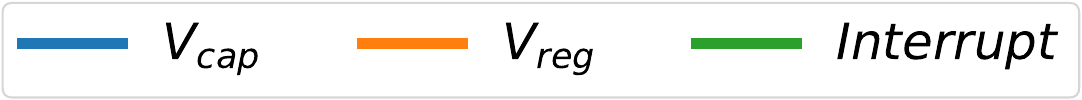}}
    \newline
    \resizebox{0.45\columnwidth}{!}{\includegraphics{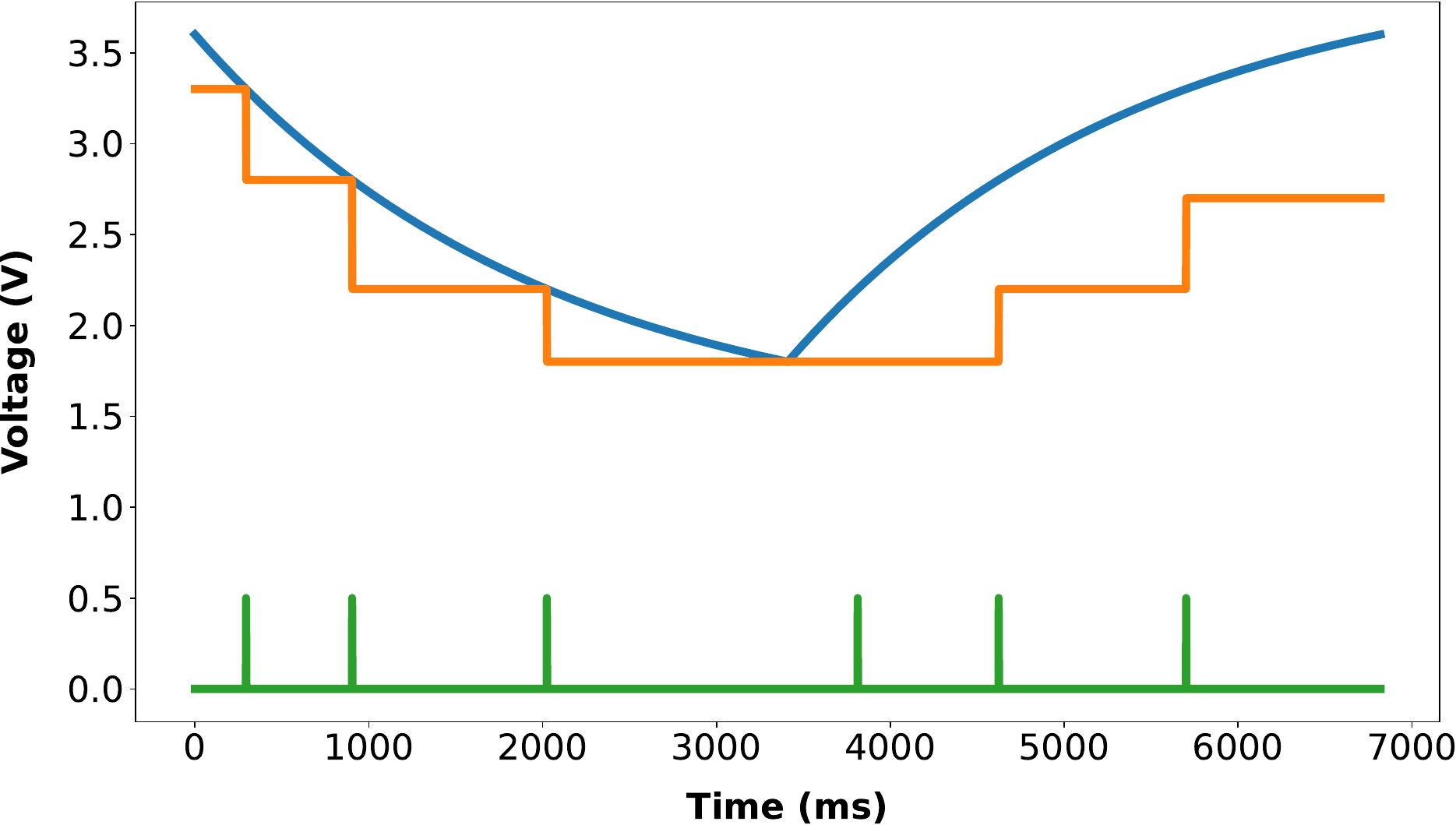}}
    \caption{Example of \dvfs behavior.}
    \label{fig:dvfs_behavior}
\end{figure}

\fakepar{Run-time behavior}
\figref{fig:dvfs_behavior} shows an example execution.
The capacitor voltage $V_{cap}$ starts at $3.6V$ and the DVFS driver sets the voltage regulator to $3.3V$ with the MCU operating at $16MHz$.
As soon as $V_{cap}$ reaches $3.3V$, the \emph{Interrupt Driver} fires an interrupt, shown in green in \figref{fig:dvfs_behavior}.
The \dvfs driver identifies the new performance window by checking the outputs of the voltage detectors and regulates supply voltage to $2.8V$ first, then sets the operating frequency to $12MHz$.
The same behavior repeats when $V_{cap}$ reaches $2.8V$ and $2.2V$, corresponding to two more interrupts.

To avoid the fluctuations mentioned in \secref{sec:design}, the \dvfs driver delays the change to the upper performance window when $V_{cap}$ increases.
Let us focus on \figref{fig:dvfs_behavior} when $V_{cap}$ is at $1.8V$ and rising.
The MCU is running at $1MHz$ and supply voltage is regulated at $1.8V$.
Whenever $V_{cap}$ reaches $2.2V$, the \emph{Interrupt Driver} fires an interrupt.
The \dvfs driver discerns the appropriate new performance window by monitoring the outputs from the voltage detectors.
To avoid the risk of fluctuations, an immediate transition to a higher performance window is deferred.
The driver awaits a subsequent interrupt to initiate this change.
Thus, when $V_{cap}$ rises to $2.8V$, the \emph{Interrupt Driver} issues a new interrupt, prompting the \dvfs driver to adjust the supply voltage to $2.2V$ and the MCU frequency to $8MHz$.

\begin{figure}[t]
    \subfigure[\fbtc logic.]{
        \centering
        \resizebox{0.47\columnwidth}{!}{\includegraphics{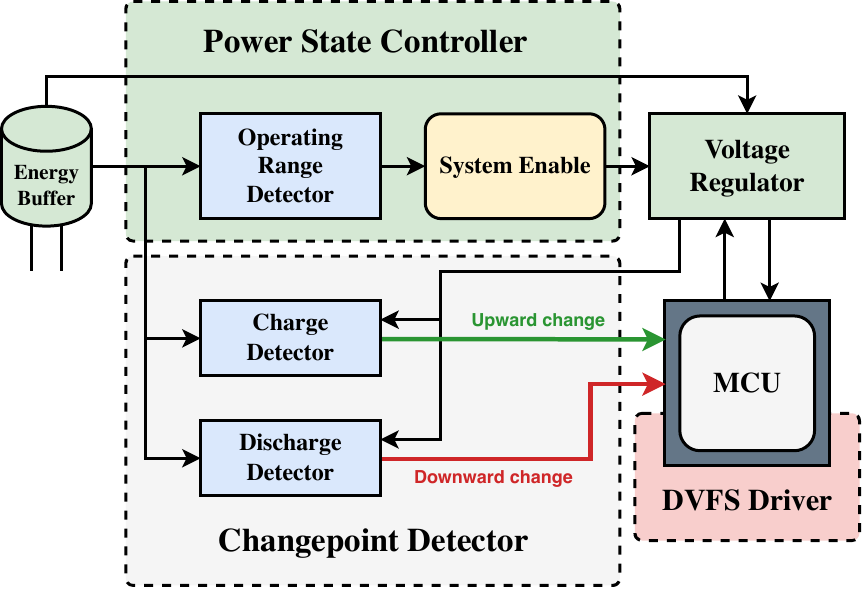}}
        \label{fig:fbtc_logic}
    }
    \subfigure[\fbtc schematics.]{
        \centering
        \resizebox{0.50\columnwidth}{!}{\includegraphics{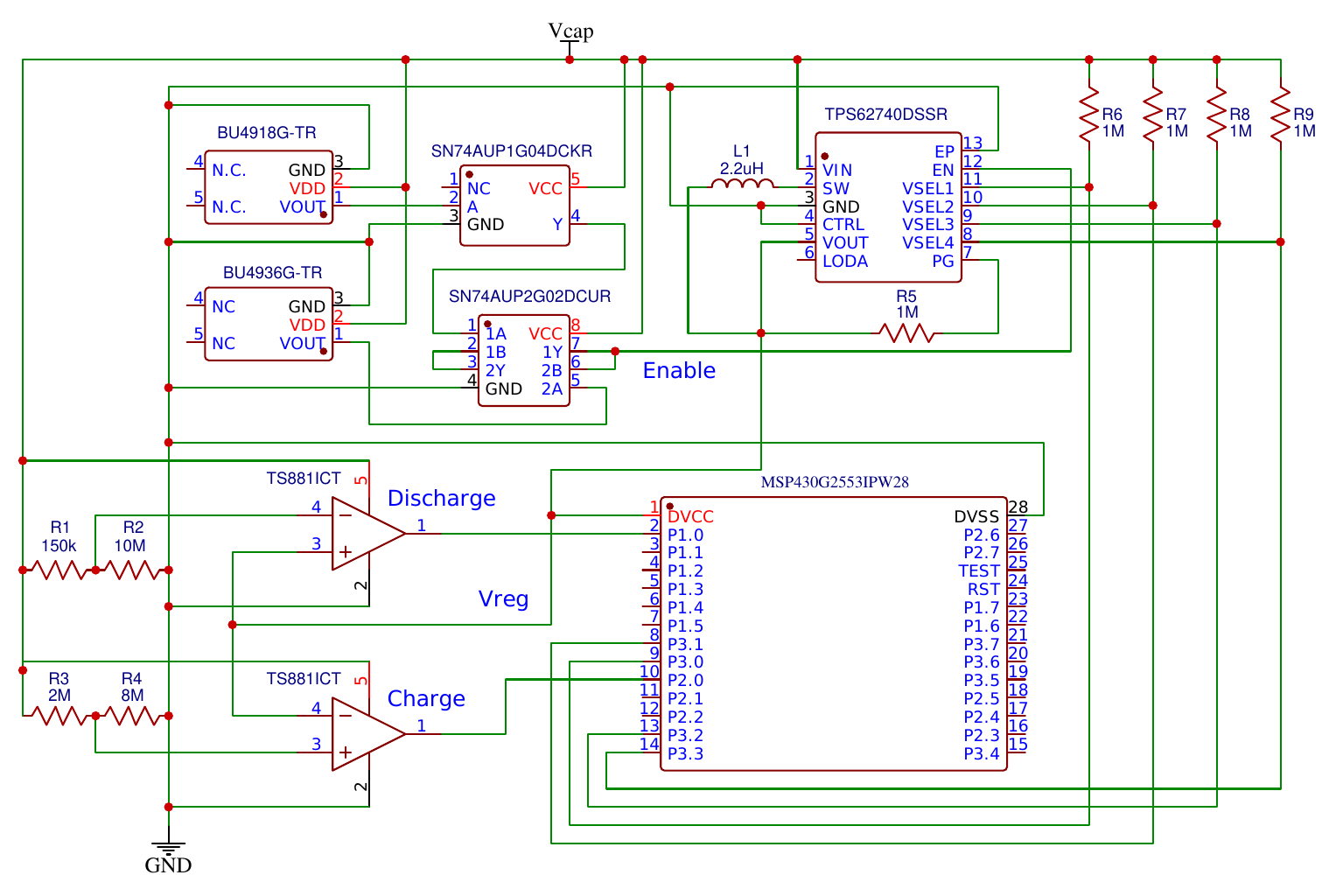}}
        \label{fig:fbtc_schematics}
    }
    \caption{\fbtc design.}
    \label{fig:fbtc}
\end{figure}

\subsection{FBTC}

\figref{fig:fbtc} shows the design of \fbtc. \figref{fig:fbtc_logic} illustrates the logic and \figref{fig:fbtc_schematics} shows the corresponding schematics.
Two macro components drive the functioning of \fbtc.
The \emph{Power State Controller} of \figref{fig:fbtc_logic} turns the system on whenever the capacitor voltage rises above a given boot threshold.
Unlike \dvfs, this threshold is hardware-configurable in \fbtc.
The \emph{Changepoint Detector}, instead, manages the detection of changes in the performance window.
We consider the same performance windows of \dvfs.

\begin{figure}[t]
    \centering
    \centering
    \parbox{0.395\columnwidth}{%
        \centering
        \vspace{3mm}
        \resizebox{0.395\columnwidth}{!}{\includegraphics[angle=90]{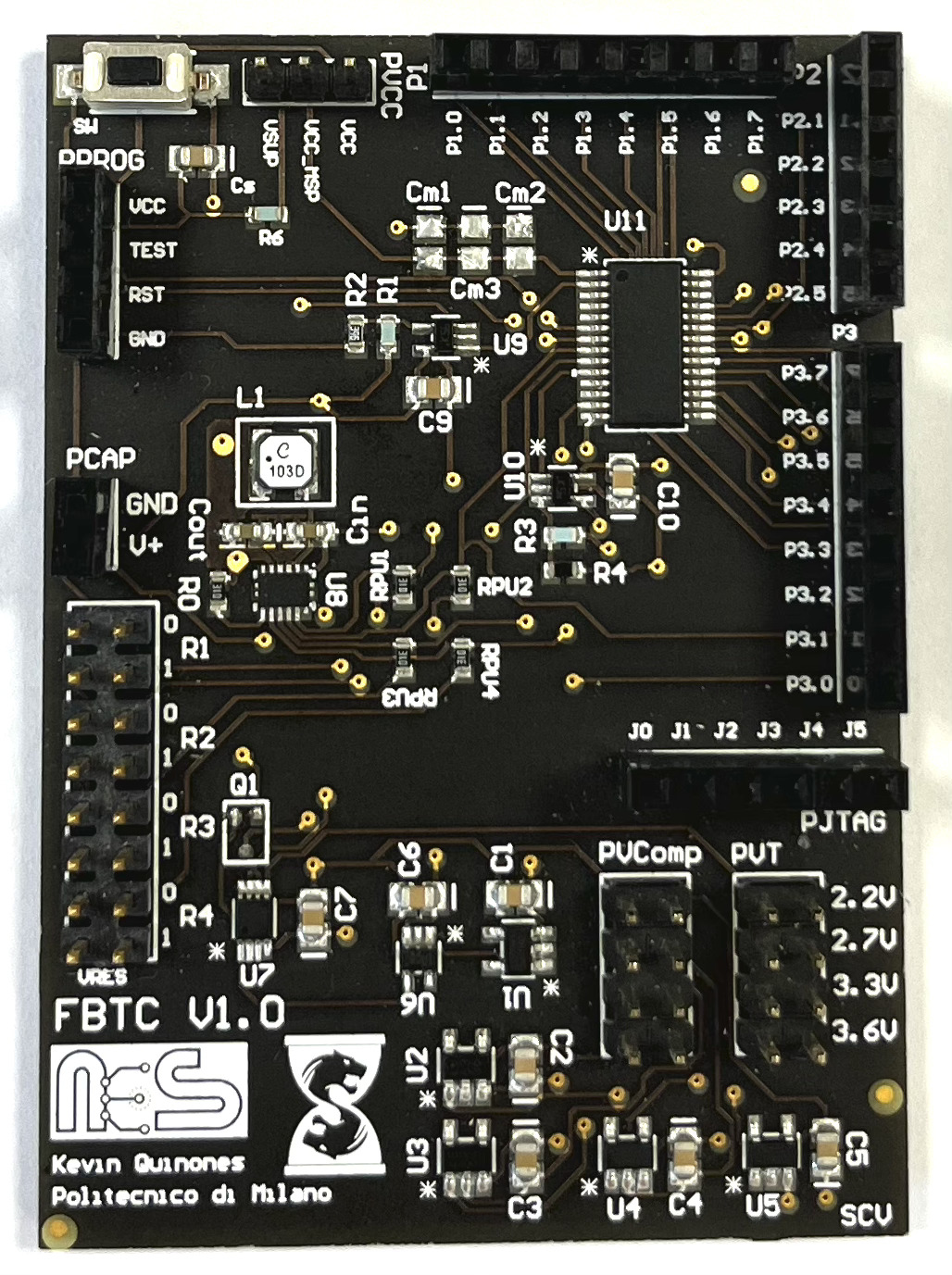}}
        \caption{\fbtc prototype.}
        \label{fig:fbtc_board}
    }%
    \qquad
    \parbox{0.55\columnwidth}{%
        \centering
        \hspace{6mm}
        \resizebox{0.45\columnwidth}{!}{\includegraphics{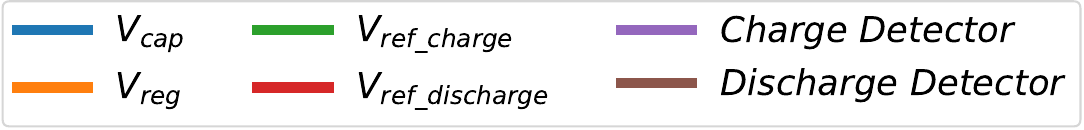}}
        \resizebox{0.55\columnwidth}{!}{\includegraphics{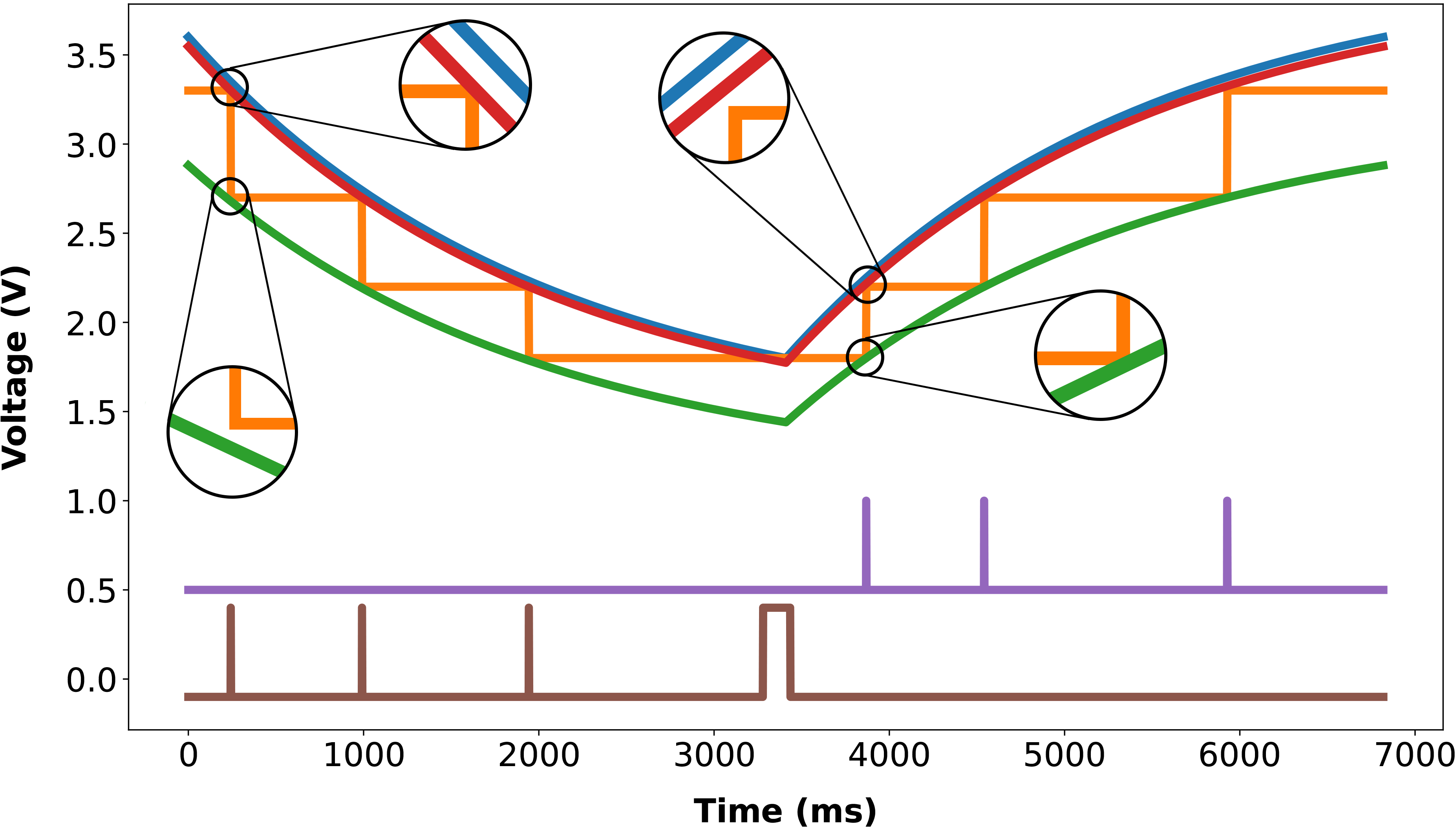}}
        \caption{Example of \fbtc behavior.}
        \label{fig:fbtc_behavior}
    }%
\end{figure}

\fakepar{Power state controller}
The \emph{Operating Range Detector}\ in \figref{fig:fbtc_logic} identifies if $V_{cap}$ is within the considered operating range.
It does so by relying on two BU49XXG~\cite{bu49xxg} voltage detectors, as shown in \figref{fig:fbtc_schematics}.
The first detector triggers when $V_{cap}$ reaches the MCU minimum operating voltage $V_{min} = 1.8V$, whereas the second detector triggers when $V_{cap}$ reaches the hardware-configurable boot threshold $V_{on}$.
Although \figref{fig:fbtc_schematics} shows a $3.6V$ setting for the second voltage detector, when fabricated, \fbtc allows users to select among four different voltage detectors to configure $V_{on}$, as indicated by the \textit{PVComp} and \textit{PVT} ports of \figref{fig:fbtc_board}.

The \emph{System Enable} function, as illustrated in \figref{fig:fbtc_logic}, determines the conditions to activate the system.
This operation utilizes a SN74AUP1G04~\cite{sn74aup1g04} NOT gate in conjunction with a SN74AUP2G02~\cite{sn74aup2g02} 2-input NOR gate, configured as a set-reset flip-flop, which is detailed in \figref{fig:fbtc_schematics}.
The NOT gate takes as input the signal of the first voltage detector, that is, the one identifying if $V_{cap}$ exceeds $V_{min}$.
The NOT gate thus verifies if $V_{cap}$ falls below $V_{min}$, resetting the flip-flop output.
Instead, the signal of the second voltage detector sets the flip-flop output.
When $V_{cap}$ exceeds the configured $V_{on}$, the flip-flop output is set to a logical high and the voltage regulator is powered on.
When $V_{cap}$ goes below $V_{min}$, the flip-flop output is reset to a logical low and the voltage regulator is powered off.

To initialize the output voltage of the voltage regulator at startup, we employ four pull-up resistors, designated as
$R6-R9$ in the schematic depicted in \figref{fig:fbtc_schematics} and as $R1-R4$ in the actual prototype shown in \figref{fig:fbtc_board}. 
This step is necessary because the voltage regulator's output is governed by the MCU, which is incapable of setting the output voltage until it has completed its startup sequence.

\fakepar{Changepoint detector}
Unlike \dvfs, \fbtc does not keep track of the current performance window in hardware; instead, it merely detects the conditions that trigger any change in the current performance window and whether this change is towards an upper or lower window.
This indication reaches the MCU through a hardware interrupt: by keeping track of the current performance window and by learning whether the change being detected is upwards or downwards, the MCU changes voltage and frequency settings.

The \emph{Interrupt Driver} of \figref{fig:fbtc_logic} provides this functionality through a \emph{Charge (Discharge) Detector} detecting upward (downward) changes in the performance window.
The two detectors are based on the same logic, which we accomplish with two components:
\begin{mylist}
    \item a voltage divider to reduce the $V_{cap}$ signal, that is, the $R1-R2$ ($R3-R4$) resistors of \figref{fig:fbtc_schematics} and
    \item a TS881~\cite{ts881} operational amplifier that compares $V_{cap}$ with the reduced signal.
\end{mylist}
To detect a discharge, the output of the voltage regulator $V_{reg}$ is connected to the non-inverting input of the operational amplifier and the reduced $V_{cap}$ signal is connected to the inverting input, as shown near the \textit{discharge} label of \figref{fig:fbtc_schematics}.
To detect the energy buffer charge, the connections to the operational amplifier are inverted.
We discuss later how to dimension $R1-R2$ and $R3-R4$, as well as the need for \emph{both} reference signals for charging and discharging.

\figref{fig:fbtc_behavior} shows an example execution.
The blue curve represents the original $V_{cap}$, whereas the orange one \mbox{represents $V_{reg}$}.
The signals representing the reference voltage for charging or discharging are $V_{ref\_charge}$ and $V_{ref\_discharge}$, shown in green and red, respectively.
Initially, the frequency is set to $16MHz$, $V_{reg}$ is set to $3.3V$, $V_{cap}$ is $3.6V$, and the capacitor is discharging.
When $V_{cap}$ reaches $3.3V$, the $V_{reg}$ signal, corresponding to the orange curve, exceeds the $V_{ref\_discharge}$ signal, corresponding to the red curve, as shown in \figref{fig:fbtc_behavior}.
The \emph{Discharge Detector} outputs a logical high, indicated with the brown line in \figref{fig:fbtc_behavior}, triggering an interrupt.
Knowing the current performance window and learning that a downward change is detected, the MCU switches to a configuration running at $12MHz$ with  $V_{reg}$ set to $2.8V$.

The same operations repeat throughout the discharge phase until the MCU switches to a configuration running at $1MHz$ with  $V_{reg}$ set to $1.8V$.
When $V_{cap}$ approaches $1.8V$, $V_{reg}$ constantly exceeds $V_{ref\_discharge}$.
This time there is no lower performance window to change to, as the MCU is already at $1MHz$ and $V_{reg}$ at $1.8V$.
To avoid unexpected behaviors, the software driver disables the interrupts from the \emph{Discharge Detector} when it sets the lowest possible performance window and enables them back whenever scaling upwards again.

The behavior when charging is dual: the \emph{Charge Detector} triggers an interrupt when $V_{reg}$ intersects $V_{ref\_charge}$.
Different than \dvfs, \fbtc need not to delay changes to the upper performance window when $V_{cap}$ increases, as the charge detector avoids bouncing between two adjacent performance windows by design, as detailed next.

\fakepar{Voltage divider configuration} The efficient operation of \fbtc rests on one key aspect: the dimensioning of $R1-R2$ and $R3-R4$.
Multiple reasons concur to this:

\begin{enumerate}
\item Properly setting the values of $R1-R2$ ensures that $V_{cap}$ never comes too close to $V_{reg}$, giving the MCU enough margin to trigger a switch to a lower performance window before $V_{cap} < V_{ref}$ for the current performance window. The $V_{ref\_discharge}$ signal exists precisely for this: if we were to compare directly $V_{cap}$ with $V_{reg}$, the time taken by the MCU to switch towards a lower performance widow would become (too) critical.
\item The reciprocal setting of $R1-R2$  and $R3-R4$ allows the system to avoid fluctuations between adjacent performance windows. For example, when switching to a lower performance window, $V_{reg}$ must not intersect $V_{ref\_charge}$, or the MCU would trigger an immediate switch back to the upper performance window. Otherwise, \fbtc may end up in a sort of livelock bouncing back and forth between adjacent performance windows.
\item By accurately tuning the $V_{ref\_charge}$ signal, that is, the values of $R3-R4$, we may ensure sufficient energy margin in the upper performance window to prevent an immediate downward transition.
This addresses the problem we discuss previously with \dvfs possibly bouncing between two adjacent performance windows when switching from a lower to an upper window.
\end{enumerate}

For a clearer illustration, we now describe the method for quantitatively determining the values for $R1-R2$, taking into account the considerations mentioned above.
The reasoning to ascertain the values for $R3-R4$ is entirely dual.
Based on the schematics of \figref{fig:fbtc_schematics}, the operational amplifiers inputs are:
\begin{equation}
    V_{ref\_discharge} = \frac{R2}{R1 + R2} \cdot V_{cap} = {\delta}_{d} \cdot V_{cap},
    \label{eq:discharge}
\end{equation}
\begin{equation}
    V_{ref\_charge} = \frac{R4}{R3 + R4} \cdot V_{cap} = {\delta}_{c} \cdot V_{cap},
    \label{eq:charge}
\end{equation}
where ${\delta}_{c}$ (${\delta}_{d}$) indicates the charge (discharge) voltage divider ratio.

Let the performance windows be ordered by ascending operating voltage and let $V_{reg}[i]$ be the voltage regulator output of the $i$-th performance window.
The interrupt signaling a change from the $i$-th performance window to the $i+1$-th performance window is triggered whenever
$V_{ref\_charge} > V_{reg}[i]$. \fbtc may, however, immediately bounce back to the $i$-th performance window if $V_{ref\_discharge} < V_{reg}[i+1]$.
In summary, we must avoid
\begin{equation}
    \textrm{when} \ \ V_{ref\_charge} > V_{reg}[i] \rightarrow V_{ref\_discharge} < V_{reg}[i+1]
    \label{eq:bouncing_a}
\end{equation}
that we can rewrite, based on \eqref{eq:discharge} and \eqref{eq:charge}, as
\begin{equation}
    \textrm{when} \ \ {\delta}_{c} \cdot V_{cap} > V_{reg}[i] \rightarrow {\delta}_{d} \cdot V_{cap} < V_{reg}[i+1]
    \label{eq:bouncing_a_ok}
\end{equation}

To avoid undesired bouncing behaviors, for any performance window $i$, \eqref{eq:bouncing_a} must never hold.
This means
\begin{equation}
    \textrm{when} \ \ {\delta}_{c} \cdot V_{cap} > V_{reg}[i] \rightarrow {\delta}_{d} \cdot V_{cap} \geq V_{reg}[i+1]
    \label{eq:constraint_a}
  \end{equation}
Say the operating range of the $i$-th performance window is ($V_{max}[i],$ $V_{min}[i]$).
To satisfy \eqref{eq:constraint_a} for any performance window $i$, we introduce a margin ${\epsilon}_{c}$ that represents the minimum voltage sensitivity we wish to obtain for the charge detector.
This means that, for a given performance window $i$, we substitute $V_{cap} = V_{cap} = V_{min}[i] + {\epsilon}_{c}\ $ as long as there exists a performance window $i-1$.

To reason quantitatively, consider the four performance windows of the MSP430-G2553~\cite{msp430g2553} introduced earlier:
\begin{enumerate}[label=\arabic*)]
    \item $1MHz$ with $V_{reg} = 1.8V$ and $V_{cap}$ in $(2.2V, 1.8V)$
    \item $8MHz$ with $V_{reg} = 2.2V$ and $V_{cap}$ in $(2.8V, 2.2V)$
    \item $12MHz$ with $V_{reg} = 2.8V$ and $V_{cap}$ in $(3.3V, 2.8V)$
    \item $16MHz$ with $V_{reg} = 3.3V$ and $V_{cap}$ in $(3.6V, 3.3V)$
\end{enumerate}
and assume ${\epsilon}_{c} = 50mV$. We return soon to how to determine ${\epsilon}_{c}$.

Consider now performance windows with $i = 1,2, 3$ and \eqref{eq:constraint_a}, obtaining the following constraints on ${\delta}_{d}$:
\begin{itemize}
    \item $V_{cap} = 2.20V + 50mV = 2.25V$, $V_{reg}[1] = 1.8V$, $V_{reg}[2] = 2.2V$ $\rightarrow$ ${\delta}_{d} \geq \frac{2.2V}{2.25V}$
    \item $V_{cap} = 2.80V + 50mV = 2.85V$, $V_{reg}[2] = 2.2V$, $V_{reg}[3] = 2.8V$ $\rightarrow$ ${\delta}_{d} \geq \frac{2.8V}{2.85V}$
    \item $V_{cap} = 3.30V + 50mV = 3.35V$, $V_{reg}[3] = 2.8V$, $V_{reg}[4] = 3.3V$ $\rightarrow$ ${\delta}_{d} \geq \frac{3.3V}{3.35V}$
\end{itemize}
These constraints collectively determine a lower bound for ${\delta}_{d}$.
To ensure all constraints are satisfied, we pick the highest value for ${\delta}_{d}$, that is, ${\delta}_{d} \geq \frac{3.3V}{3.35V} = 0.9851$.
Because ${\delta}_{d} = \frac{R2}{R1+R2}$, a possible selection is $R1 = 150k \Omega$ and $R2 = 10M \Omega$.

Determining the values for $R3-R4$ requires dual reasoning, where the resulting constraints identify an upper bound for ${\delta}_{c}$.
Therefore, we pick the lowest value for ${\delta}_{c}$, that is, ${\delta}_{c} \geq \frac{1.8V}{2.25V} = 0.8$.
Similarly to the previous case, we consider a margin ${\epsilon}_{d} = 50mV$ that represents the minimum voltage sensitivity we wish to obtain for the discharge detector.
Because ${\delta}_{c} = \frac{R4}{R3+R4}$, a possible selection is $R3 = 2M \Omega$ and $R4 = 8M \Omega$.

\fakepar{Selecting $\mathbf{{\epsilon}_{c}}$}
To prevent an immediate transition back to a lower performance window, we must ensure that the capacitor stores sufficient energy to sustain the computation in the upper performance window for a reasonable amount of time.
An extra voltage of ${\epsilon}_{c}$ in a capacitor corresponds to $\frac{1}{2} C {{\epsilon}_{c}}^2$ energy.
Say the maximum energy consumption per clock cycle is $e_{cc}$, the number of extra clock cycles $n_{clock\_cycles}$ that an extra voltage ${\epsilon}_{c}$ allows the MCU to execute is
\begin{equation}
    n_{clock\_cycles} = \frac{\frac{1}{2} C {{\epsilon}_{c}}^2}{e_{cc}}
    \label{eq:extra_instr}
\end{equation}

The software driver of \fbtc requires $18$ machine-code instructions to change the performance window, that is, $18$ clock cycles.
To justify switching to an upper performance window, we must satisfy
\begin{equation}
    n_{clock\_cycles} * p_{\mathit{lower}} \geq 18 + n_{clock\_cycles}
    \label{eq:n_instr}
\end{equation}
where $p_{lower}$ represents the energy consumption increase at a lower operating frequency compared to the higher one, sustained at the same voltage level.
For the MSP430-G2553~\cite{msp430g2553}, the average $p_{lower}$ between the three switching points, that is, $1MHz-8MHz$, $8MHz-12MHz$, and $12MHz-16MHz$ is $1.17$.
This means that switching to a higher frequency provides, on average, a $17\%$ better energy efficiency; hence $n_{instr} \geq 106$ clock cycles.

\fbtc sets the MCU to operate at the minimum possible voltage for each performance window.
To identify the highest energy consumption per clock cycle of the MCU, we consider the operating frequency with the highest energy consumption at the corresponding minimum operating voltage, that is, $16MHz$ with a $3.3V$ voltage supply, corresponding to $0.85 nJ$ energy consumption per clock cycle, as shown in \figref{fig:energy_per_cc}.
By substituting these values in \eqref{eq:extra_instr} and by considering a target capacitor of $100 \mu F$, ${\epsilon}_{c}$ must be at least $0.042V$.

%% file: evaluation.tex
\section{Evaluation}
\label{sec:eval}

We evaluate the performance of \dvfs and \fbtc under different system settings and energy harvesting scenarios.
We describe next the experiments and system setup, the considered energy scenarios, and the results of the experiments.

Our setup is designed to investigate a broad spectrum of energy conditions, ranging from energy-rich sources that prevent energy failures to energy-poor sources that result in frequent energy failures, with various intermediate scenarios in between.
Benchmarks comprise a diverse array of embedded programs, each exposing a variety of programming structures and workloads.
Our evaluation includes more than $500k$ data points.
Despite the extreme diversity of the setup and the quantity of experimental data at hand, the results allow us to conclude that:

\begin{enumerate}
  \item \fbtc and \dvfs significantly surpass all static configurations \emph{at both extremes}—with energy-rich or energy-poor sources—as their capacity to maximize the number of instructions executed per active cycle results in substantially reduced energy consumption and completion times;
  \item with setups lying between the two extremes, the performance of \fbtc and \dvfs is on par with the best-performing static configuration;
  \item The best performing static configuration \emph{differs} across setups; for instance, the static 16 $MHz$ configuration excels with an energy-rich source but turns into the least effective baseline with an energy-poor one;
  \item \fbtc outperforms \dvfs in diverse contexts with its energy-efficient design that diminishes external circuitry overhead, reducing energy use and quiescent current.
\end{enumerate}

Our primary conclusion from the above is that given the variable nature of ambient energy, \fbtc either significantly outperforms or matches static configurations in most scenarios.
Real-world deployments often show drastic changes in energy supply~\cite{batteryless-future, soil-termoelectric, bridge-sensor, traffic-flow-sensor, sensys20deployment, flicker, permadaq}, and may even be approximated to either of the two extremes we consider \emph{at different times} of the system lifetime.
Deploying \fbtc enables the system to adapt to prevailing energy conditions, maximizing the amount of useful work derived from a given energy budget.

\input{evaluation-setting}

\input{evaluation-fbtc-model.tex}

\input{evaluation-results-solar.tex}

\input{evaluation-results-rf.tex}

\input{evaluation-results-discharge.tex}

%% file: evaluation-setting.tex
\subsection{Setting}
\label{sec:evaluation-setting}

Accurately measuring the performance of \dvfs and \fbtc is a challenge per se.
Reproducing ambient energy sources is indeed extremely difficult, as their behavior is non-deterministic~\cite{EKHO, SIREN}.
We thus opt for software-based system emulation, as this not only enables fine-grained control of experiments but most importantly ensures reproducibility \emph{by us and others}.
The code, documentation, and datasets we use are publicly available~\cite{dvfs-sceptic}.

We describe next the experimental setting, the benchmarks we run, the baselines we compare with, and the energy environment that systems are exposed to.

\fakepar{Platform and emulation} We employ \sceptic~\cite{maioli21ewsn}, an extendable emulator for intermittent programs previously utilized in various studies~\cite{maioli21ewsn, maioli19lctes, ALFRED}
We extend \sceptic to emulate the functioning and energy consumption of the circuitry enabling \dvfs or \fbtc functionality.
We emulate ambient energy sources by replaying voltage traces~\cite{Mementos, EKHO, D2VFS} that are either synthetic or gathered from a real harvester.
Throughout program execution, \sceptic monitors the capacitor voltage, taking into account the total device energy consumption and harvested energy.
Whenever the capacitor voltage falls below a threshold, \sceptic emulates an energy failure.

We emulate the MSP430-G2553~\cite{msp430g2553} MCU from the MSP430 family~\cite{msp430family}, attached to a $8Kbyte$ MB85RC64V~\cite{MB85RC64V} non-volatile FRAM chip through $I^2C$ operating at $1MHz$.
We incorporate an energy model of the MCU into \sceptic, which considers the various operating modes, and leverages established experimental data~\cite{epic} to simulate active mode behavior.
Evidence exists that during active mode this MCU experiences fluctuations in power consumption that are not represented in its datasheet~\cite{epic}.
We instead rely on the latter~\cite{msp430g2553} to model its energy consumption in low-power mode as well as the energy consumption and latency of peripheral accesses.

We model the latency and energy consumption of the FRAM chip and of the additional components in \dvfs and \fbtc using a combination of datasheet information and real measures taken from the fabricated board for \fbtc.
To validate the accuracy of our model, we experimentally verify, for \fbtc, that the discharge patterns observed by relying on datasheet information mirror those of the fabricated board.
Further details about these aspects are available in \secref{sec:fbtc_model}.
We also note that the ADC minimum operating voltage is $2.2V$ on the MSP430-G2553 we consider.
Should $V_{cap}$ be lower than $2.2V$, the ADC may return unreliable values, causing unexpected system behaviors, including unnecessary state-save operations.
To account for this, we consider three possible settings for Mementos:
\begin{mylist}
  \item \default, where every function call performs a state-save operation as soon as $V_{cap}$ is lower than $2.2V$, yet the execution continues until $V_{cap}<1.8V$,
  \item \noadcoff, where we pretend the ADC can operate in the same voltage range of the MCU, and
  \item \twovmin, where we set the MCU to power off at $2.2V$
\end{mylist}.

We consider two well-established techniques to ensure forward progress: Hibernus~\cite{Hibernus} and Mementos~\cite{Mementos}.
Both save the program state on the FRAM chip, including the register file, special registers, and main memory, whenever $V_{cap}$ falls below a specified threshold $V_{save}$.
Hibernus relies on system interrupts that fire whenever the $V_{save}$ is reached;
Mementos relies on special function calls, statically placed at specific program locations, that probe $V_{cap}$ through the ADC and accordingly determine whether to save the state.
We use \sceptic itself to determine an efficient setting for $V_{save}$, empirically exploring different possible values and eventually settling on the one providing the best energy efficiency to complete a given workload.

For Hibernus, we consider an external voltage divider of $200 K \ohm$ as in the original setup~\cite{Hibernus} and we use \sceptic to model the execution of state-save operations whenever $V_{cap}$ falls below $V_{save}$.
For Mementos, we use the \textit{loop-latch} placement strategy~\cite{Mementos} to insert function calls in the source code that probe the value of $V_{cap}$ and compare it with $V_{save}$.
In line with the behavior of a real deployment, we also assume that Hibernus operations to save the system state only cover the \emph{used} portion of main memory, that is, the one delimited by the stack pointer, instead of the whole memory content~\cite{Hibernus} including unused segments.

\fakepar{Benchmarks, metrics, and baselines}
Battery-less devices usually run a periodic sense-process-transmit loop to gather data from the environment and relay that to a collection point~\cite{sensys20deployment}.
Sensing and data transmission employ external peripherals, such as sensors and radio transceivers; their performance is thus not a function of MCU behavior.
Therefore, we focus on benchmarks that represent processing, which execute entirely on the MCU.

We have chosen a suite of benchmarks that exemplify the diverse processing tasks typical in intermittent computing environments~\cite{alpaca, ratchet, clank, chain, Hibernus, Hibernus++, Mementos, QuickRecall, ALFRED}:
\begin{mylist}
  \item the Dijkstra algorithm for computing the shortest path between two nodes of a graph,
  \item a Fast Fourier Transform (FFT) for signal analysis, and
  \item the RSA for data encryption
\end{mylist}.
We consider the open-source implementation of each benchmark available in the MiBench2~\cite{MiBench2, MiBench} benchmark suite and we compile them using Clang~\cite{llvm} version $8.0.1$ with default compiler settings.

We prioritize the metrics of \textit{completion time}—the duration to finish a workload—and \textit{energy consumption}, which are directly influenced by the voltage and frequency adjustments in \dvfs and \fbtc.
We compare them against a baseline that uses static frequency configurations for the MSP430-G2553~\cite{msp430g2553}, including $1MHz$, $8MHz$, $12MHz$, and $16MHz$.

When quantifying the duration to complete a workload, we distinguish between \textit{execution time} for active periods and \textit{recharge time} for inactive periods.
This allows us to identify
\begin{mylist}
  \item whether performance is lost or gained in either or both of the phases,
  \item how different configurations of voltage and frequency affect the execution time, and
  \item how the external circuitry of \dvfs and \fbtc affect the recharge time
\end{mylist}.
This separation also allows us to identify how different voltage operational ranges affect performance, as different frequencies have different voltage ranges that affect both the execution and recharge time.

We also track the \textit{number of energy failures} occurring while completing a workload.
We consider this metric as an indicator showing how energy consumption affects performance.
Given the same initial energy budget, a higher energy consumption leads to shorter energy cycles and thus the system experiences more energy failures.
This increases both the execution and recharge time due to additional restore operations and capacitor recharges.

\begin{figure}[t]
    \subfigure[Solar (walking outdoor)]{
        \centering
        \resizebox{0.319\columnwidth}{!}{\includegraphics{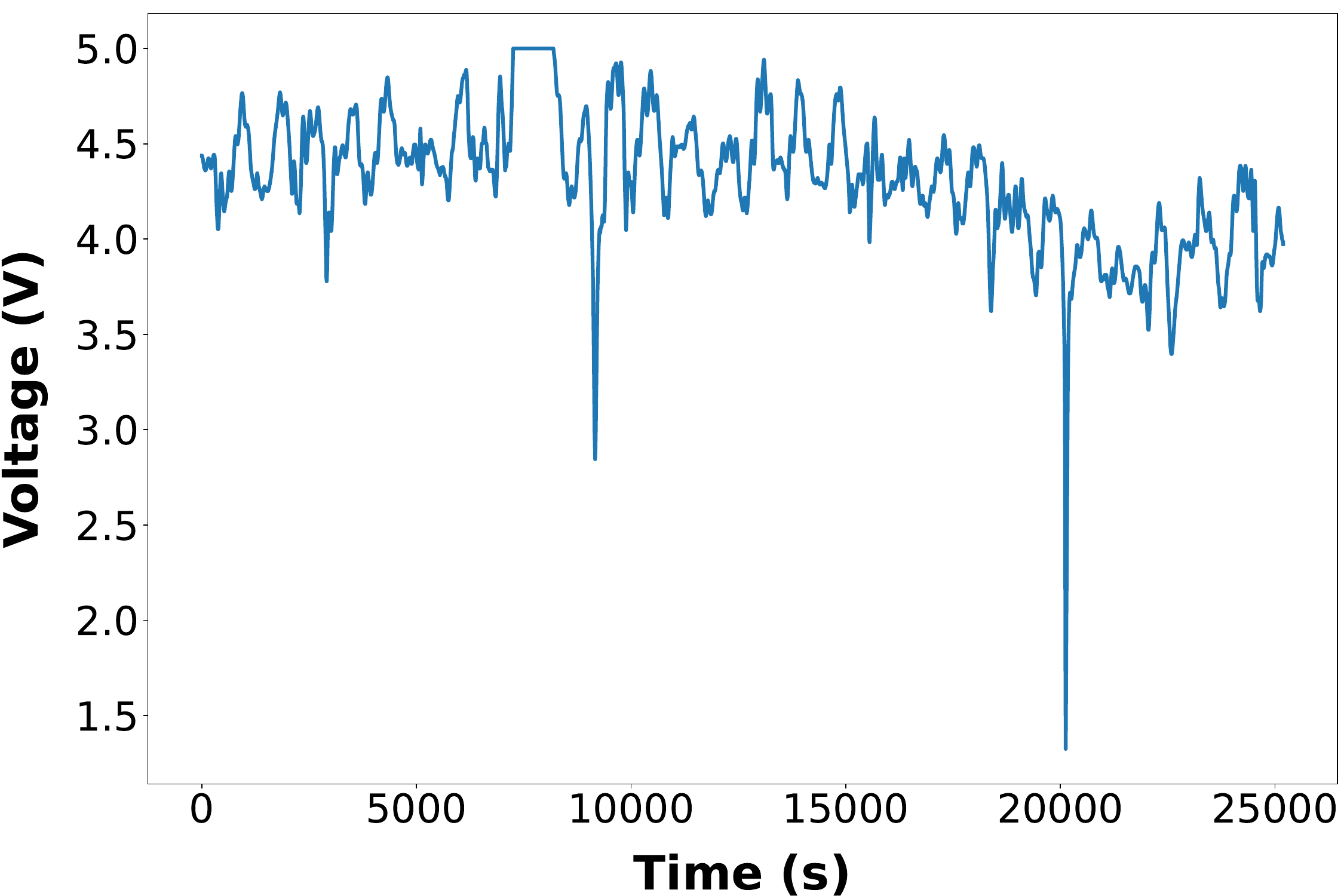}}
        \label{fig:energy_sources_solar}
    }
    \subfigure[Synthetic]{
        \centering
        \resizebox{0.319\columnwidth}{!}{\includegraphics{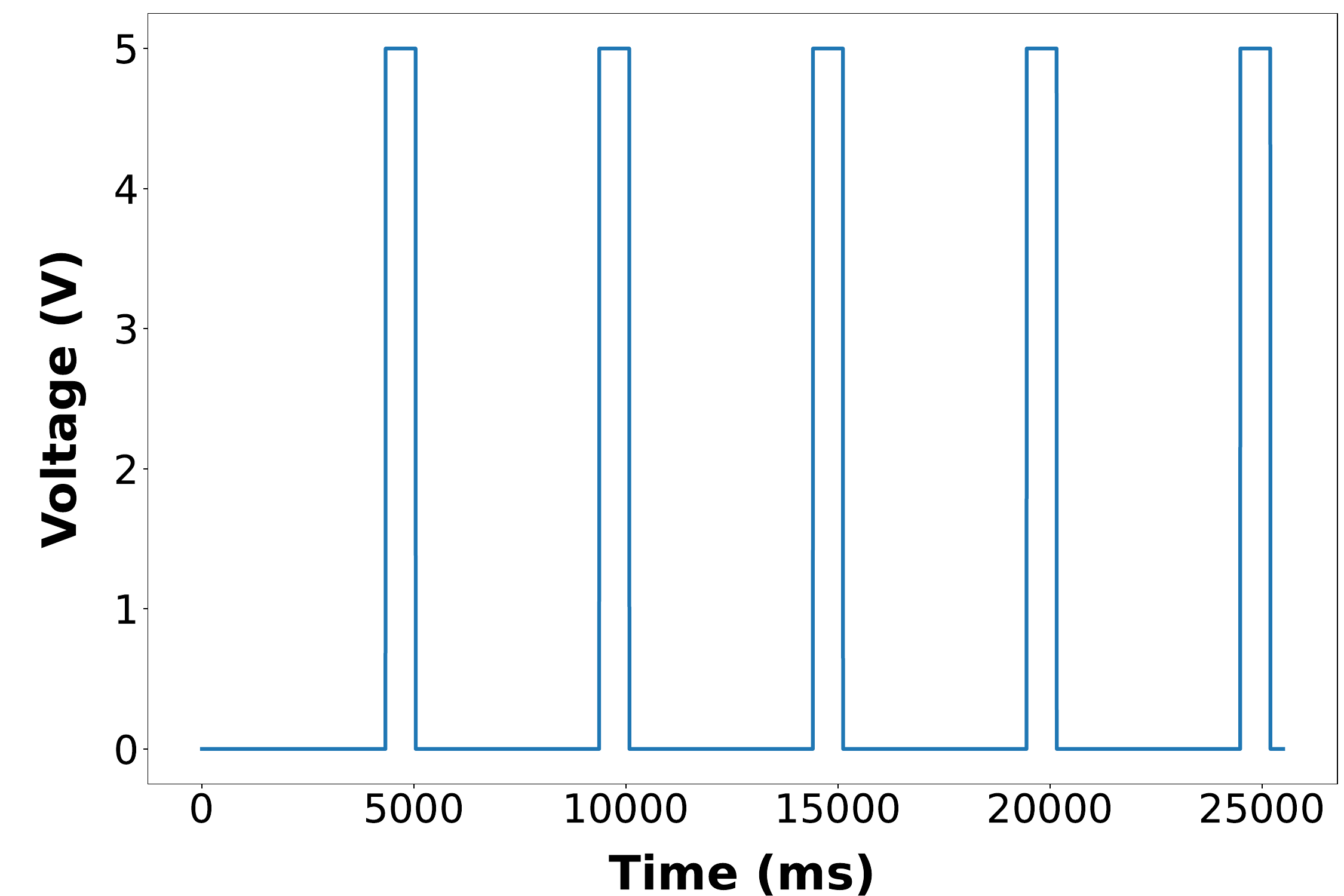}}
        \label{fig:energy_sources_discharge}
    }
    \subfigure[RF]{
        \centering
        \resizebox{0.319\columnwidth}{!}{\includegraphics{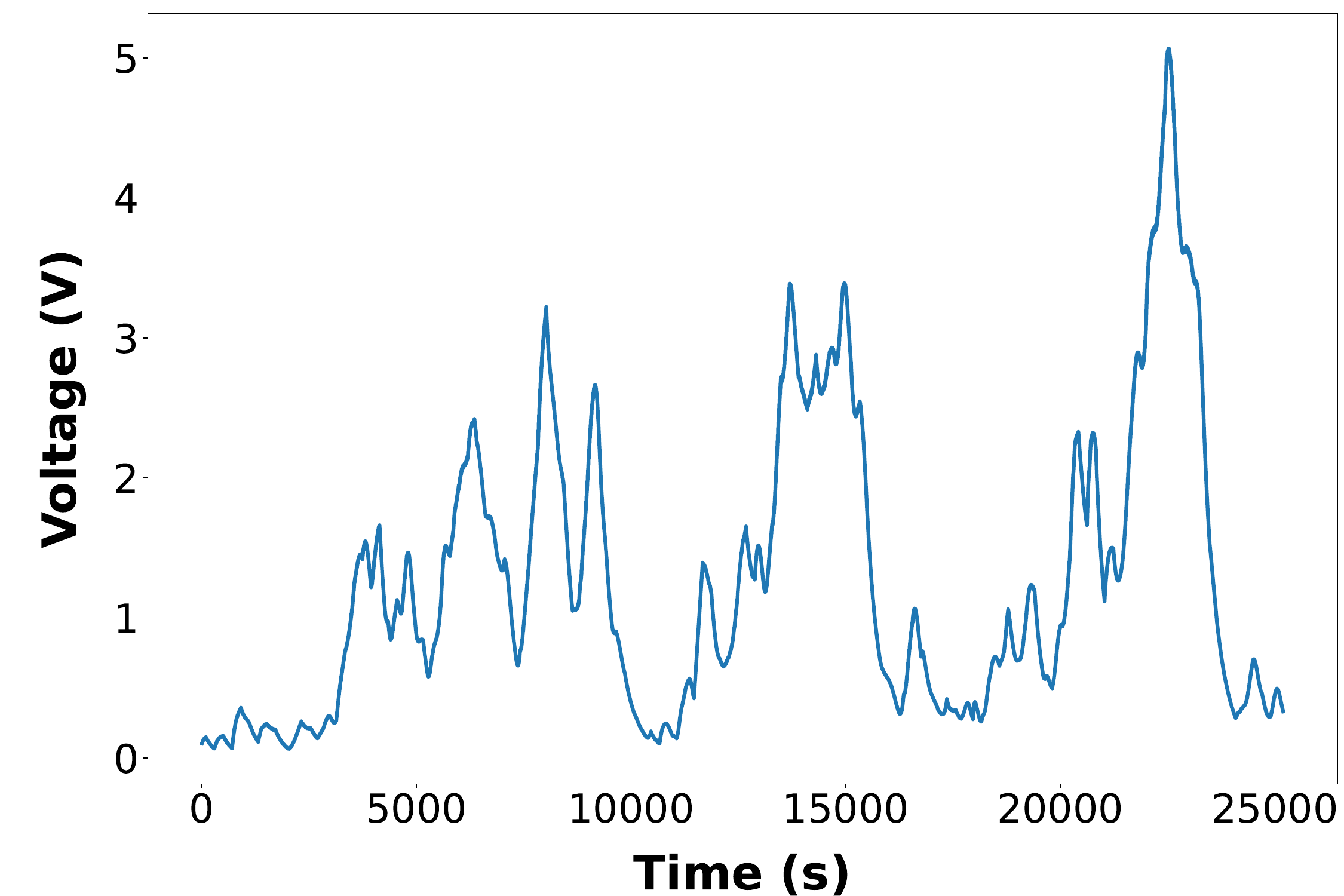}}
        \label{fig:energy_sources_rf}
    }
    \caption{Voltage traces of the considered energy sources.}
    \label{fig:energy_sources}
\end{figure}

\fakepar{Energy sources and system settings} The characteristics of the energy source largely determine the system's performance.
We investigate the system performance with three diverse energy sources.
\begin{enumerate}
\item An \textit{energy-rich} source, whose trace is shown in \figref{fig:energy_sources_solar}, which enables long energy cycles and yields a low energy failure rate.
We reproduce this scenario with the voltage trace of a solar energy source, measured from a solar panel outside our lab while walking~\cite{D2VFS}.
\item At the opposite extreme, we consider an \textit{energy-poor} source, whose trace is shown in \figref{fig:energy_sources_discharge}, which only produces short energy cycles and yields a high energy failure rate.
Similar to previous works~\cite{charge-scheduling}, we reproduce this scenario with a synthetic $5V$ energy source that supplies energy only when the device is powered off.
\item The \textit{energy-moderate} source, whose trace is found in \figref{fig:energy_sources_rf}, represents a middle point between the two extremes.
We reproduce this scenario by considering the voltage trace of an RF energy source, taken from Mementos~\cite{Mementos, Mementos-source-repo}.
\end{enumerate}

\begin{figure}[t]
    \subfigtopskip = -2pt
    \subfigure{
        \hspace{70pt}
        \resizebox{0.38\columnwidth}{!}{\includegraphics{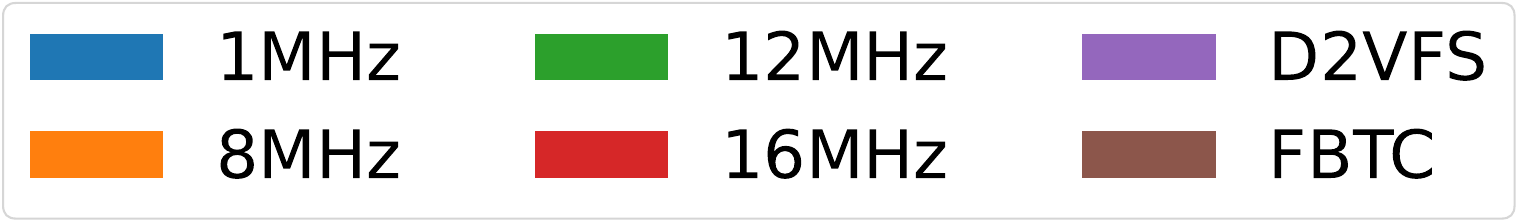}}
    }
    \newline
    \setcounter{subfigure}{0}
    \subfigure[Hibernus]{
        \centering
        \resizebox{0.35\columnwidth}{!}{\includegraphics{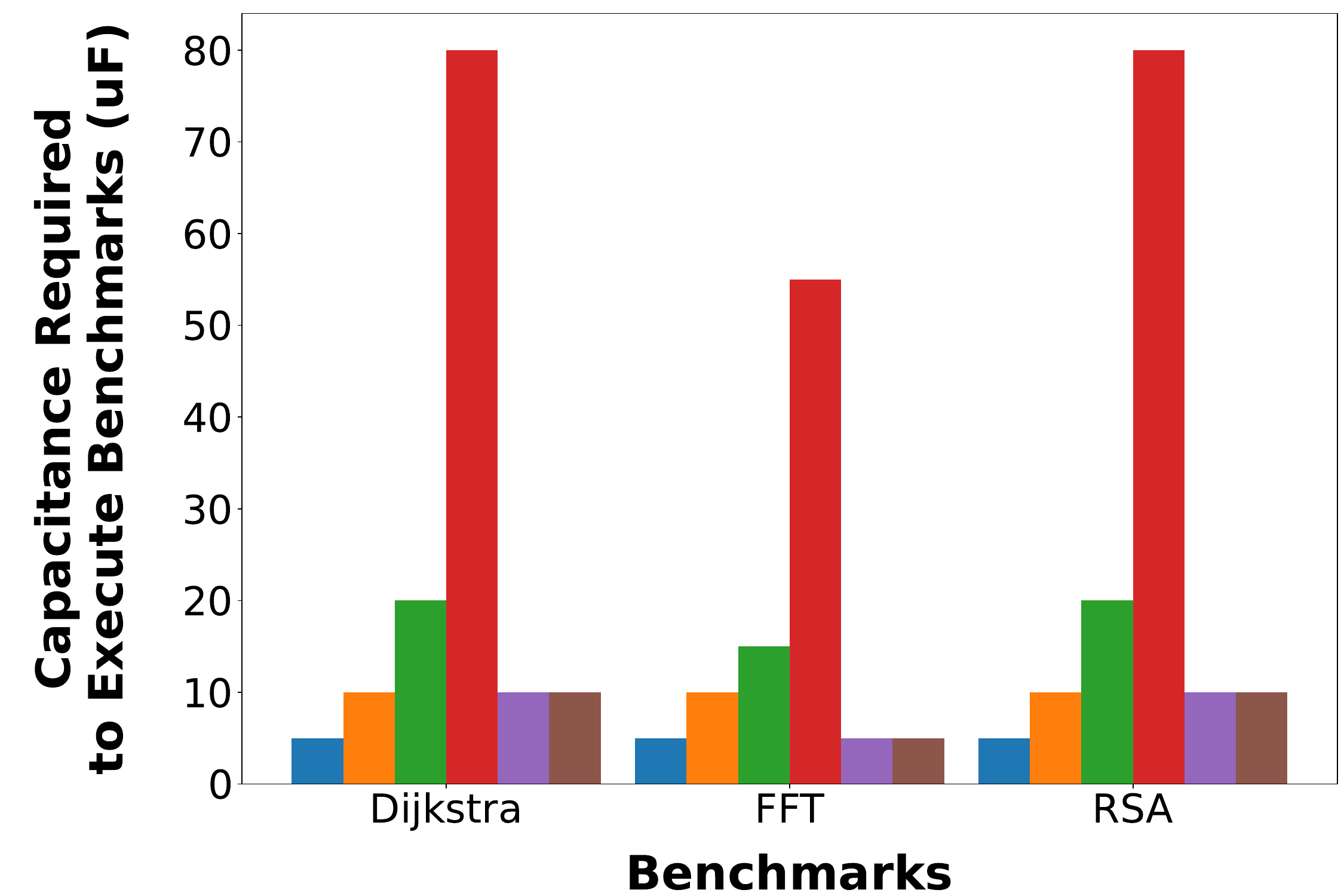}}
        \label{fig:min_cap_hibernus}
    }
    \subfigure[Mementos]{
        \centering
        \resizebox{0.35\columnwidth}{!}{\includegraphics{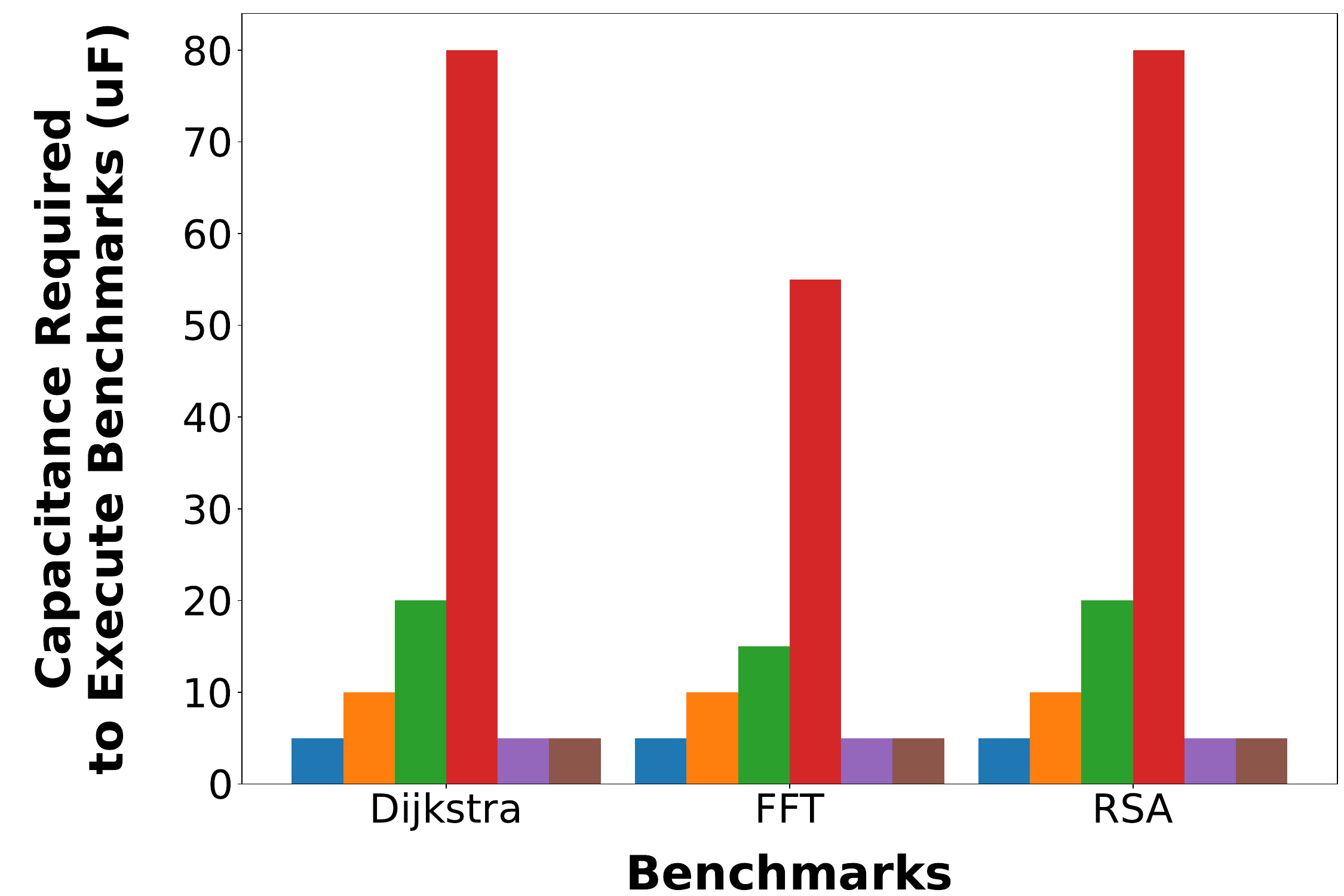}}
        \label{fig:min_cap_mementos}
    }
    \caption{Minimum capacitance required to execute benchmarks at a given frequency.}
    \label{fig:min_cap}
\end{figure}

Capacitor size $C$ and boot threshold $V_{boot}$ determine the length of energy cycles and the time required to recharge after an energy failure.
Large capacitors increase the duration of an energy cycle, as they store more energy, yet they also increase the time to \mbox{reach $V_{boot}$}.
Similarly, a high $V_{boot}$ extends the duration of an energy cycle by providing a larger initial energy budget, but it also increases the recharge time.
There also exist lower bounds for $C$ and $V_{boot}$, depending on frequency setting and workload.
Their setting determines the energy available in an energy cycle, which we call $e_{active}$, which must be strictly larger than the sum of the energy consumed by state-save and state-restore operations.
Otherwise, a device would not achieve forward progress across energy failures.

To evaluate the performance of \dvfs and \fbtc under different conditions, we consider multiple combinations of lower bounds for $C$ and $V_{boot}$.
We use \sceptic to determine these settings, running repeated experiments to measure the performance of the various possible configurations.
\figref{fig:min_cap} shows the lower bound for $C$ for the systems we consider and across all benchmarks and system support configurations.
The execution of benchmarks at a static frequency of $16MHz$ or $12MHz$ requires at least a $80 \mu F$ or $20 \mu F$ capacitor, respectively.
Instead, the static setting at $1MHz$ or $8MHz$, \dvfs, and \fbtc require no more than a $10 \mu F$ capacitor, that is, the minimum decoupling capacitance of the MSP430-G2553 suggested by TI~\cite{msp430-hw-design-tips}.

\begin{figure}[t]
    \subfigtopskip = -2pt
    \subfigure{
        \hspace{70pt}
        \resizebox{0.35\columnwidth}{!}{\includegraphics{figures/evaluation/min_cap/legend.pdf}}
        \vspace{2mm}
    }
    \newline
    \setcounter{subfigure}{0}
    \subfigure[Hibernus - Dijkstra]{
        \centering
        \resizebox{0.319\columnwidth}{!}{\includegraphics{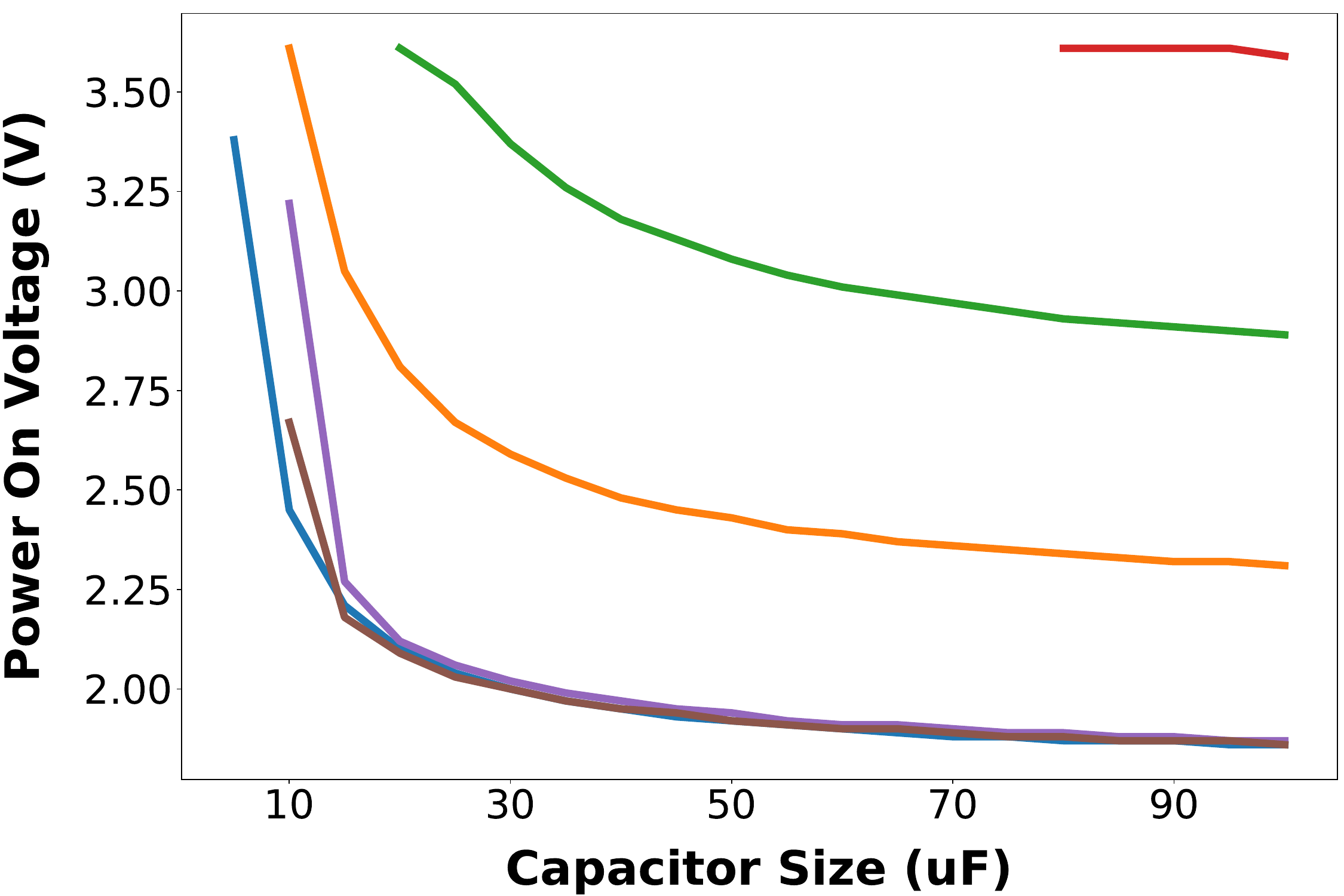}}
        \label{fig:min_v_on_hibernus_dijkstra}
    }
    \subfigure[Mementos - Dijkstra]{
        \centering
        \resizebox{0.319\columnwidth}{!}{\includegraphics{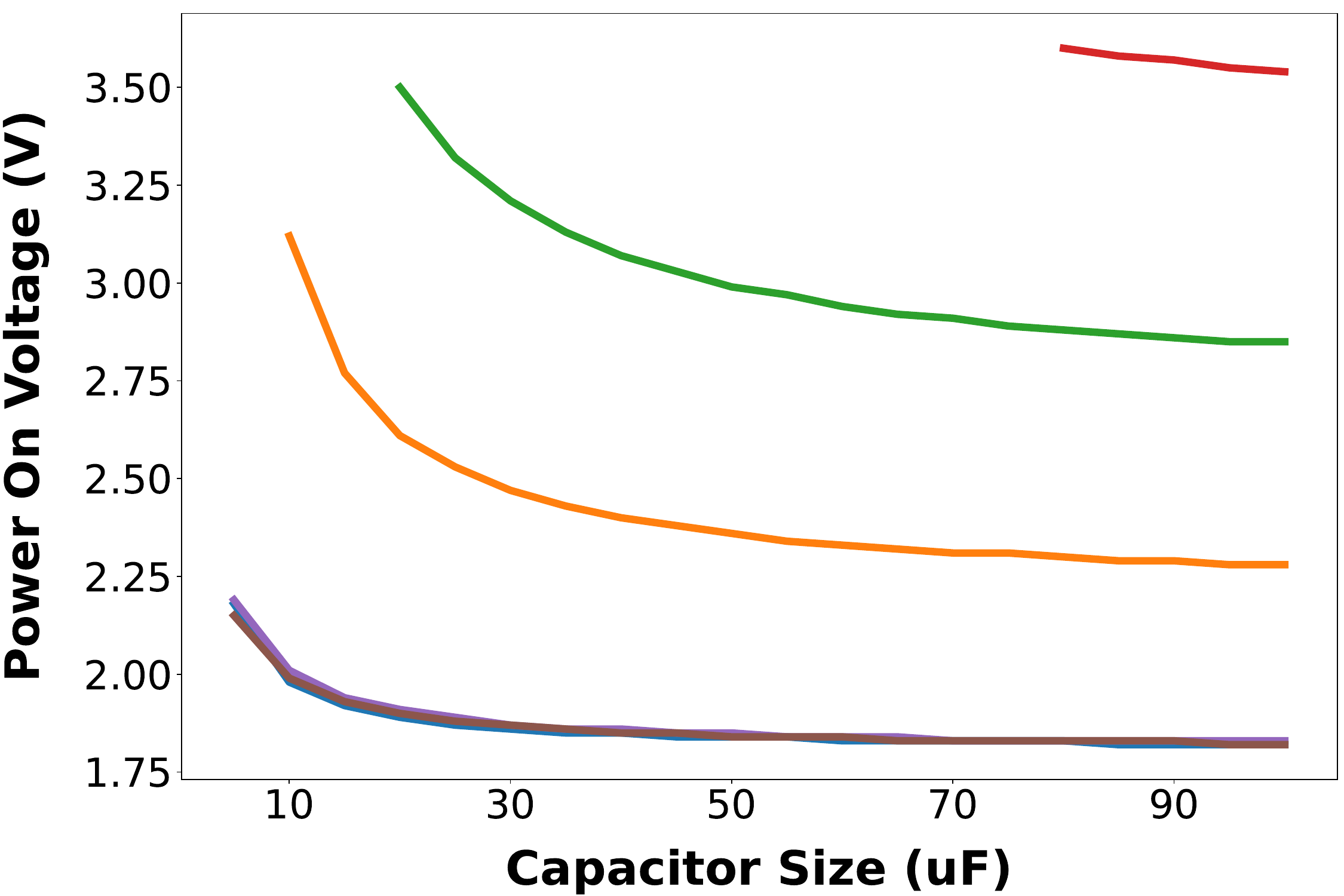}}
        \label{fig:min_v_on_mementos_dijkstra}
    }
    \subfigure[Hibernus - FFT]{
        \centering
        \resizebox{0.319\columnwidth}{!}{\includegraphics{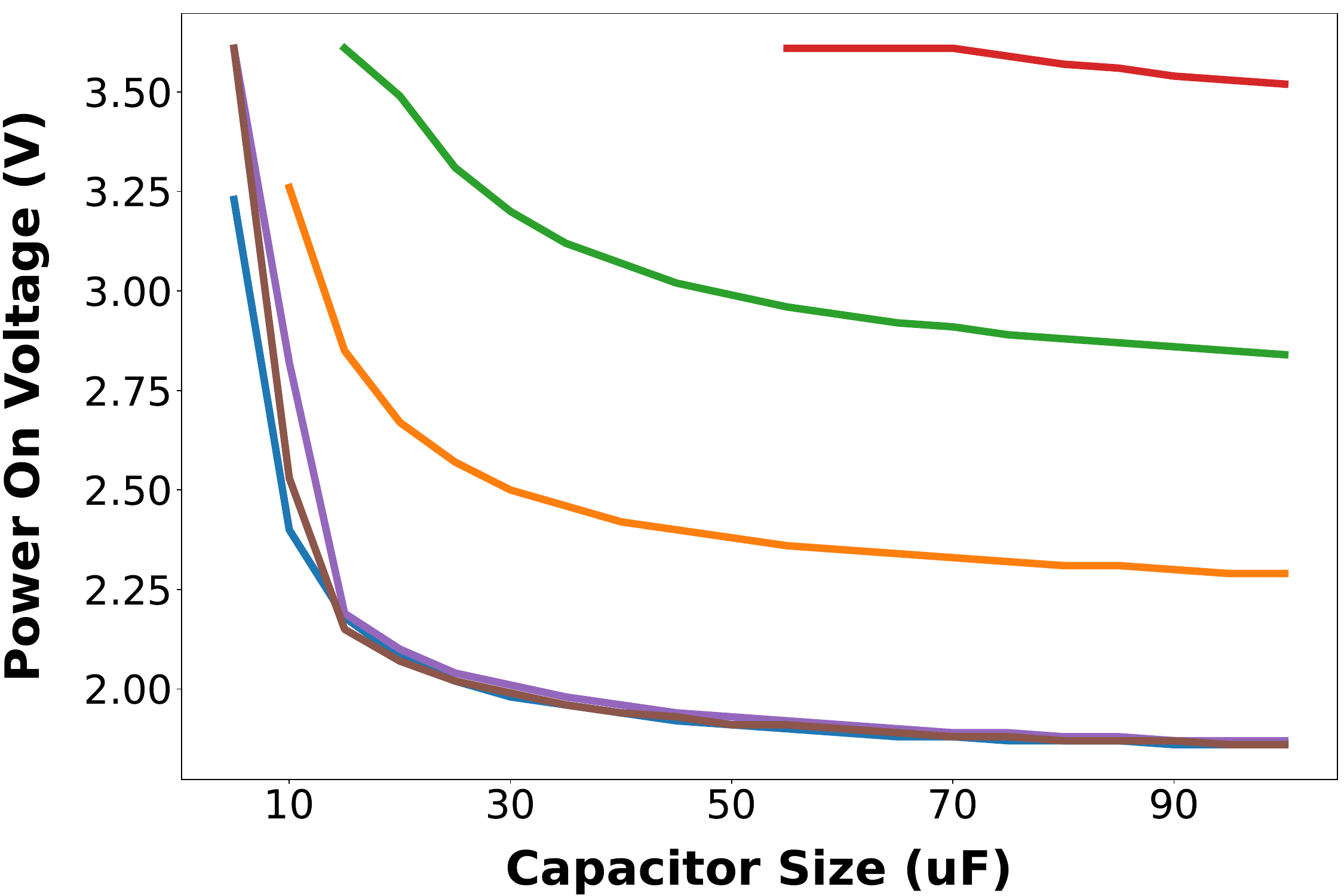}}
        \label{fig:min_v_on_hibernus_fft}
    }
    \subfigure[Mementos - FFT]{
        \centering
        \resizebox{0.319\columnwidth}{!}{\includegraphics{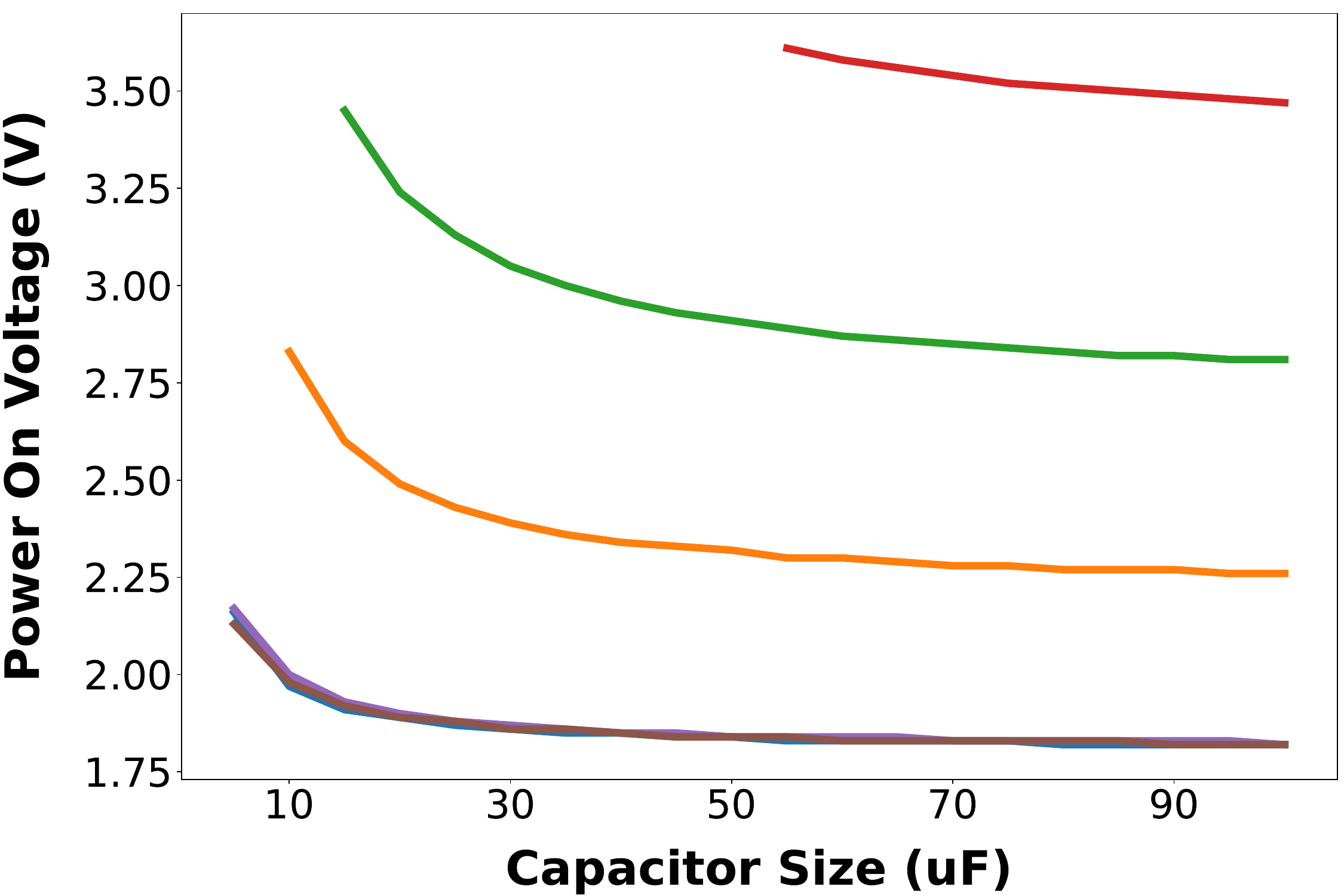}}
        \label{fig:min_v_on_mementos_fft}
    }
    \subfigure[Hibernus - RSA]{
        \centering
        \resizebox{0.319\columnwidth}{!}{\includegraphics{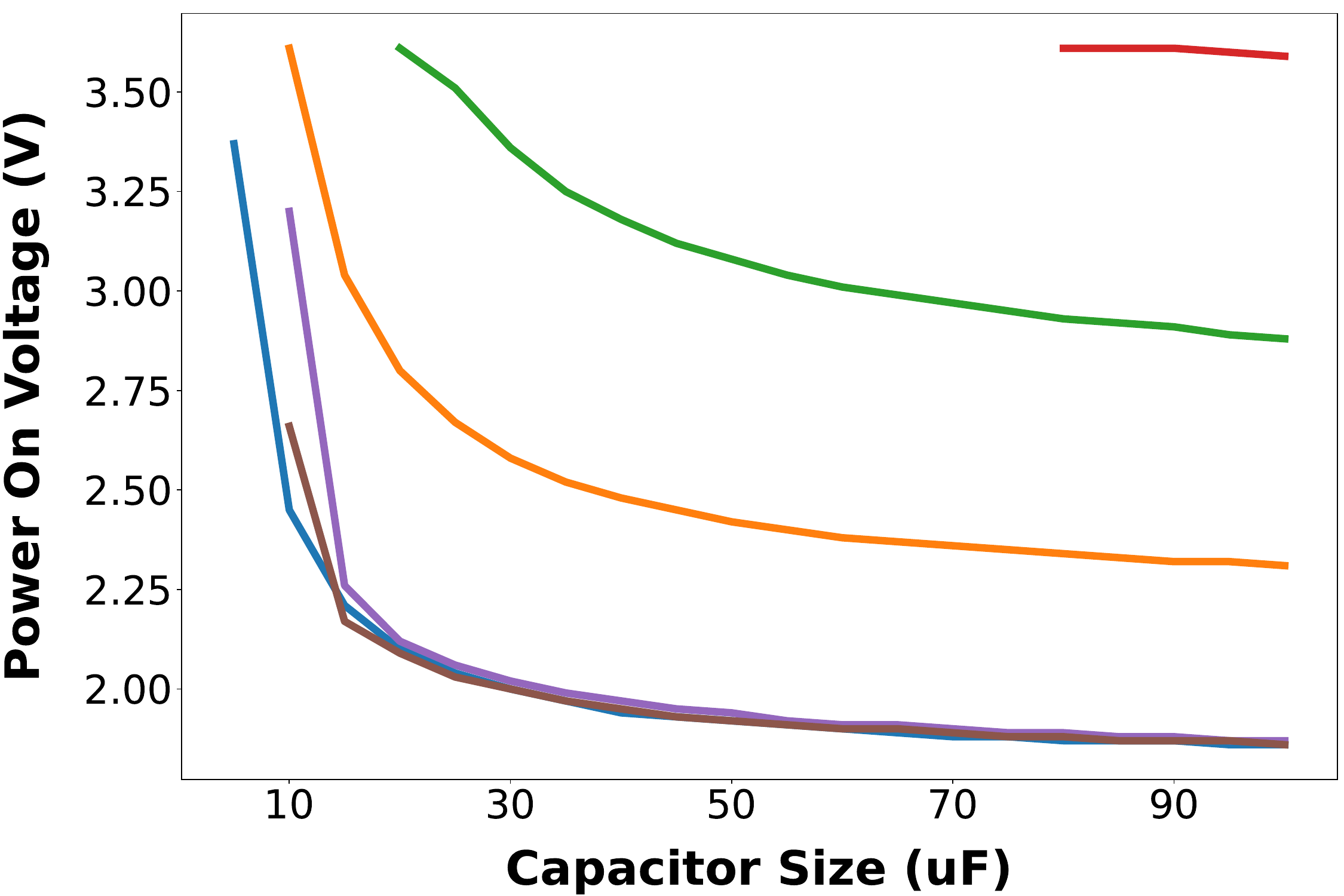}}
        \label{fig:min_v_on_hibernus_rsa}
    }
    \subfigure[Mementos - RSA]{
        \centering
        \resizebox{0.319\columnwidth}{!}{\includegraphics{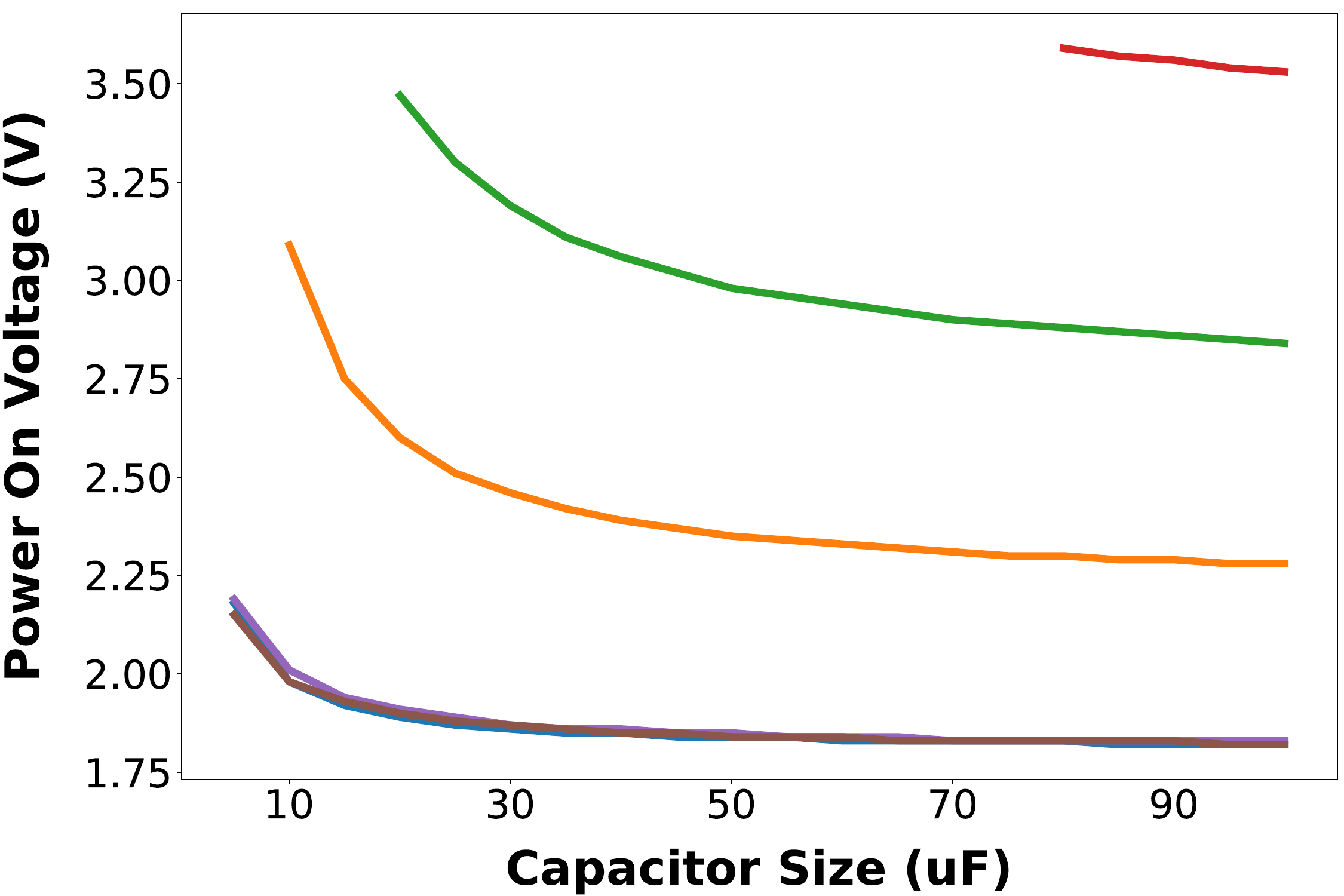}}
        \label{fig:min_v_on_mementos_rsa}
    }
    \caption{Minimum $V_{boot}$ required for benchmark execution.}
    \label{fig:min_v_on}
\end{figure}

Based on these results, we use two capacitor sizes:
\begin{mylist}
  \item $80 \mu F$ to run experiments for all baselines and settings, and
  \item $20 \mu F$ to run experiments using all baselines except $16MHz$
\end{mylist}.
Then, we identify the minimum $V_{boot}$ for each possible capacitor size.
\figref{fig:min_v_on} shows the $V_{boot}$ setting across benchmarks and capacitor sizes.
In general, the trend is consistent with the voltage operating range at a given frequency: the $16MHz$ configuration has the highest $V_{boot}$, whereas the $1MHz$ configuration has the lowest.
Note that the curves for \dvfs and \fbtc closely align with that of the $1MHz$ configuration, due to their similar voltage operating ranges..

\fakepar{Quiescent current}
Our models in \sceptic account for the quiescent current $I_{quiescent}$ due to external circuitry,  which causes the capacitor to discharge even when the MCU is off.
This applies to Hibernus~\cite{Hibernus}, \dvfs, and \fbtc.
Note that we ignore the capacitor leakage current, as it is negligible compared to $I_{quiescent}$.
Due to $I_{quiescent}$ and depending on the other system parameters, the energy source may be unable to make the system eventually reach $V_{boot}$, potentially leading to a scenario where the device never powers on.
This is the case of the \emph{energy-poor} source with $C = 100 \mu F$ and $V_{boot} = 3.6V$.
The short energy bursts rarely exceed the capacitor voltage $V_{cap}$, and contribute no additional charge.

To address this issue, we integrate into \sceptic a model of a voltage doubler between the energy harvester and the capacitor, as used in the WISP platform~\cite{wisp1, wisp6}.
Using the voltage doubler, energy bursts exceeding $V_{cap}$ are both more frequent and longer, allowing the capacitor to eventually reach $V_{boot}$ despite the influence of $I_{quiescent}$.
This addition is unnecessary for the \emph{energy-rich} and \emph{energy-moderate} sources, but mandatory for the \emph{energy-poor} one when using $20 \mu F$ capacitors.
Using a voltage doubler may not always be an option, because
\begin{mylist}
    \item voltage doublers usually require AC input currents~\cite{voltage-doubler}, whereas an energy harvester may output DC current~\cite{harvesting-survey}, and
    \item similarly to voltage regulators, voltage doublers never have a $100\%$ efficiency~\cite{voltage-doubler} and thus waste energy
\end{mylist}.

%% file: evaluation-fbtc-model.tex
\subsection{Energy Model Validation}
\label{sec:fbtc_model}

\begin{figure}[t]
    \subfigtopskip = -2pt
    \subfigure{
        \centering
        \hspace{70pt}
        \resizebox{0.35\columnwidth}{!}{\includegraphics{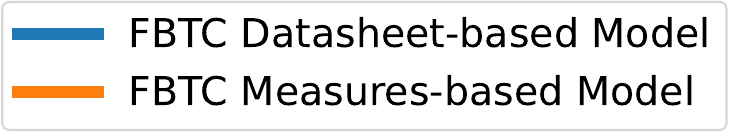}}
    }
    \newline
    \setcounter{subfigure}{0}
    \subfigure[Energy per cycle]{
        \centering
        \resizebox{0.48\columnwidth}{!}{\includegraphics{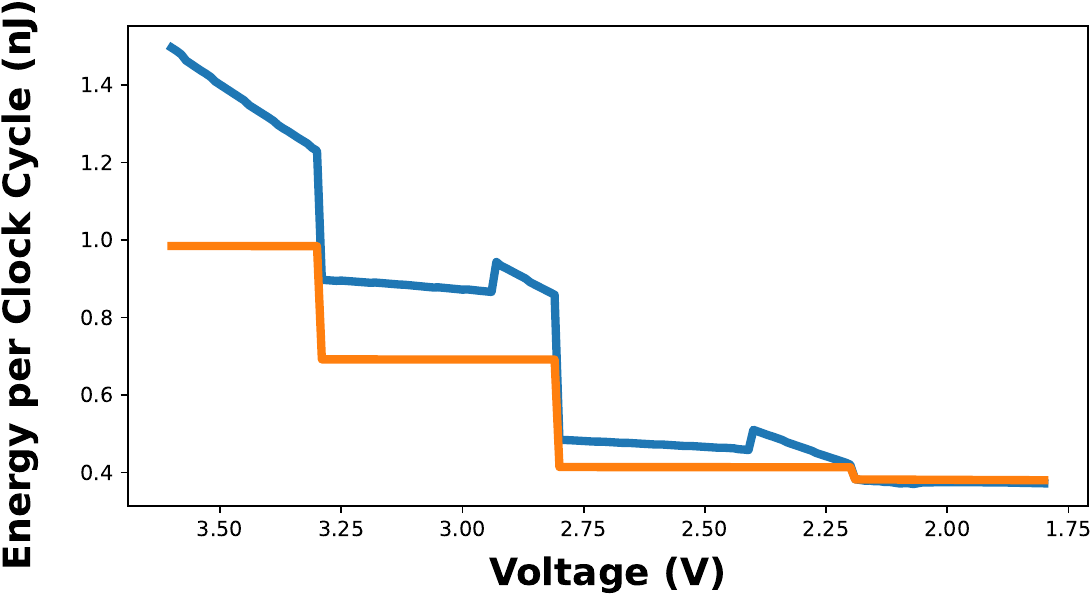}}
        \label{fig:fbtc_real_vs_model_energy_cc}
    }
    \subfigure[Capacitor discharge]{
        \centering
        \resizebox{0.48\columnwidth}{!}{\includegraphics{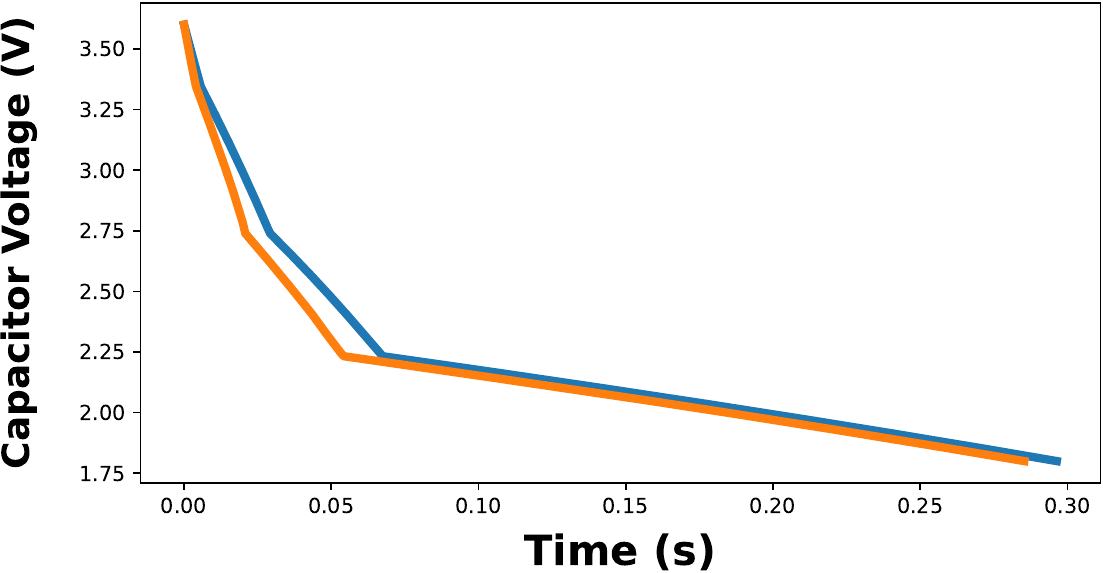}}
        \label{fig:fbtc_real_vs_model_discharge}
    }
    \caption{Comparison of \fbtc datasheet-based model against \fbtc measures-based model.}
    \label{fig:fbtc_real_vs_model}
\end{figure}

We model \dvfs and \fbtc energy consumption using real measures of the MSP430-G2553~\cite{msp430g2553} MCU and the datasheet information for the various circuitry components of \dvfs and \fbtc.
To validate the model, we measure the energy consumption of the \fbtc board we fabricated.
We use a PeakTech 6225A~\cite{peaktech-psu} variable power supply to vary the voltage of the \fbtc board between $3.6V$ and the minimum operating voltage for the considered clock frequency, using steps of $0.01V$.
We measure the \fbtc board current draw using a UNI-T UT61E multimeter~\cite{unit-multimeter}.
We repeat the measures for each operating frequency we consider, namely, $16MHz$, $12MHz$, $8MHz$, and $1MHz$.

\figref{fig:fbtc_real_vs_model} compares \fbtc datasheet-based model against the fabricated \fbtc board.
\figref{fig:fbtc_real_vs_model_energy_cc} compares the energy consumption per clock cycle of the datasheet-based \fbtc model against our measures.
Our model considers an average efficiency of $90\%$ for the TPS62740~\cite{tps62740} voltage regulator~\cite{tps62740}.
However, this does not represent the actual behavior of the voltage regulator.
The measures of \figref{fig:fbtc_real_vs_model_energy_cc} show that the voltage regulator has a non-linear behavior and its efficiency depends on the input/output voltages.
In particular, between $3.6V$ and $3.3V$, that is, the operating voltage range of the $16MHz$ configuration, our model underestimates the energy consumption by up to $50\%$ and, on average, by $38\%$.
This discrepancy decreases down to $34\%$ ($23\%$) in the voltage range associated to $12MHz$ ($8MHz$), that is, between $3.3V$ ($2.8V$) and $2.8V$ ($2.2V$), with an average underestimation of $28\%$ ($13\%$).
Conversely, between $2.2V$ and $1.8V$, that is, the voltage range associated to the $1MHz$ configuration, our model overestimates the energy consumption by up to $2\%$.

To evaluate the impact of these inaccuracies, we compare the workload
achieved in a single discharge of a $100 \mu F$ capacitor between the
fabricated board the the \fbtc model.
The lower energy consumption of the model results in the execution of $16\%$ more clock cycles.
Interestingly, the capacitor discharge time depicted in \figref{fig:fbtc_real_vs_model_discharge} shows an interesting behavior.
The significant difference in the energy estimation between $3.6V$ and $3.3V$ barely affects the discharge time.
The overall difference between the discharge times is only $4\%$, which is mainly caused by the differences in the energy estimation between $3.3V$ and $2.2V$.
This is due to the non-linear relation between the capacitor voltage and the capacitor energy, which makes the MCU sustain lower frequencies for longer periods.
Consequently, the discrepancy in the energy estimation of higher
frequencies bears a very limited impact.

For these reasons, despite the energy estimation difference, there is
essentially no difference in the performance trend of the \fbtc models against static frequencies and \dvfs across our experiments.
Therefore, the results we report next are obtained using the
datasheet-based \fbtc model, making the results also comparable with
those of \dvfs and enabling a per-component analysis of the \fbtc
energy consumption, which would be unfeasible otherwise.

%% file: evaluation-results-solar.tex
\subsection{Results $\rightarrow$ Energy-rich Source}
\label{sec:results-rich}

Experiments with the energy-rich source experience no energy failures, as sufficient energy is available to complete the workload in a single energy cycle in any configuration.
Thus, we do not report on the number of energy failures and the recharge times.
Similarly, we do not report on the execution time, as it corresponds to the completion time.
In these experiments, the energy source always keeps the capacitor at its maximum voltage, independently of size.
We discuss only the experiments with a $80 \mu F$ capacitor, as the $20 \mu F$ capacitor produces the same results.

\begin{figure}[t]
    \subfigtopskip = -2pt
    \subfigure{
        \centering
        \hspace{70pt}
        \resizebox{0.35\columnwidth}{!}{\includegraphics{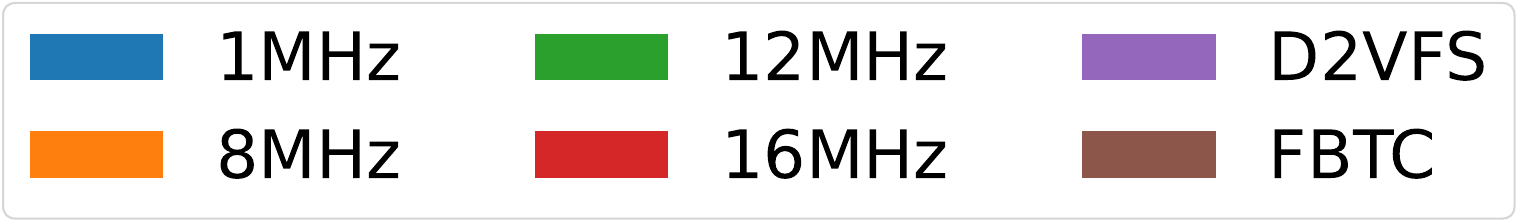}}
    }
    \newline
    \setcounter{subfigure}{0}
    \subfigure[Completion time]{
        \centering
        \resizebox{0.319\columnwidth}{!}{\includegraphics{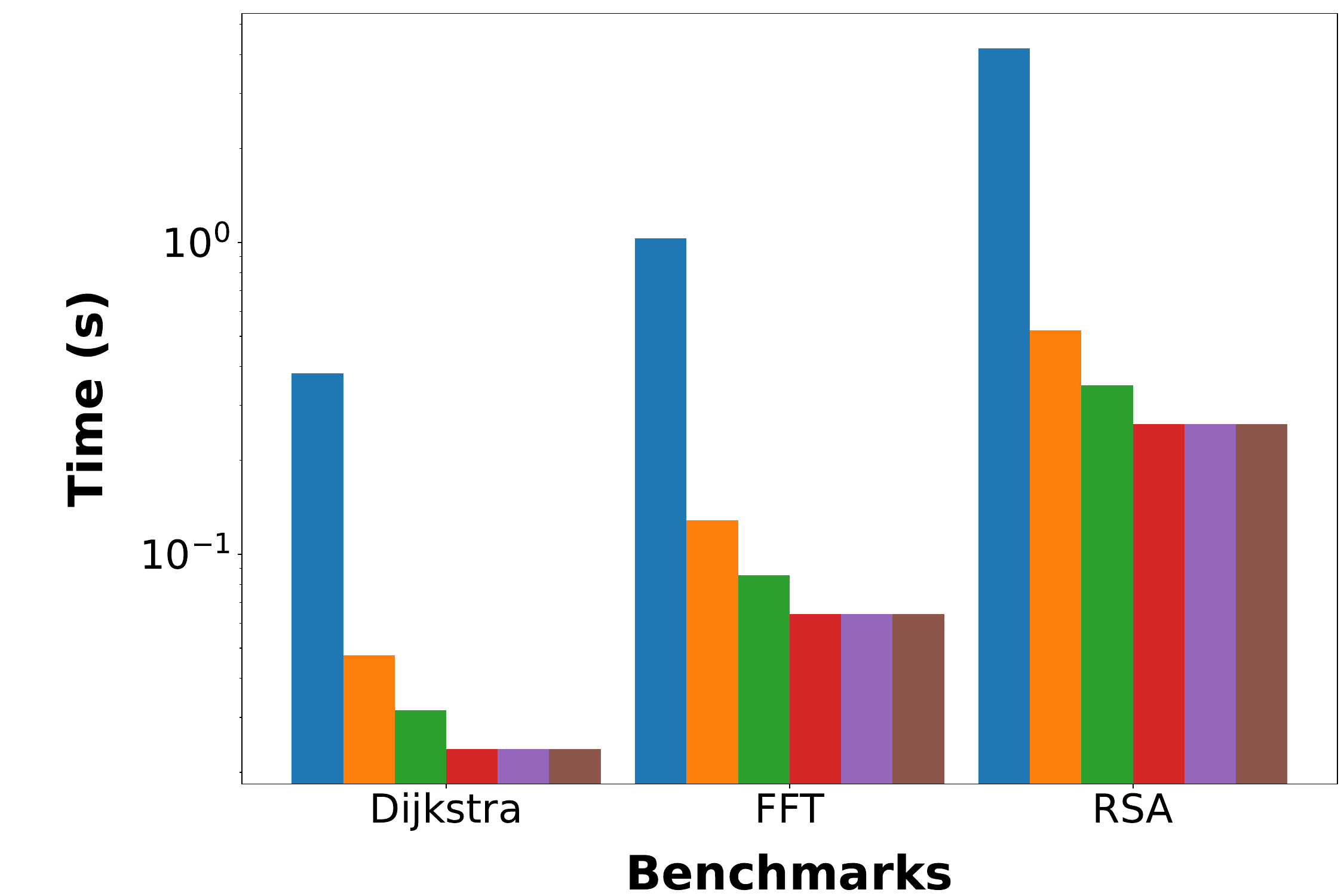}}
        \label{fig:results_solar_80u_36v_hibernus_time}
    }
    \subfigure[Energy consumption]{
        \centering
        \resizebox{0.319\columnwidth}{!}{\includegraphics{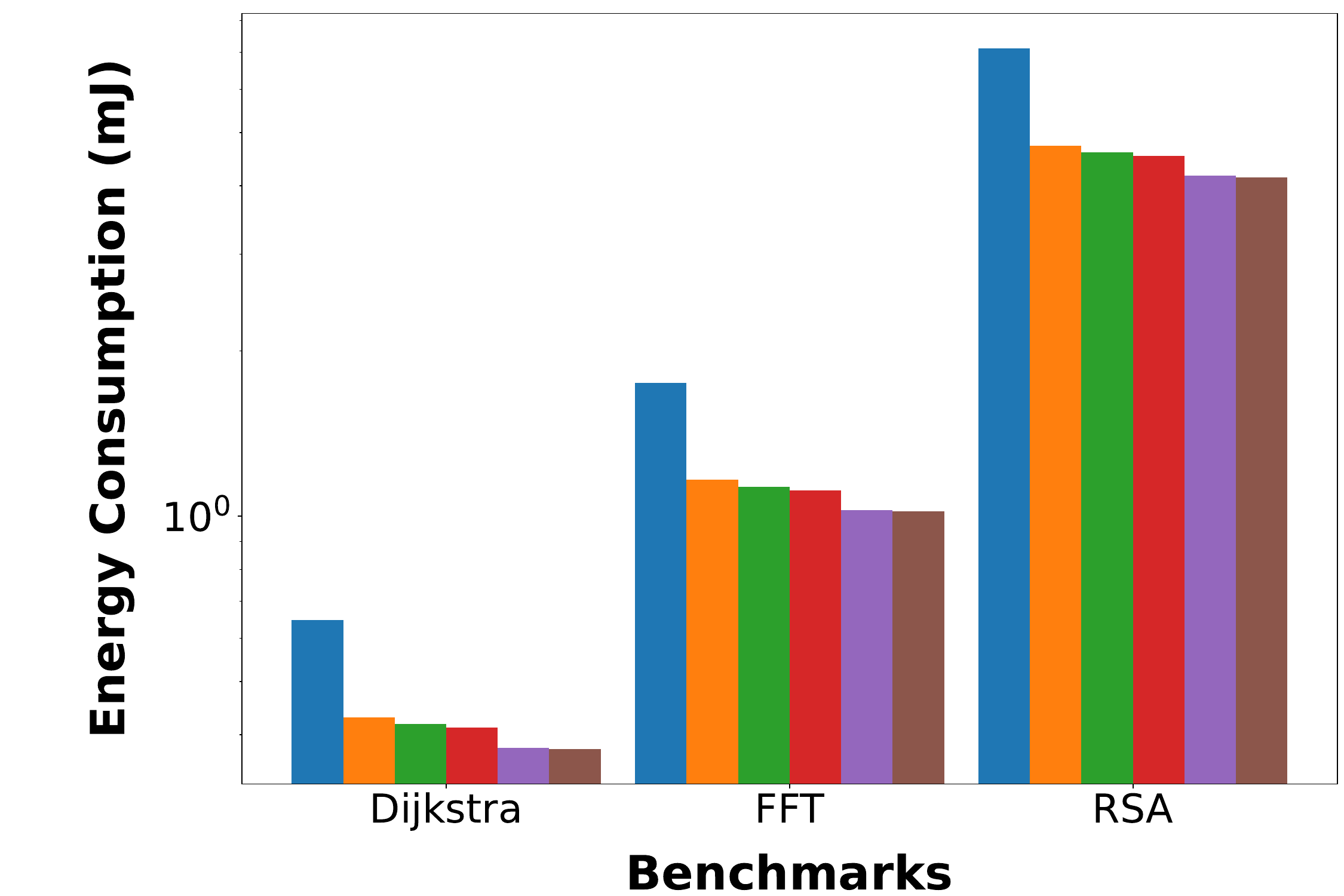}}
        \label{fig:results_solar_80u_36v_hibernus_energy}
    }
    \subfigure[Impact of external circuitry]{
        \centering
        \resizebox{0.319\columnwidth}{!}{\includegraphics{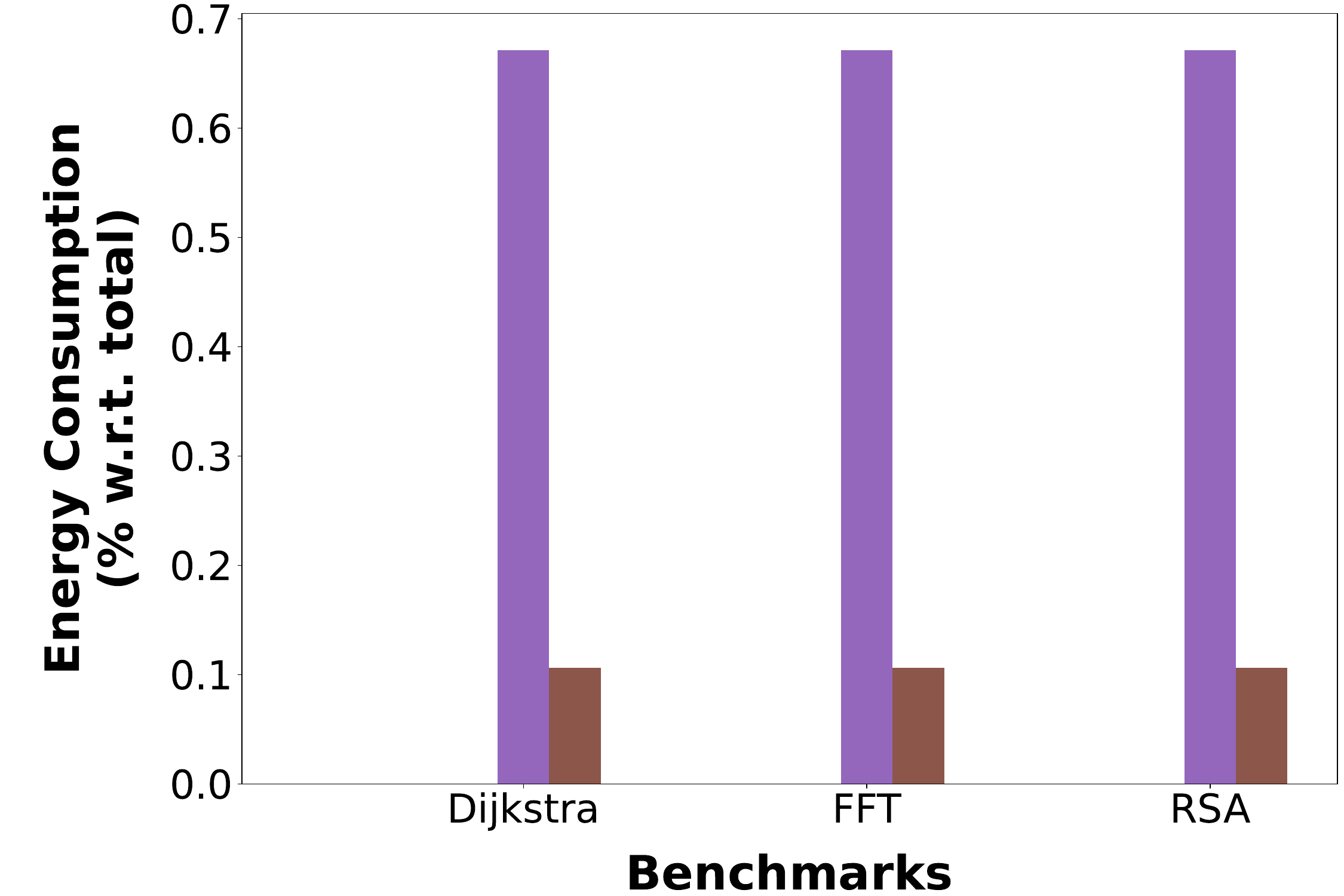}}
        \label{fig:results_solar_80u_36v_hibernus_energy_dvfs_ratio}
    }
    \caption{Results with the energy-rich source and Hibernus, $\mathbf{C = 80 \mu F}$, and $\mathbf{V_{boot} = 3.6V}$.}
    \label{fig:results_solar_80u_36v_hibernus}
\end{figure}

\fakepar{Hibernus}
\figref{fig:results_solar_80u_36v_hibernus} shows the results with Hibernus.
\figref{fig:results_solar_80u_36v_hibernus_time} depicts the completion time of each benchmark.
\dvfs and \fbtc require the same time of the static $16MHz$ configuration and are up to $16x$ faster than the other baselines.
Under conditions where the harvested energy maintains the capacitor fully charged throughout the experiment, both \dvfs and \fbtc consistently select the $16MHz$ frequency for its optimal speed and energy efficiency.
This results in up to $1.7x$ lower energy consumption, as \figref{fig:results_solar_80u_36v_hibernus_energy} shows.

Despite constantly executing at $16MHz$, we note that \dvfs and \fbtc show a $9\%$ lower energy consumption than the static $16MHz$ configuration.
Both \dvfs and \fbtc regulate the supply voltage to the lower bound of the current performance window, that is, $3.3V$.
The static configuration running at $16MHz$, instead, does not regulate the supply voltage and provides energy in the range $3.6V-3.3V$ as the capacitor discharges, ultimately consuming more energy despite the energy overhead of
\begin{mylist}
    \item the voltage regulator and
    \item the circuitry of \dvfs and \fbtc
\end{mylist}.

\dvfs and \fbtc custom circuitry also bears a negligible impact.
Across all benchmarks, \figref{fig:results_solar_80u_36v_hibernus_energy_dvfs_ratio} shows that it is responsible for just $0.67\%$ and $0.1\%$ of the overall energy consumption, respectively.
\fbtc has a $0.57\%$ lower energy impact than \dvfs while achieving the same completion time.

\begin{figure}[t]
    \subfigtopskip = -2pt
    \subfigure{
      \centering
      \hspace{70pt}
      \resizebox{0.35\columnwidth}{!}{\includegraphics{figures/evaluation/simulations/80u_3.6v/legend.pdf}}
    }
    \newline
    \setcounter{subfigure}{0}
    \subfigure[Completion time]{
        \centering
        \resizebox{0.319\columnwidth}{!}{\includegraphics{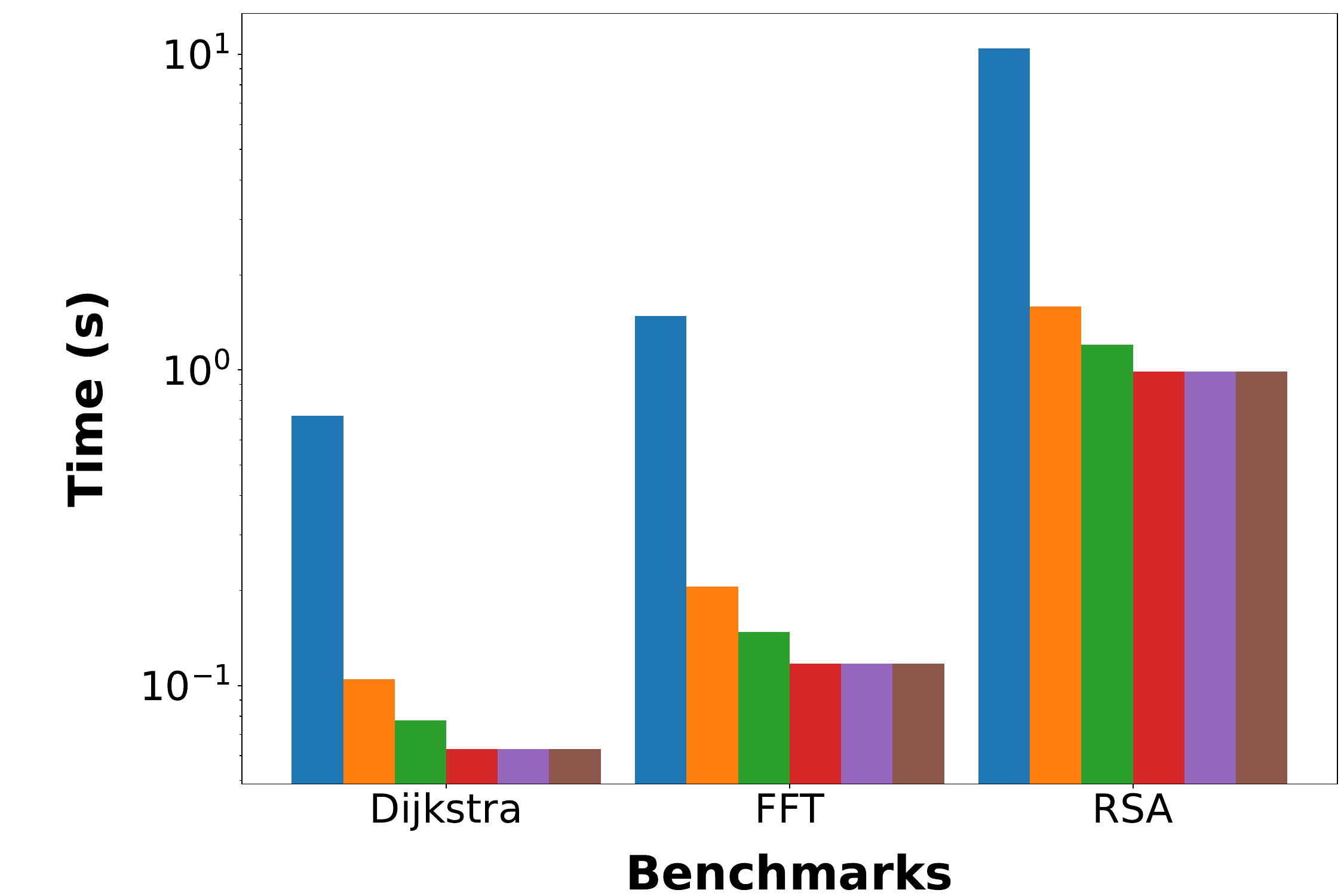}}
        \label{fig:results_solar_80u_36v_mementos_time}
    }
    \subfigure[Energy consumption]{
        \centering
        \resizebox{0.319\columnwidth}{!}{\includegraphics{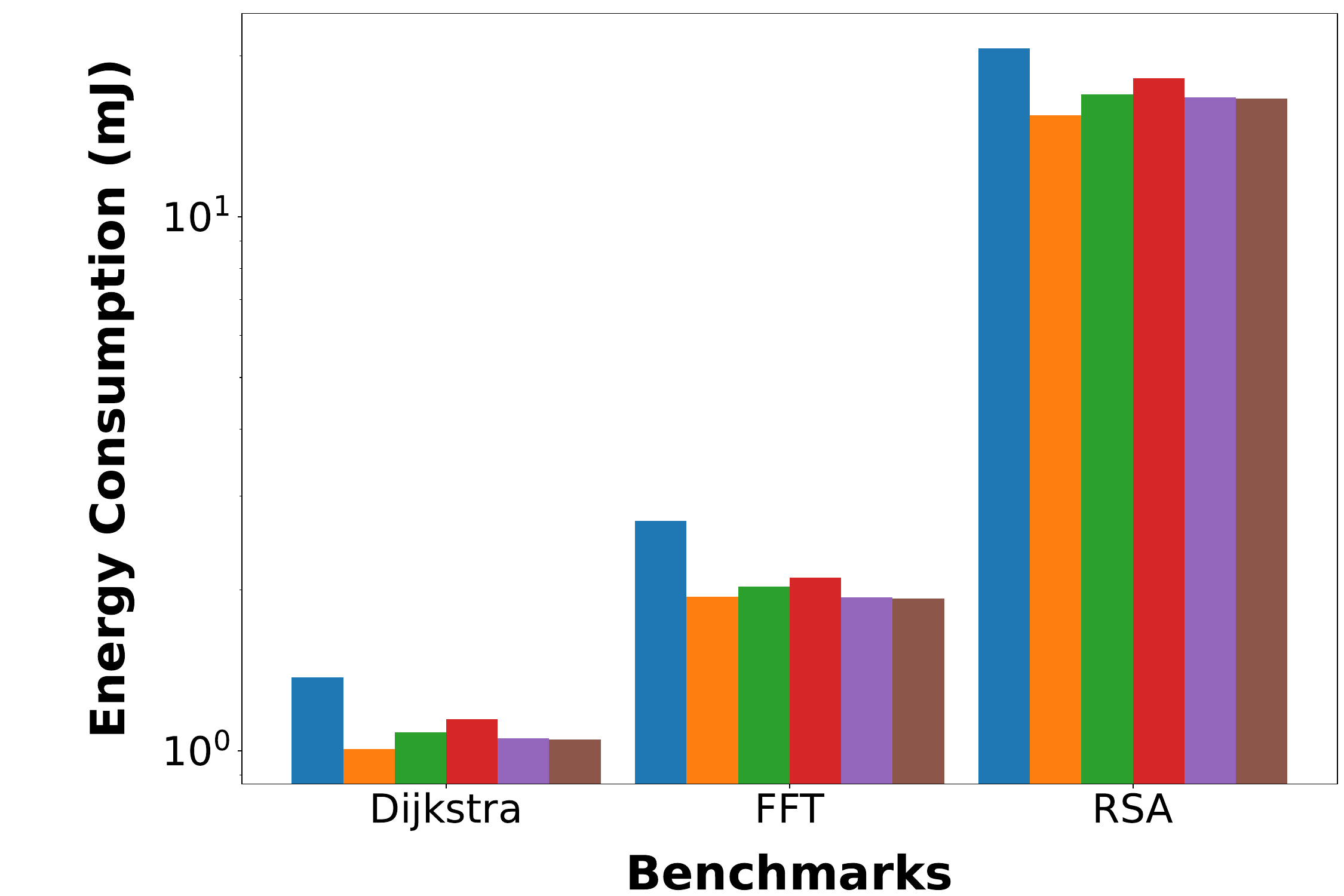}}
        \label{fig:results_solar_80u_36v_mementos_energy}
    }
    \subfigure[Impact of external circuitry]{
        \centering
        \resizebox{0.319\columnwidth}{!}{\includegraphics{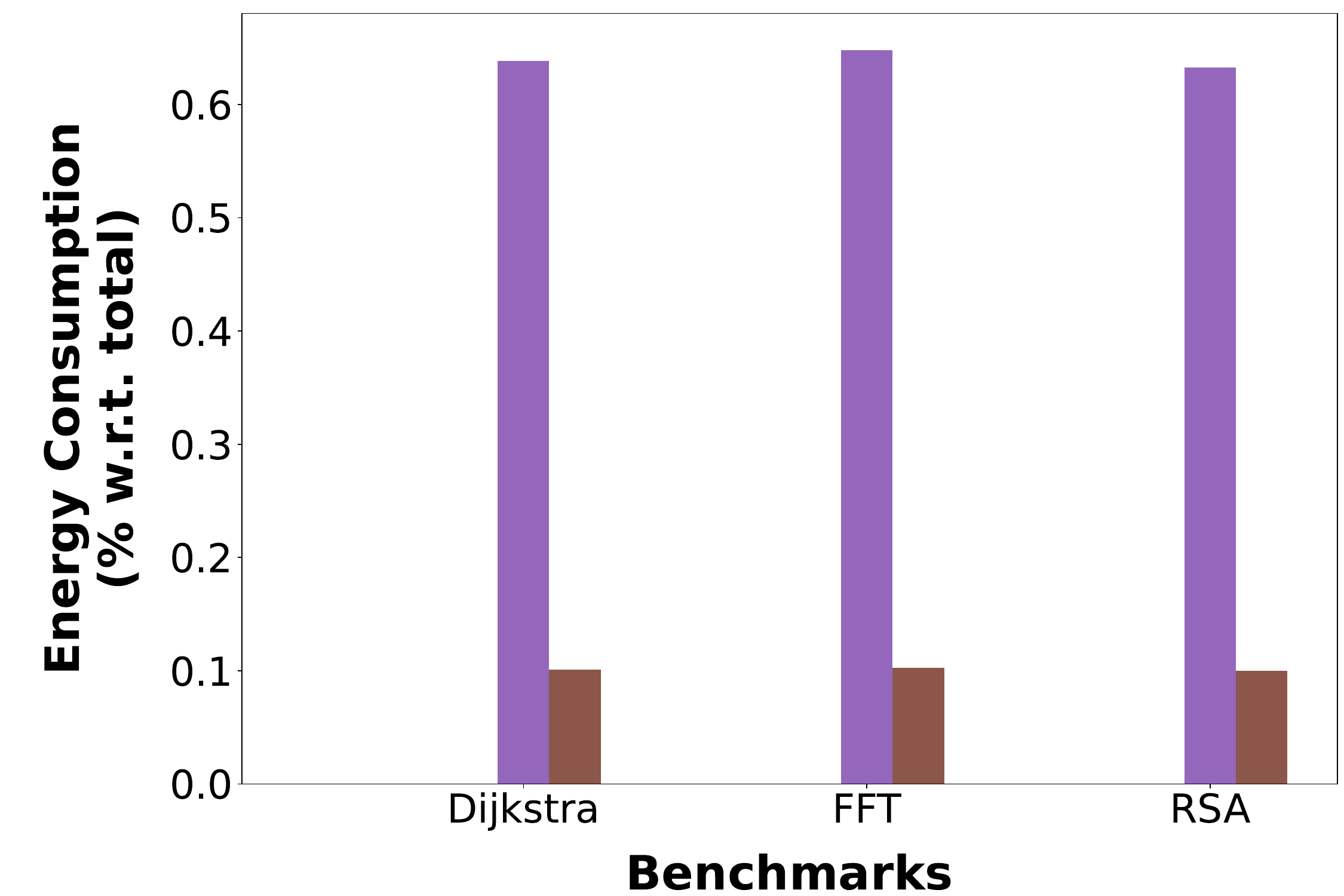}}
        \label{fig:results_solar_80u_36v_mementos_energy_dvfs_ratio}
    }
    \caption{Results with the energy-rich source and Mementos, $\mathbf{C = 80 \mu F}$, and $\mathbf{V_{boot} = 3.6V}$.}
    \label{fig:results_solar_80u_36v_mementos}
\end{figure}

\fakepar{Mementos}
As the energy-rich source never yields energy failures, the three ADC configurations for Mementos produce the same results, as the voltage is always in the correct ADC operating voltage range.
We report only the results of the \default~configuration, shown in \figref{fig:results_solar_80u_36v_mementos} with the $80 \mu F$ capacitor.

\figref{fig:results_solar_80u_36v_mementos_time} shows the same patterns of the experiments with Hibernus: \dvfs and \fbtc require the same time of the static $16MHz$ configuration to complete the benchmarks and they are up to $12x$ faster than the other baselines.
However, as \figref{fig:results_solar_80u_36v_mementos_energy} shows, \dvfs and \fbtc no longer show the same marked improvement in energy consumption as with Hibernus.
This is due to Mementos' probe function, which turns the ADC on, waits for a sample of capacitor voltage, and turns the ADC back off.
These operations introduce an overhead consisting of mandatory wait states that the MCU fills up by executing null operations (NOPs).
The number of NOPs is proportional to the MCU operating frequency, thus higher frequencies are subject to a higher penalty.
The cost for these NOPs partially outweighs the gains due to regulating the input voltage at the lower bound of the performance window.

Despite the penalty of ADC accesses, \dvfs and \fbtc still consume less energy than the static $1MHz$, $12MHz$, and $16MHz$ configurations across all benchmarks, as \figref{fig:results_solar_80u_36v_mementos_energy} shows.
This is again mainly due to the voltage regulation.
Instead, \fbtc (\dvfs) consumes, on average, $3.7\%$ ($4.29\%$) more energy than the static $8MHz$ configuration, with a maximum of $7.6\%$ ($8.22\%$) more in RSA.
Here again, the cost of ADC accesses at higher frequencies, that is, $16MHz$ compared to $8MHz$, represents a cost that makes the static $8MHz$ configuration more efficient.
However, \fbtc and \dvfs are, on average, $67\%$ faster than the static $8MHz$ configuration.
The decrease in completion time may compensate for the small increase in energy consumption, especially considering that the energy source supplies more energy than the device can buffer anyways.
Therefore, an increase in energy consumption does not cause any energy failure.

\dvfs and \fbtc custom circuitry bear negligible impact as in the case of Hibernus, that is, $0.64\%$ and $0.1\%$ of the total energy consumption, respectively.
\fbtc again has a $0.55\%$ lower energy consumption than \dvfs, with the same completion time.

%% file: evaluation-results-rf.tex
\subsection{Results $\rightarrow$ Energy-moderate Source}
\label{sec:results-rf}

\begin{figure}[t]
    \subfigtopskip = -2pt
    \subfigure{
        \centering
        \hspace{70pt}
        \resizebox{0.35\columnwidth}{!}{\includegraphics{figures/evaluation/simulations/80u_3.6v/legend.pdf}}
    }
    \newline
    \setcounter{subfigure}{0}
    \subfigure[Completion time]{
        \centering
        \resizebox{0.319\columnwidth}{!}{\includegraphics{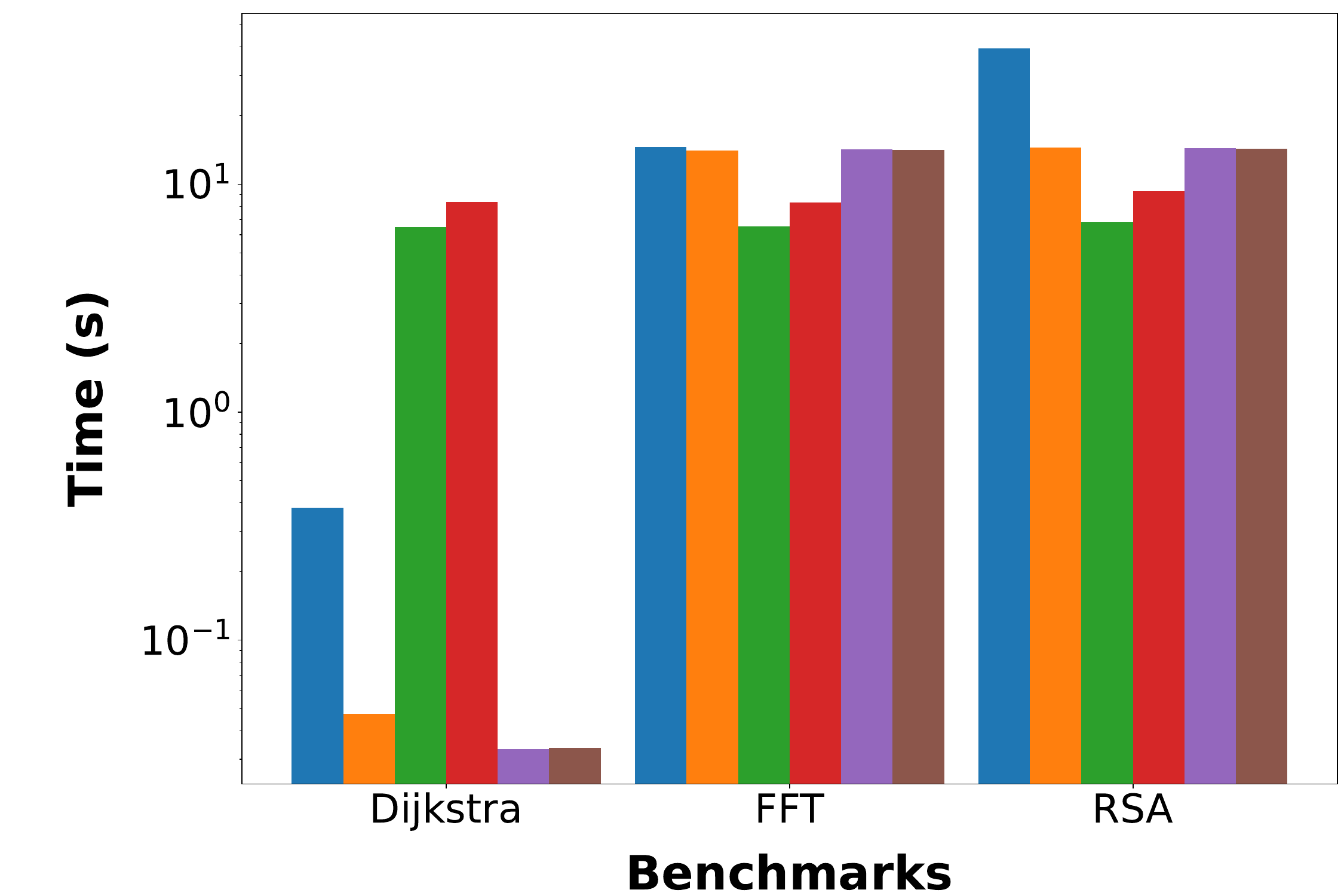}}
        \label{fig:results_rf_80u_36v_hibernus_time}
    }
    \subfigure[Recharge time]{
        \centering
        \resizebox{0.319\columnwidth}{!}{\includegraphics{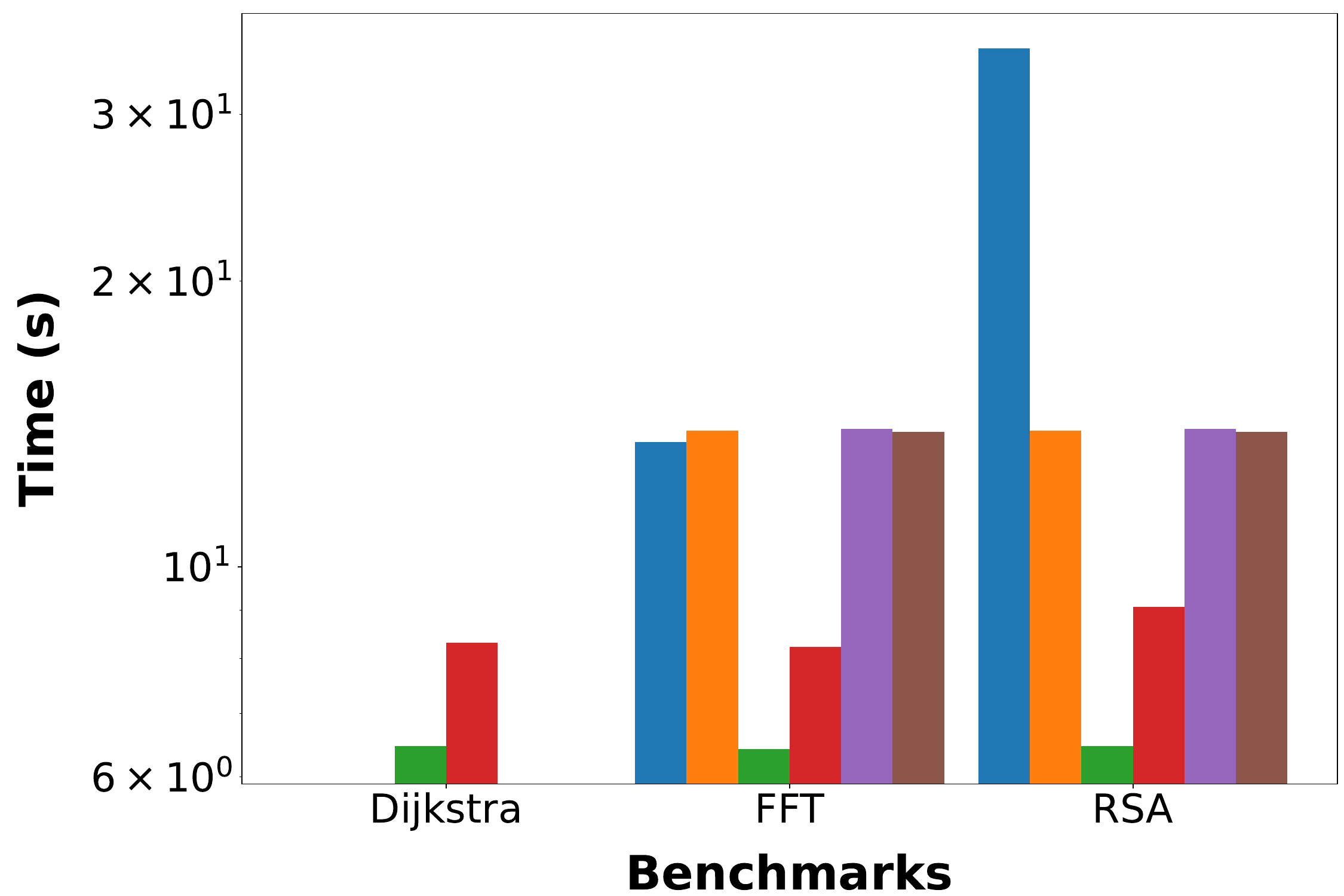}}
        \label{fig:results_rf_80u_36v_hibernus_time_recharge}
    }
    \subfigure[Execution time]{
        \centering
        \resizebox{0.319\columnwidth}{!}{\includegraphics{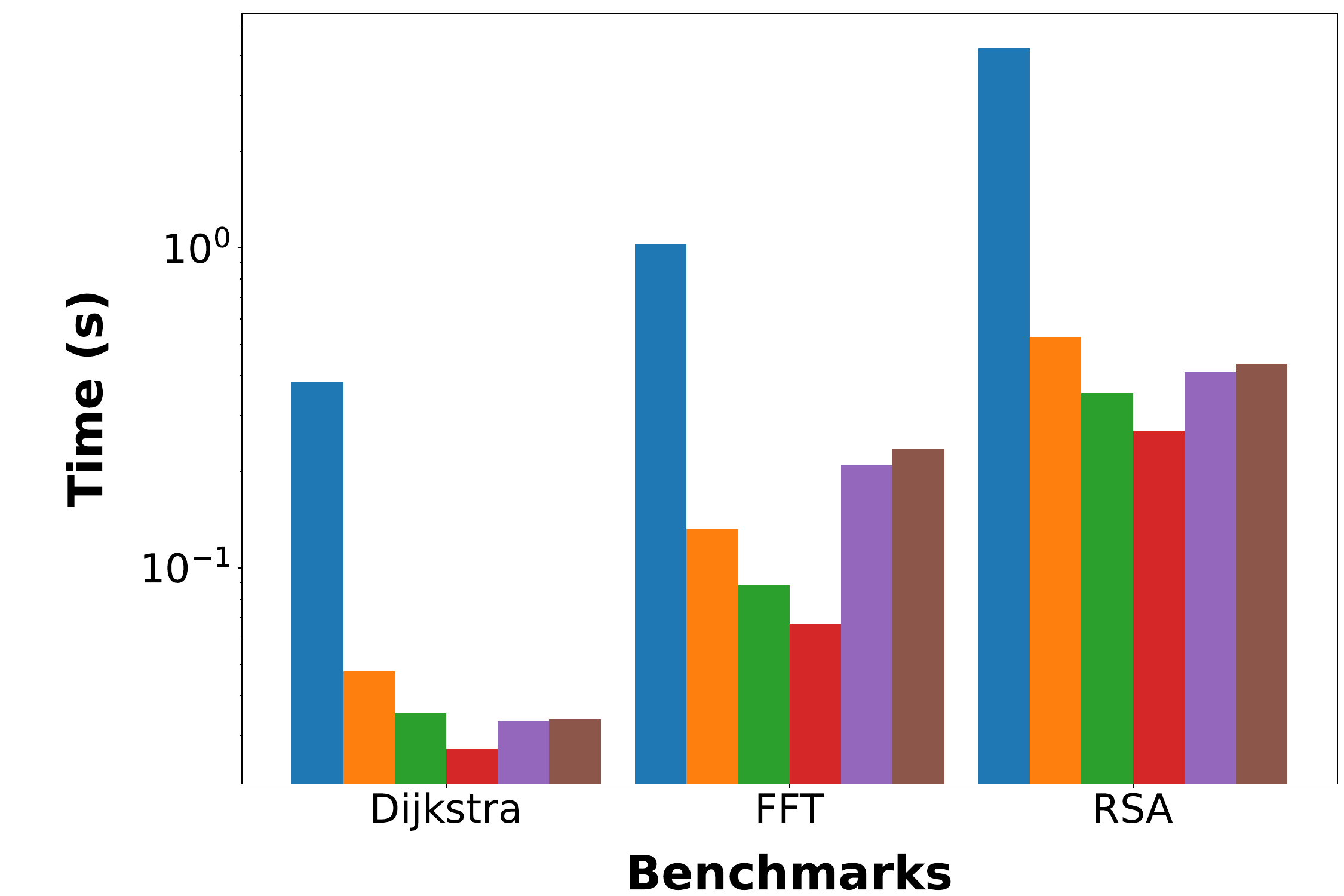}}
        \label{fig:results_rf_80u_36v_hibernus_time_execution}
    }
    \subfigure[Energy consumption]{
        \centering
        \resizebox{0.319\columnwidth}{!}{\includegraphics{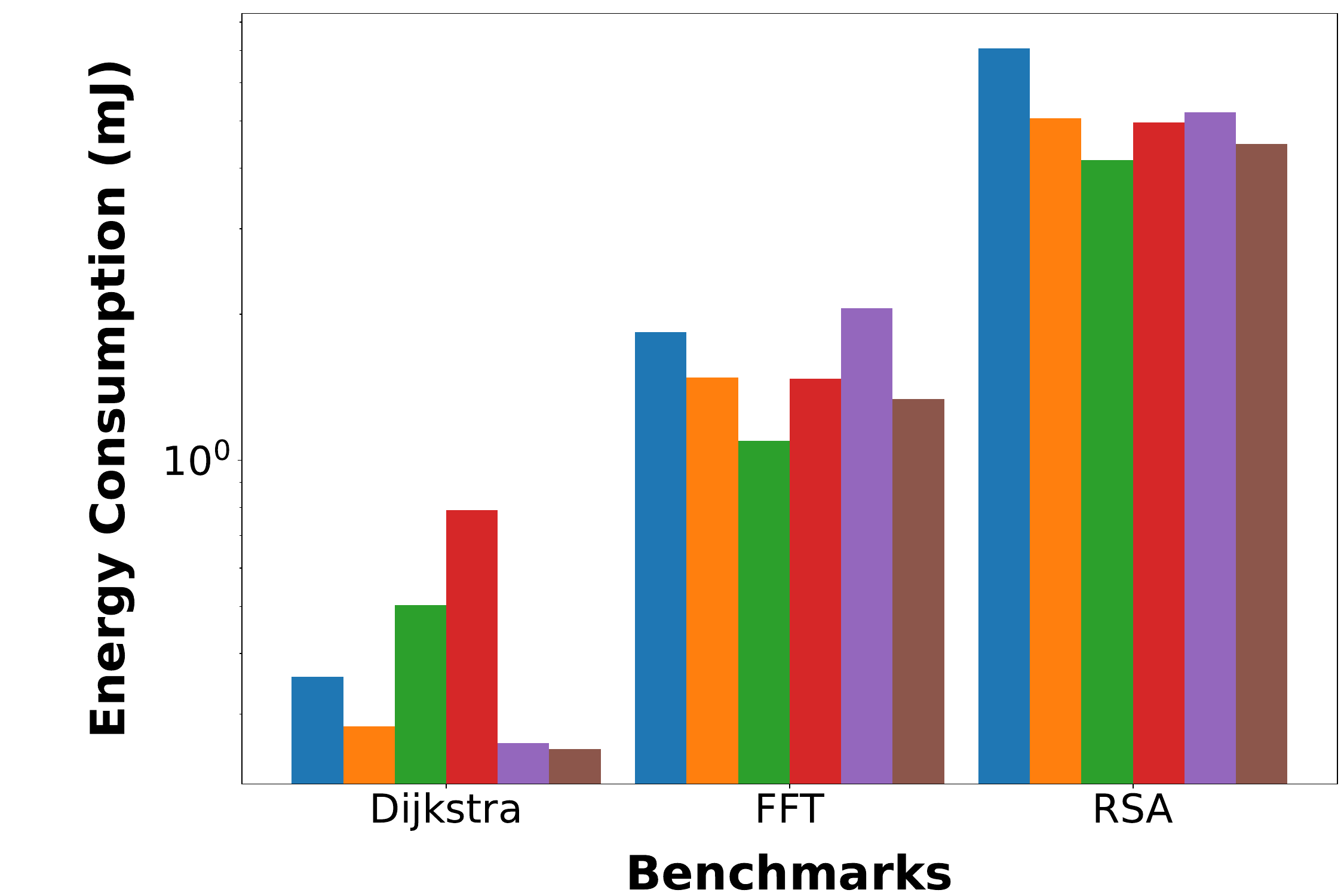}}
        \label{fig:results_rf_80u_36v_hibernus_energy}
    }
    \subfigure[Impact of external circuitry]{
        \centering
        \resizebox{0.319\columnwidth}{!}{\includegraphics{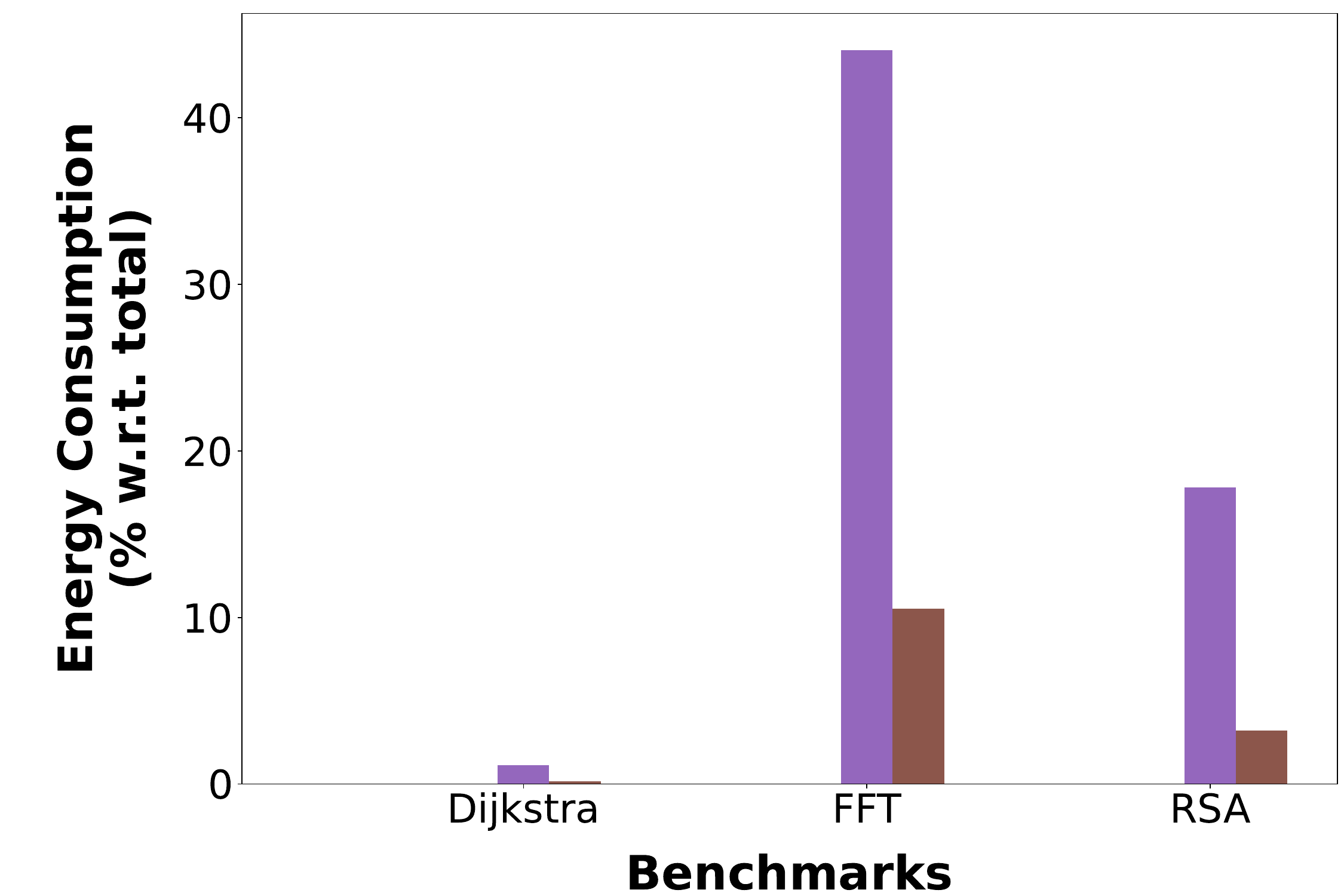}}
        \label{fig:results_rf_80u_36v_hibernus_energy_dvfs_ratio}
    }
    \subfigure[Number of energy failures]{
        \centering
        \resizebox{0.319\columnwidth}{!}{\includegraphics{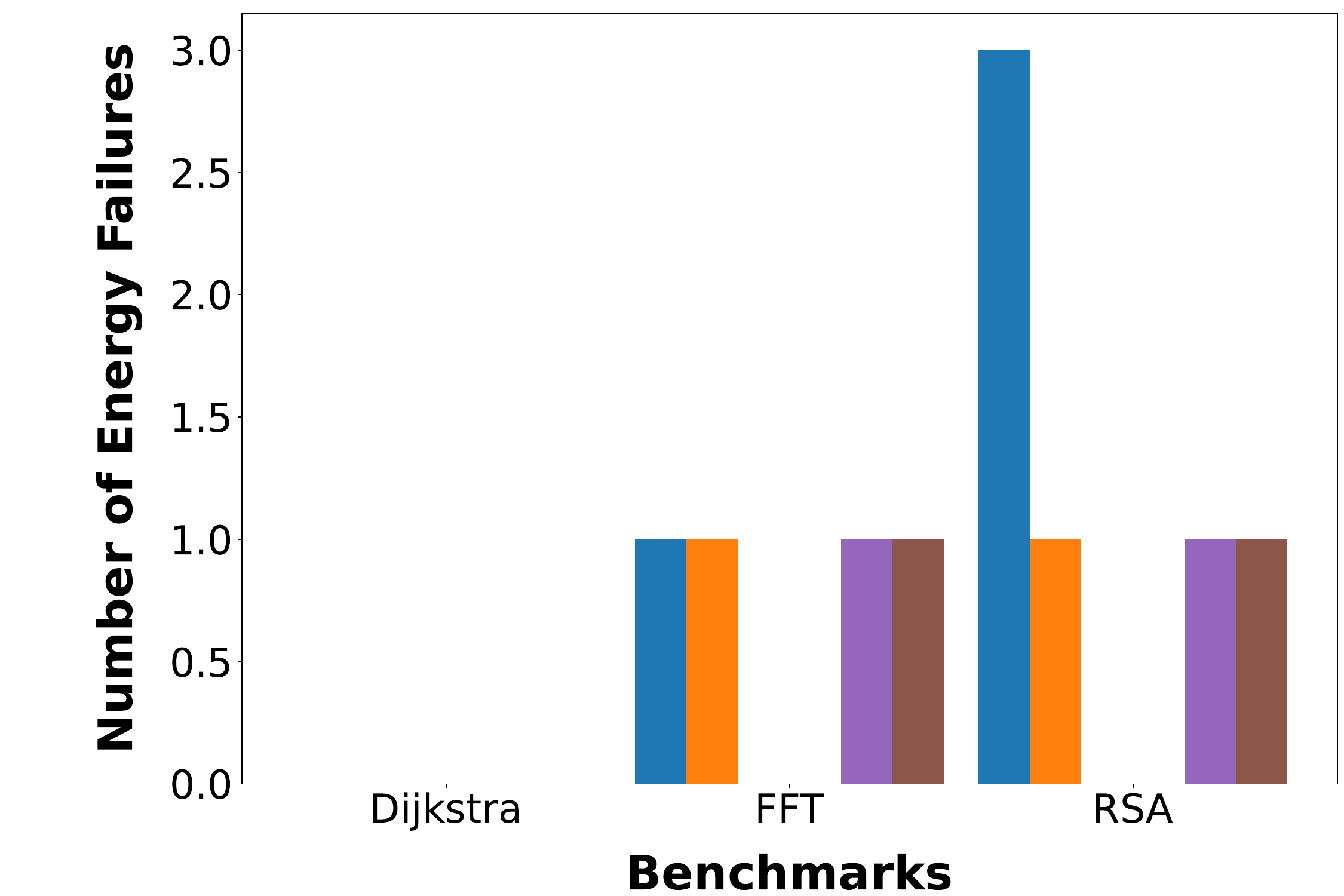}}
        \label{fig:results_rf_80u_36v_hibernus_power_failures}
    }
    \caption{Results with the energy-moderate source and Hibernus, $\mathbf{C = 80 \mu F}$, and $\mathbf{V_{boot} = 3.6V}$.}
    \label{fig:results_rf_80u_36v_hibernus}
\end{figure}

We discuss next the results for the experiments with the energy-moderate source, obtained using the voltage traces of an RF energy harvesting system~\cite{Mementos, Mementos-source-repo}.
We set $V_{boot} = 3.6V$ in these experiments.

\fakepar{Hibernus with $\mathbf{C=80 \mu F}$}
\figref{fig:results_rf_80u_36v_hibernus} shows the results.
The completion times shown in \figref{fig:results_rf_80u_36v_hibernus_time} indicate two different trends.
With the implementation of the Dijkstra algorithm,  \dvfs and \fbtc outperform all baselines, whereas with the implementation of FFT and RSA they are on par with the baselines.
The two trends deserve separate discussions.

When executing the Dijkstra algorithm, both \dvfs and \fbtc surpass the highest-performing static benchmark—the $8MHz$ configuration.
They offer a $42\%$ and $41\%$ improvement in completion time, as shown in \figref{fig:results_rf_80u_36v_hibernus_time}, and consume $8\%$ and $11\%$ less energy, as \figref{fig:results_rf_80u_36v_hibernus_energy} demonstrates, respectively.
Moreover, \dvfs and \fbtc are up to two orders of magnitude faster than the baselines and consume up to $3x$ less energy than the static frequency configurations.

The enhanced performance can be attributed to the voltage and frequency scaling capabilities of \dvfs and \fbtc.
\figref{fig:results_rf_80u_36v_hibernus_time_execution} shows that scaling the frequency grants \dvfs and \fbtc a shorter execution time than the static $8MHz$ configuration, as they can execute a portion of the code faster.
Additionally, by transitioning to the most efficient performance window based on the current capacitor voltage, they maximize the number of instructions executed in each energy cycle.
With this, \dvfs and \fbtc show lower energy consumption than the baselines, as shown in \figref{fig:results_rf_80u_36v_hibernus_energy}, allowing both to complete the execution in a single energy cycle, as \figref{fig:results_rf_80u_36v_hibernus_power_failures} shows.
Note that the static $1MHz$ and $8Mhz$ configurations show a similar behavior.
However, due to frequency scaling, \dvfs and \fbtc execute faster.

Unlike the Dijkstra algorithm, the FFT and RSA implementations encompass a significantly larger number of machine instructions.
This prevents \dvfs and \fbtc from completing their execution in a single energy cycle, despite frequency and voltage scaling.
As a result, they no longer perform better than all static configurations.
Compared to the best-performing baseline, that is, $12MHz$, \dvfs and \fbtc are $2.1x$ slower, as shown in \figref{fig:results_rf_80u_36v_hibernus_time}, and consume, on average, $56\%$ and $15\%$ more energy, as shown in \figref{fig:results_rf_80u_36v_hibernus_energy}, respectively.

The efficacy of \dvfs and \fbtc arises from the nature of the energy source coupled with their limited voltage span when activating \emph{hibernation mode}. 
This mode, unique to Hibernus, transitions the system to a low-power state without full shutdown, allowing for energy accumulation before a checkpoint is imperative. 
The initiation of hibernation mode is contingent on the minimum voltage required for MCU operation, which is in turn determined by the operating frequency of the MCU.

The higher static frequency configurations, such as $12MHz$ and $16MHz$, enter hibernation mode at a higher voltage level than \dvfs and \fbtc.
In contrast, \dvfs and \fbtc enter hibernation mode with a lower energy reserve.
The energy source supplies short energy bursts that are $5s$ apart from each other, as shown in \figref{fig:energy_sources_rf}, which is insufficient to let \dvfs and \fbtc wait in hibernation mode, as the bursts are too far from each other, eventually causing an energy failure.
Instead, the $12MHz$ and $16MHz$ static configurations have sufficient energy to wait for the next energy burst and therefore experience no energy failures.
\figref{fig:results_rf_80u_36v_hibernus_power_failures} provides evidence for this analysis.

\figref{fig:results_rf_80u_36v_hibernus_time}, \figref{fig:results_rf_80u_36v_hibernus_time_execution}, and \figref{fig:results_rf_80u_36v_hibernus_time_recharge} also indicate that the recharge times represent most of the completion time, whereas the execution times contribute in a limited way.
In RSA, the recharge times of the best static frequency configuration, that is, $12MHz$, are $95\%$ of the total completion time, whereas in \dvfs and \fbtc the recharge times are $97\%$ of the completion time.
The increase in recharge times is another consequence of \dvfs and \fbtc entering hibernation mode with lower energy compared to the $12MHz$ static configuration.
\dvfs and \fbtc show $2.1x$ higher recharge times than the latter configuration, as both must recharge the capacitor to $V_{boot}$ starting from a lower voltage.

On average, \fbtc achieves a $0.01\%$ faster completion time and exhibits a $24\%$ reduction in energy usage compared to \dvfs across all evaluated benchmarks.
\figref{fig:results_rf_80u_36v_hibernus_energy_dvfs_ratio} shows that \dvfs external circuitry bears a higher impact on overall energy consumption than in the case of \fbtc.
\dvfs external circuitry is indeed responsible for up to $44\%$ of the total energy consumption, whereas this figure is limited to $11\%$ for \fbtc.

\begin{figure}[t]
    \subfigtopskip = -2pt
    \subfigure{
        \centering
        \hspace{70pt}
        \resizebox{0.35\columnwidth}{!}{\includegraphics{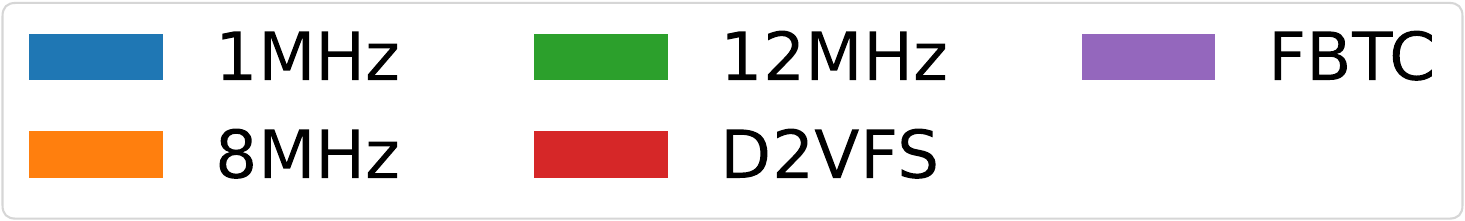}}
    }
    \newline
    \setcounter{subfigure}{0}
    \subfigure[Completion time]{
        \centering
        \resizebox{0.319\columnwidth}{!}{\includegraphics{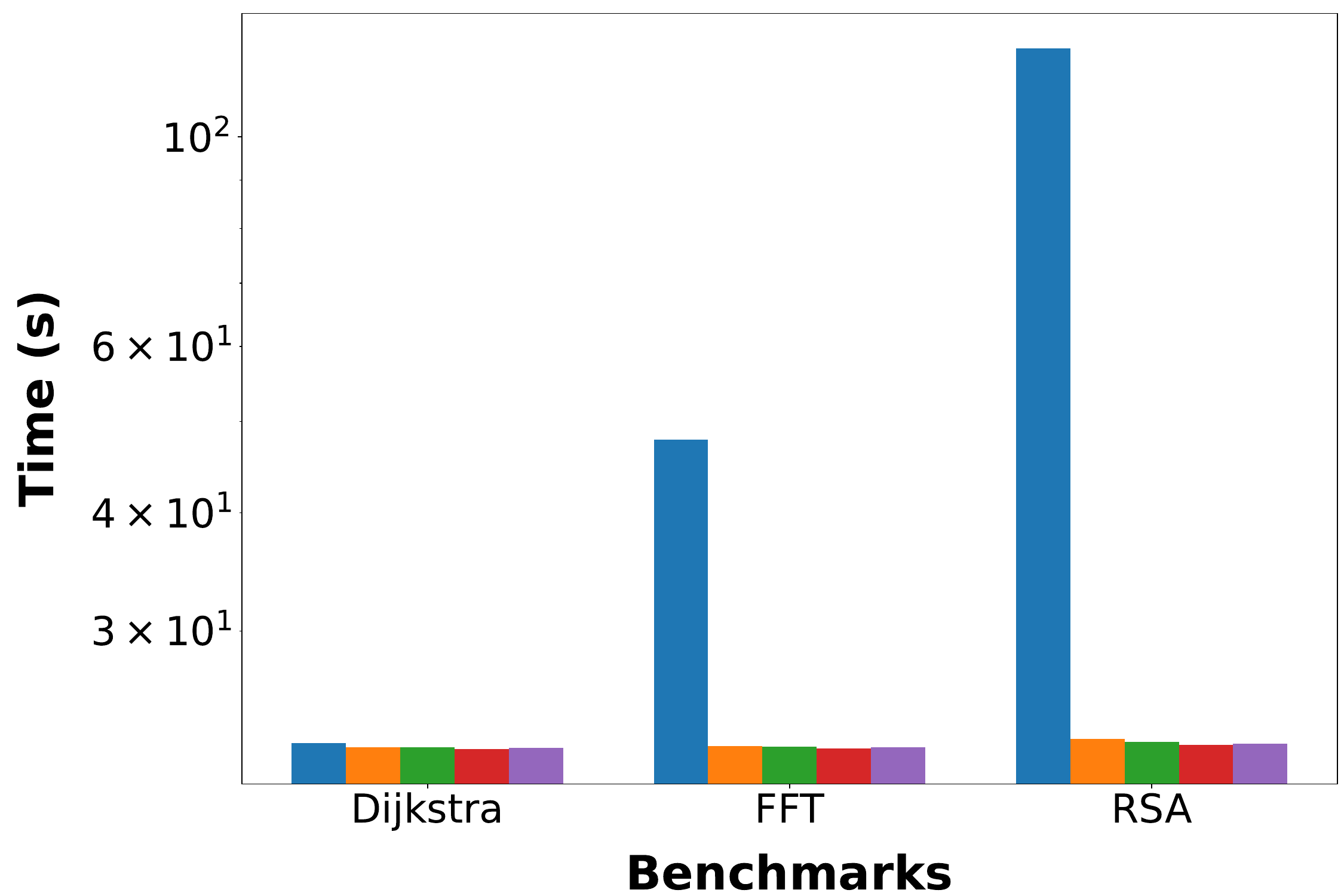}}
        \label{fig:results_rf_20u_36v_hibernus_time}
    }
    \subfigure[Recharge time]{
        \centering
        \resizebox{0.319\columnwidth}{!}{\includegraphics{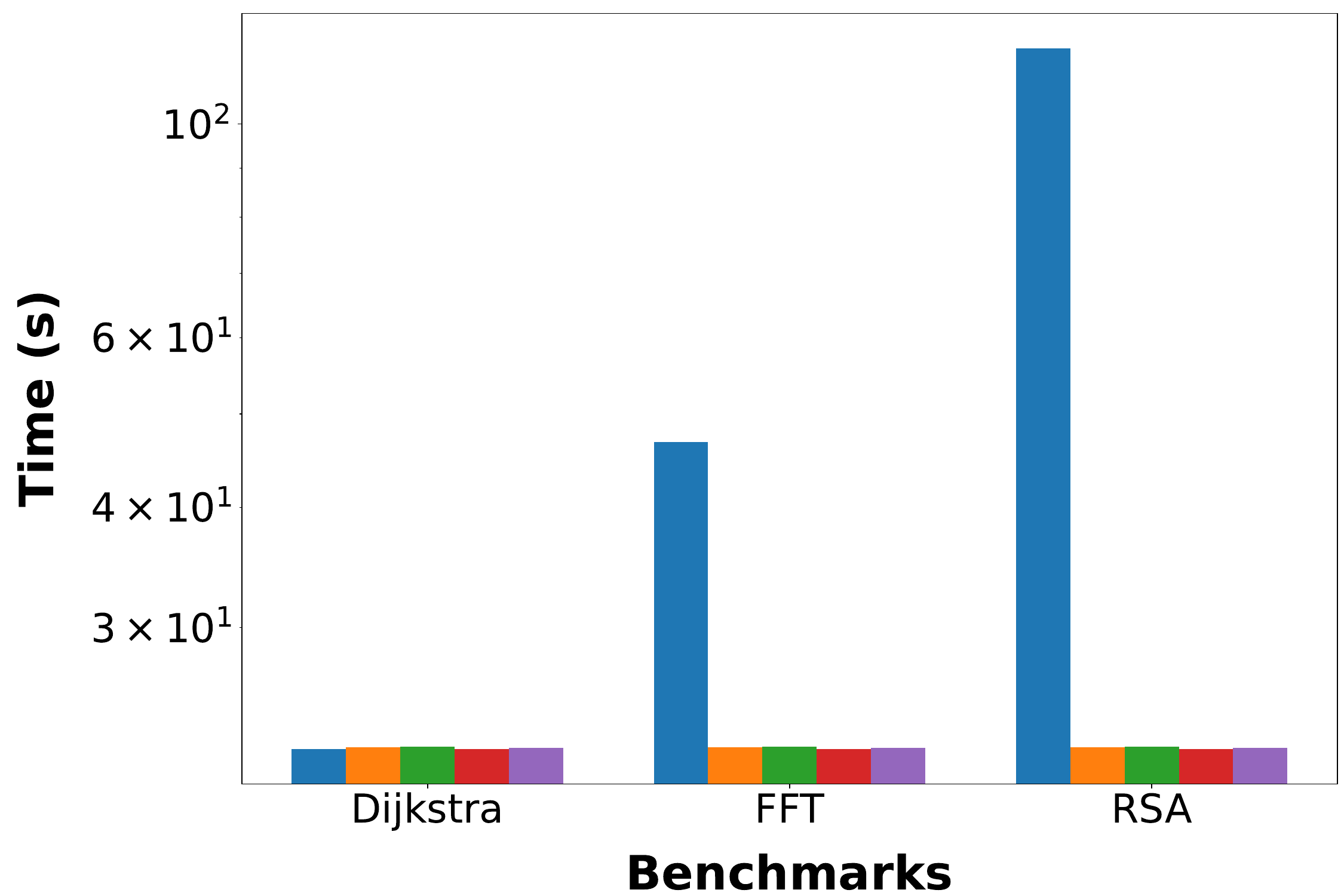}}
        \label{fig:results_rf_20u_36v_hibernus_time_recharge}
    }
    \subfigure[Execution time]{
        \centering
        \resizebox{0.319\columnwidth}{!}{\includegraphics{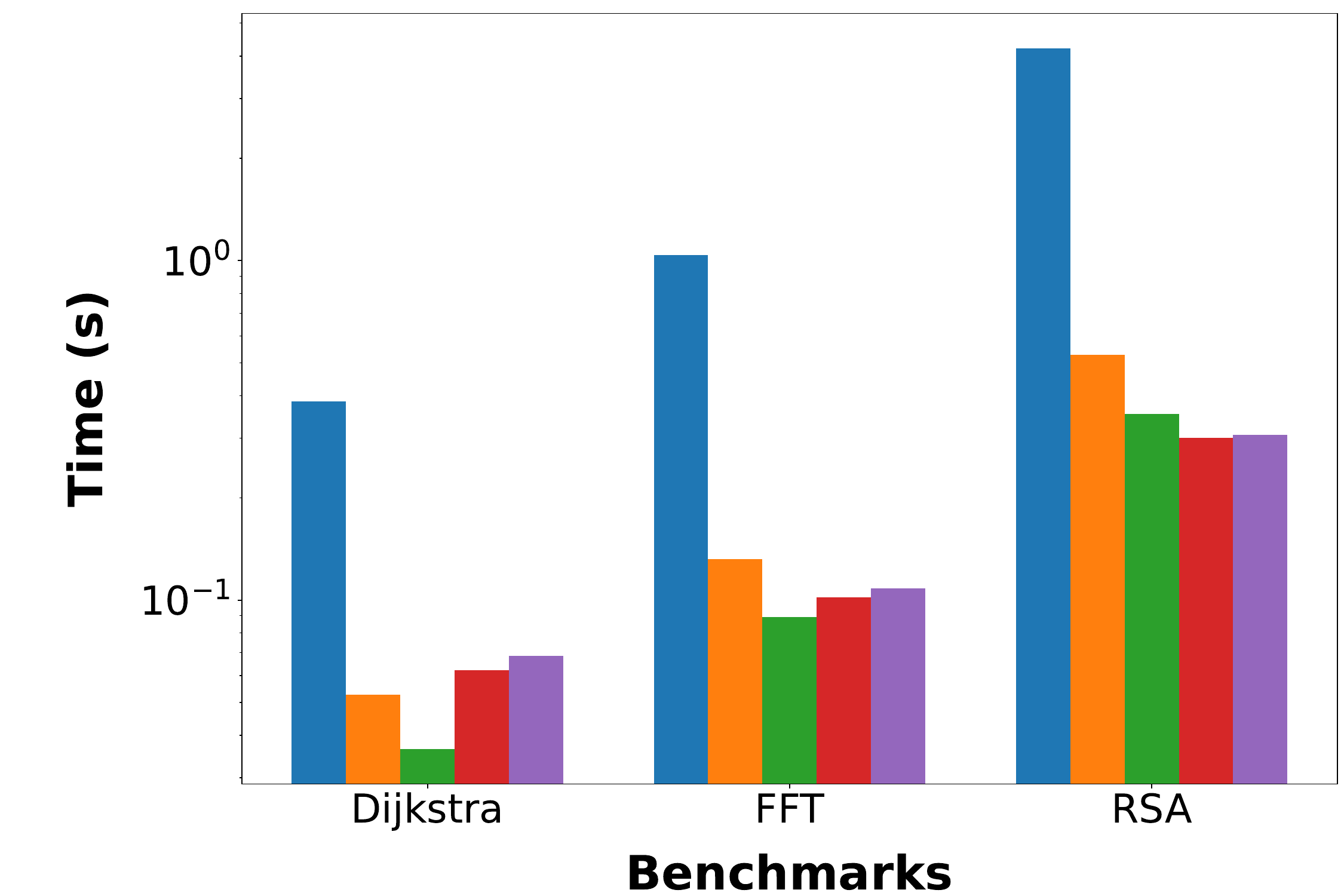}}
        \label{fig:results_rf_20u_36v_hibernus_time_execution}
    }
    \subfigure[Energy consumption]{
        \centering
        \resizebox{0.319\columnwidth}{!}{\includegraphics{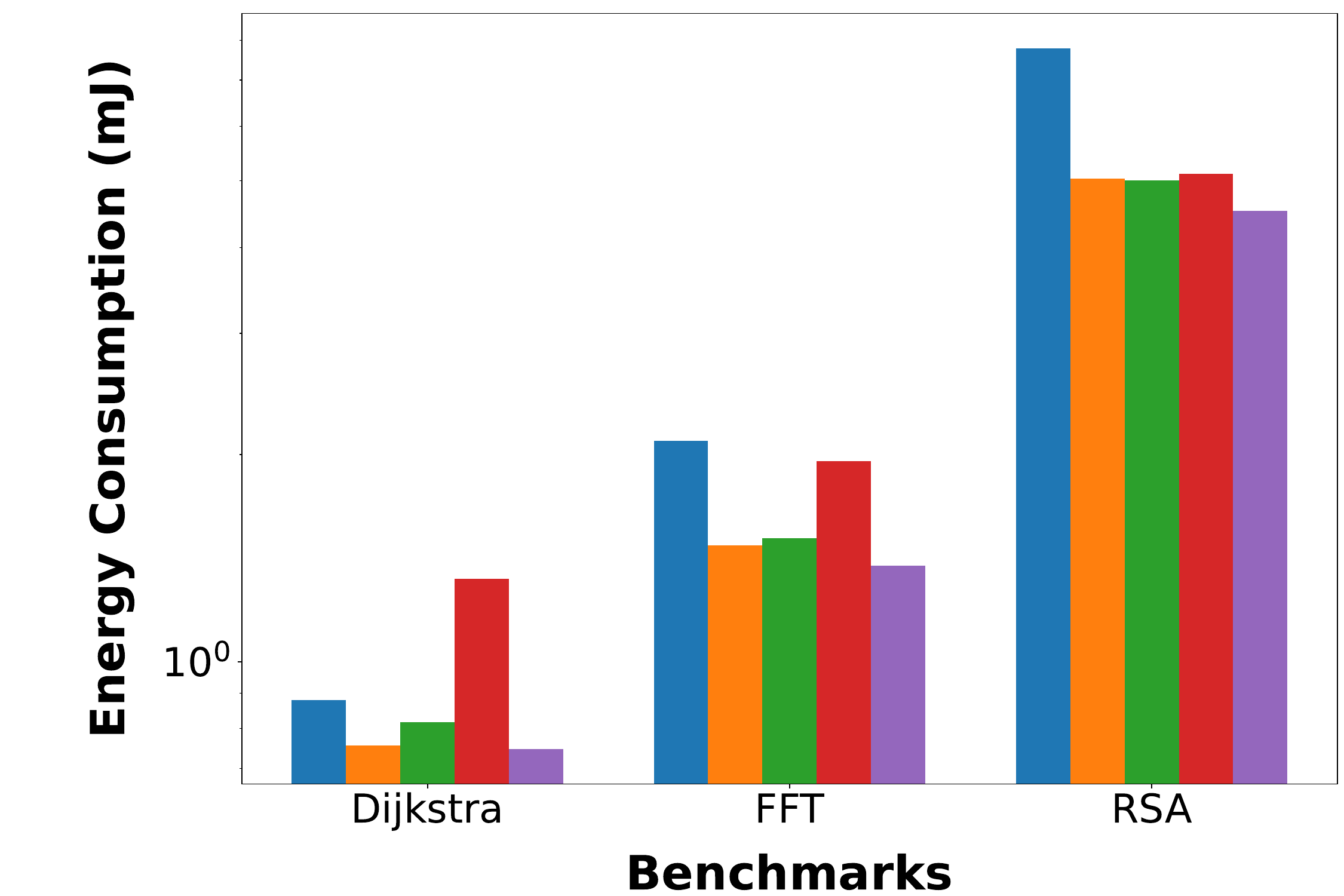}}
        \label{fig:results_rf_20u_36v_hibernus_energy}
    }
    \subfigure[Impact of external circuitry]{
        \centering
        \resizebox{0.319\columnwidth}{!}{\includegraphics{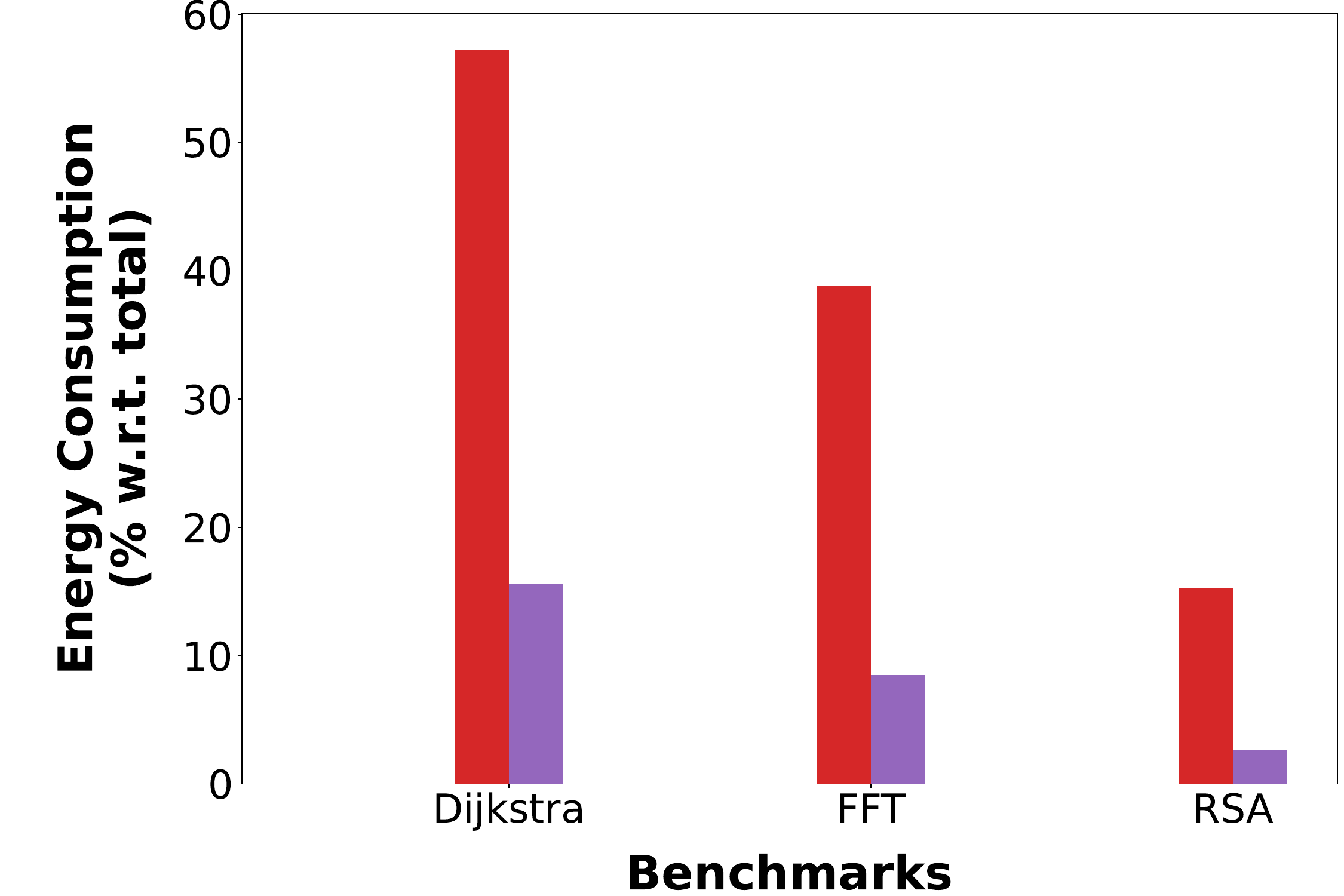}}
        \label{fig:results_rf_20u_36v_hibernus_energy_dvfs_ratio}
    }
    \subfigure[Number of energy failures]{
        \centering
        \resizebox{0.319\columnwidth}{!}{\includegraphics{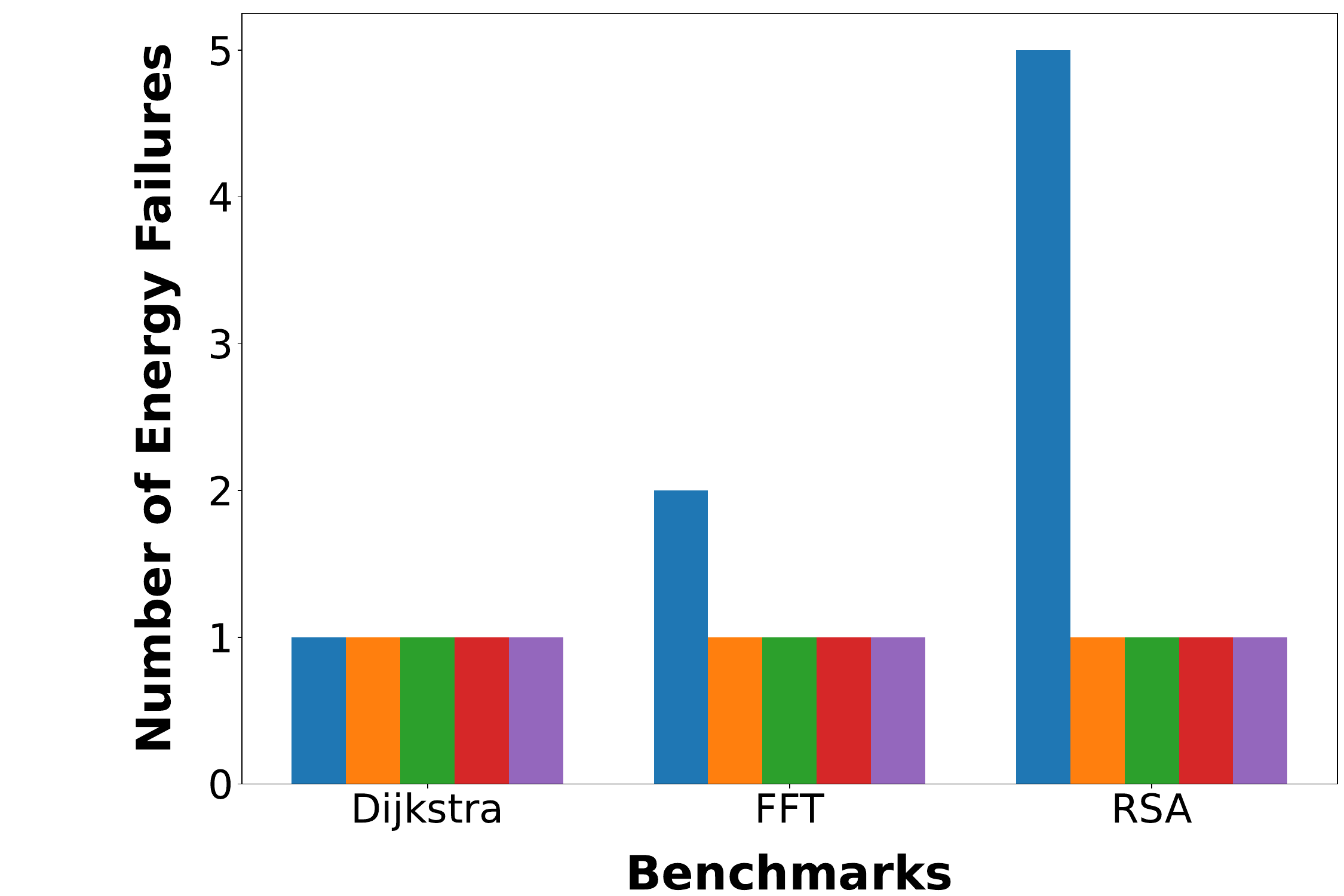}}
        \label{fig:results_rf_20u_36v_hibernus_power_failures}
    }
    \caption{Results with the energy-moderate source and Hibernus, $\mathbf{C = 20 \mu F}$, and $\mathbf{V_{boot} = 3.6V}$.}
    \label{fig:results_rf_20u_36v_hibernus}
\end{figure}

\fakepar{Hibernus with $\mathbf{C=20 \mu F}$}
The smaller $20 \mu F$ capacitor setting allows us to run tests with RF energy harvesting without using a voltage doubler, as discussed in \secref{sec:evaluation-setting}.
We do not consider the static $16MHz$ configuration here, as it cannot complete the workload with such a small capacitor size.

\figref{fig:results_rf_20u_36v_hibernus} shows the results.
Unlike the case with $C=80 \mu F$, \dvfs and \fbtc outperform all baselines.
The capacitor size determines this performance, as it causes all systems to enter hibernation mode with little energy.
In fact, as \figref{fig:results_rf_20u_36v_hibernus_time_recharge} shows, the recharge times of \dvfs and \fbtc are close to the best-performing baseline and overall account for up to $99\%$ of the total completion time.

\figref{fig:results_rf_20u_36v_hibernus_time} shows that \dvfs and
\fbtc complete the benchmarks $5.4x$ times faster than the static
$1MHz$ configuration, with a performance similar to the two static $8MHz$ and $12MHz$ configurations.
\fbtc also shows the lowest energy consumption across the board, as shown in \figref{fig:results_rf_20u_36v_hibernus_energy}: it consumes \emph{at least} $22\%$ less energy than the baselines.
Instead, \dvfs higher quiescent current results, on average, in a $22\%$ higher energy consumption than the baselines.

These results are due to voltage and frequency scaling, as \dvfs and \fbtc can temporarily set the MCU to run at $16MHz$, operating in a more efficient condition than the baselines.
Compared to the $80 \mu F$ case, this produces a shorter execution time, as \figref{fig:results_rf_20u_36v_hibernus_time_execution} shows.
Further, the higher the number of clock cycles in the workload, the faster \dvfs and \fbtc complete the benchmarks compared to static configurations.
This is the case in the RSA implementation, as opposed to Dijkstra and FFT implementations.

The smaller capacitor impacts the number of energy failures the system is subject to, shown in \figref{fig:results_rf_20u_36v_hibernus_power_failures}: all the baselines now experience an energy failure, whereas with a $80 \mu F$ capacitor the static $12MHz$ configuration did not.
No system can now complete the Dijkstra implementation in one energy cycle.

The same performance difference of the $80 \mu F$ capacitor case remains here between \dvfs and \fbtc.
On average, \fbtc is $0.27\%$ slower than \dvfs, while consuming $44\%$ less energy.
However, there is now an increase in the overall energy consumption of \dvfs and \fbtc components due to higher recharge times.
\figref{fig:results_rf_20u_36v_hibernus_energy_dvfs_ratio} shows that \dvfs circuitry is now responsible for up to $57\%$ of the total energy consumption, whereas \fbtc circuitry is responsible only for up to $15\%$ of it.

\begin{figure}[t]
    \subfigtopskip = -2pt
    \subfigure{
        \centering
        \hspace{70pt}
        \resizebox{0.35\columnwidth}{!}{\includegraphics{figures/evaluation/simulations/80u_3.6v/legend.pdf}}
    }
    \newline
    \setcounter{subfigure}{0}
    \subfigure[Completion time]{
        \centering
        \resizebox{0.319\columnwidth}{!}{\includegraphics{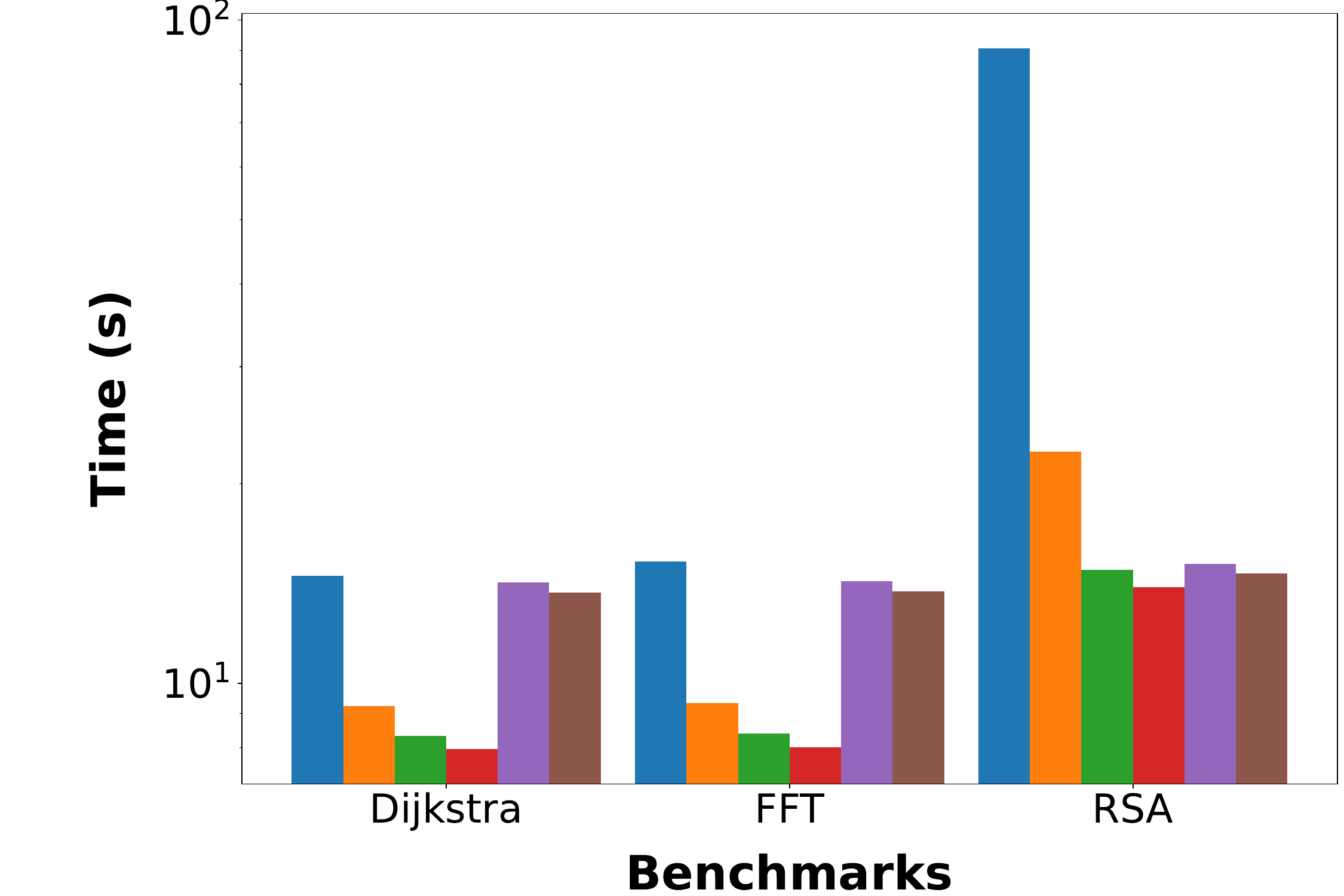}}
        \label{fig:results_rf_80u_36v_mementos_time}
    }
    \subfigure[Recharge time]{
        \centering
        \resizebox{0.319\columnwidth}{!}{\includegraphics{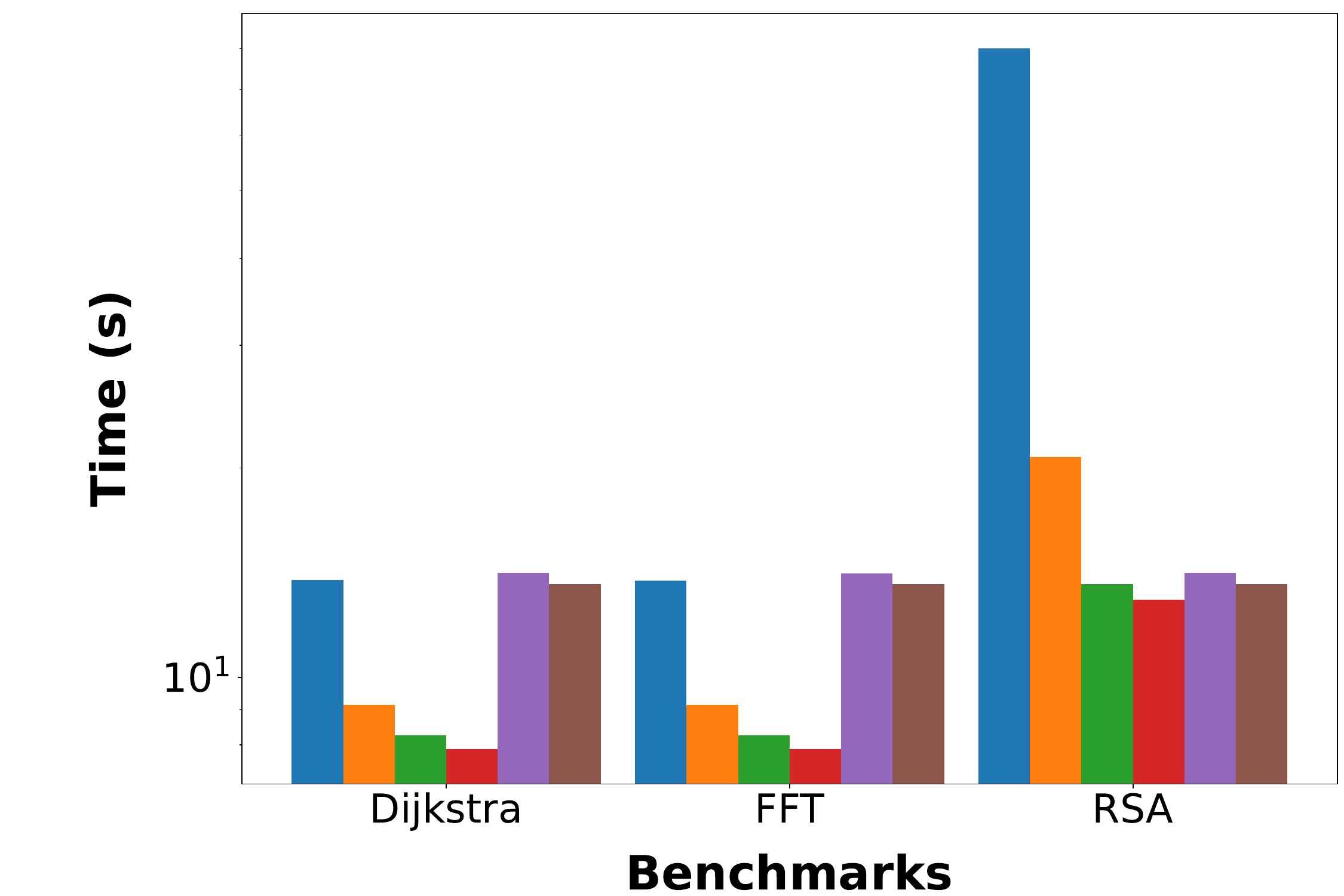}}
        \label{fig:results_rf_80u_36v_mementos_time_recharge}
    }
    \subfigure[Execution time]{
        \centering
        \resizebox{0.319\columnwidth}{!}{\includegraphics{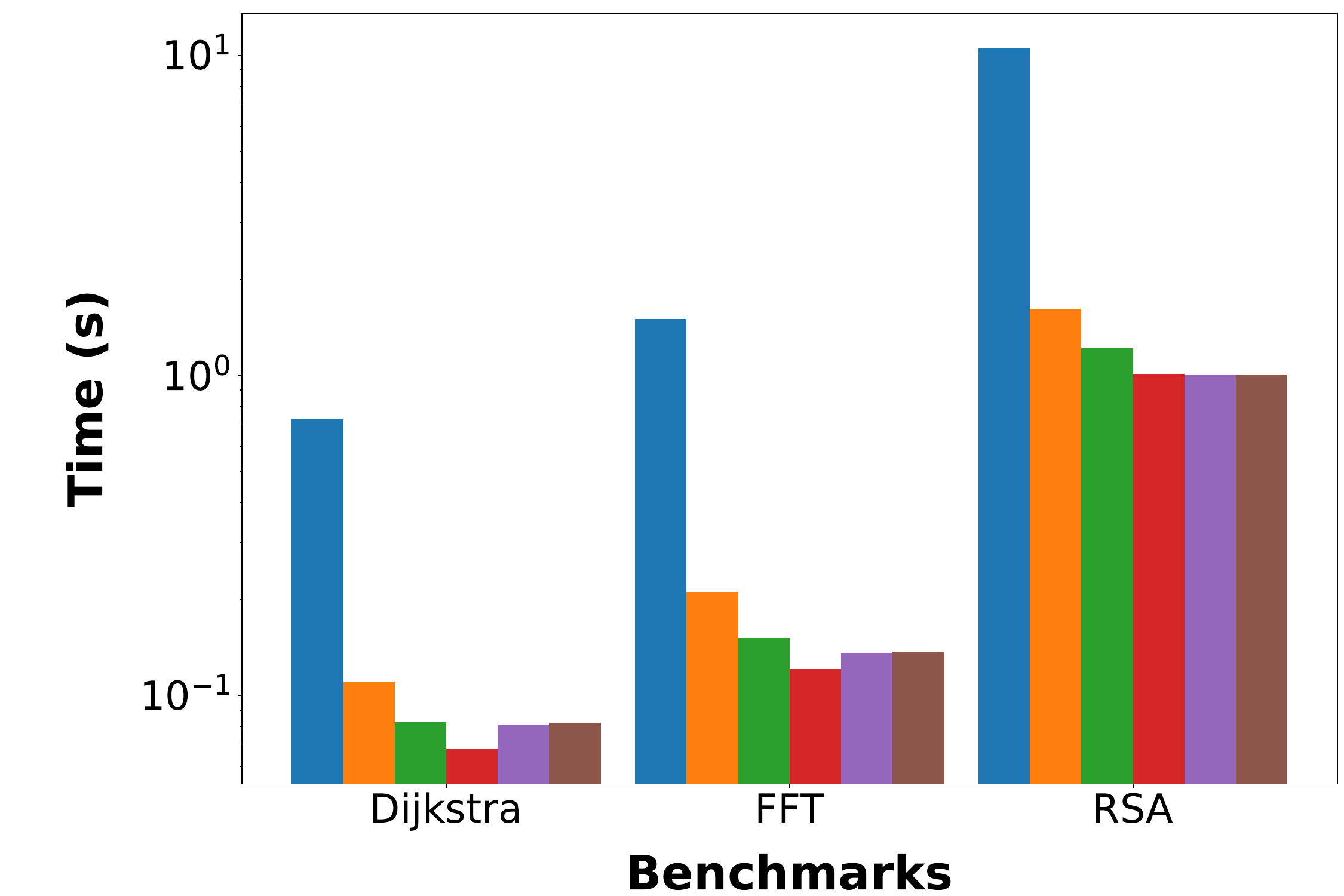}}
        \label{fig:results_rf_80u_36v_mementos_time_execution}
    }
    \subfigure[Energy consumption]{
        \centering
        \resizebox{0.319\columnwidth}{!}{\includegraphics{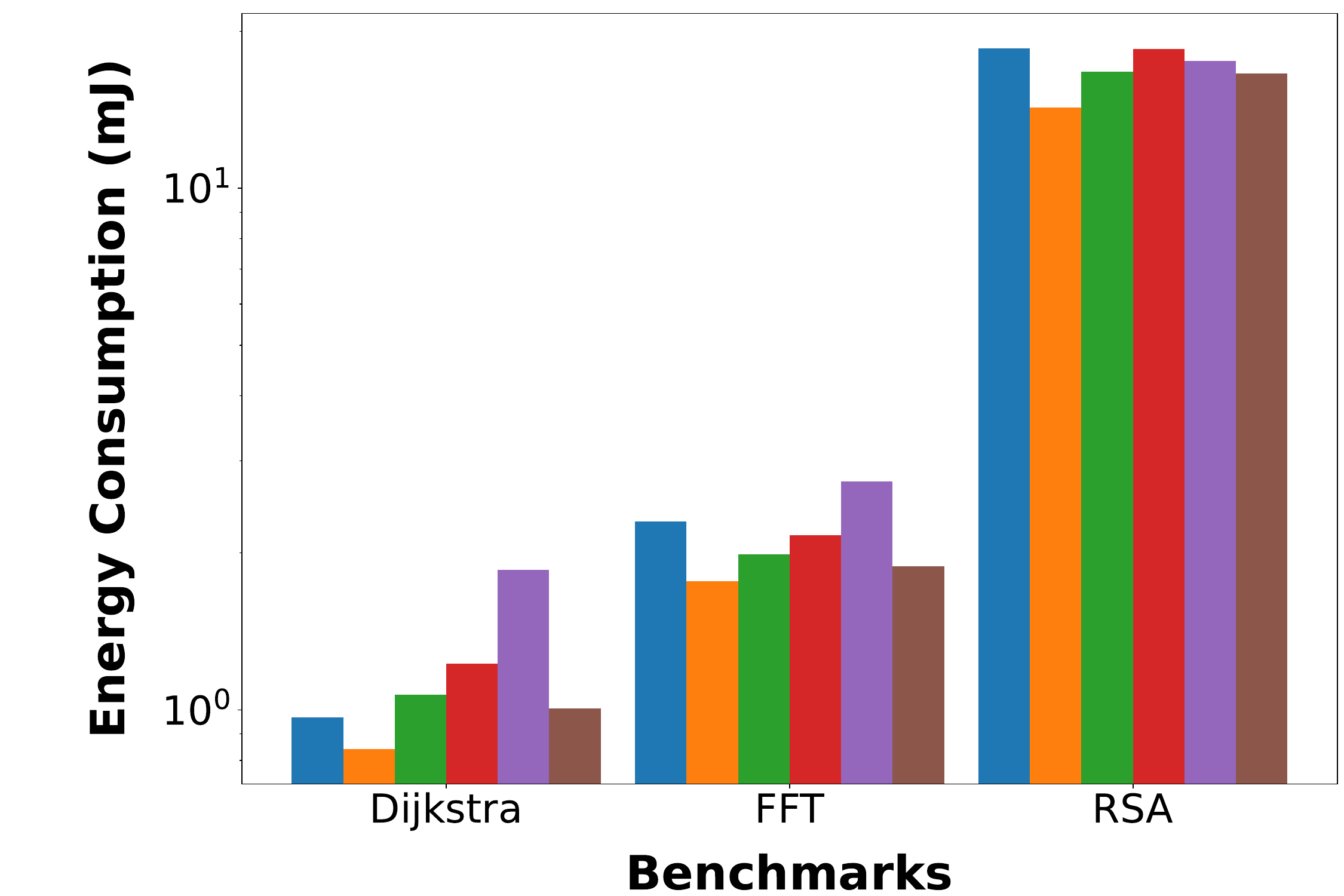}}
        \label{fig:results_rf_80u_36v_mementos_energy}
    }
    \subfigure[Impact of external circuitry]{
        \centering
        \resizebox{0.319\columnwidth}{!}{\includegraphics{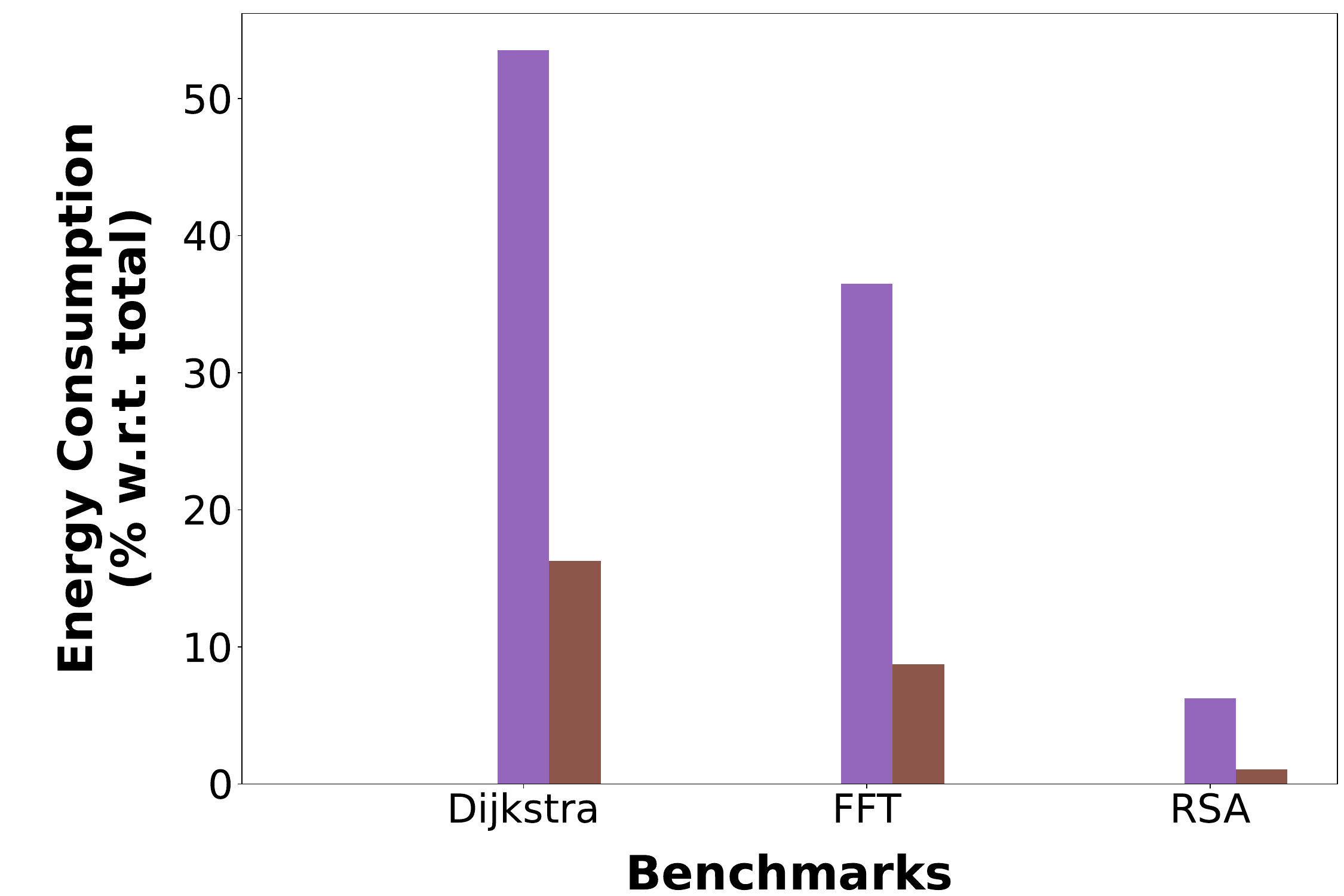}}
        \label{fig:results_rf_80u_36v_mementos_energy_dvfs_ratio}
    }
    \subfigure[Number of energy failures]{
        \centering
        \resizebox{0.319\columnwidth}{!}{\includegraphics{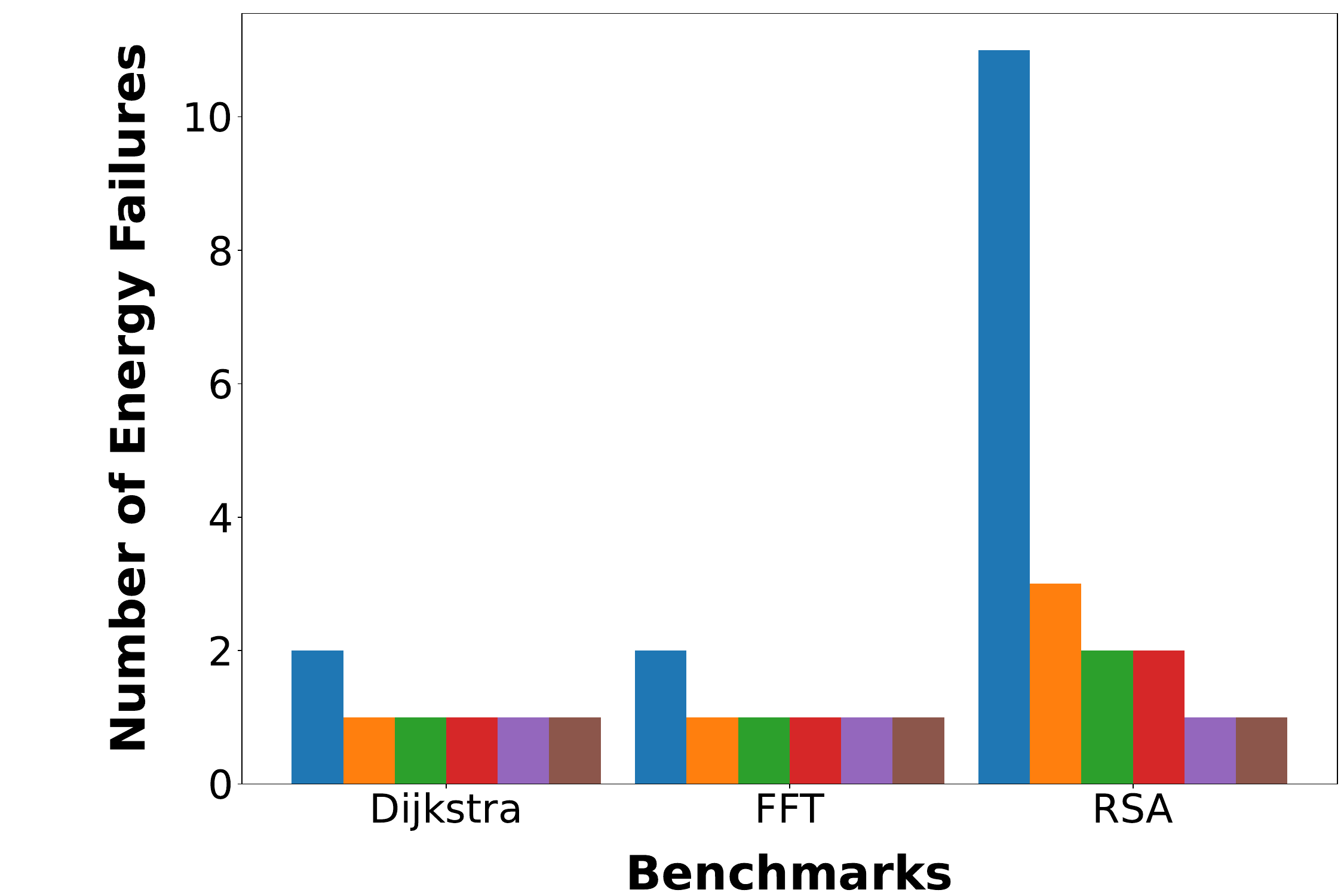}}
        \label{fig:results_rf_80u_36v_mementos_power_failures}
    }
    \caption{Results with the energy-moderate source and Mementos with \twovmin, $\mathbf{C = 80 \mu F}$, and $\mathbf{V_{boot} = 3.6V}$.}
    \label{fig:results_rf_80u_36v_mementos}
\end{figure}

\begin{figure}[t]
    \subfigtopskip = -2pt
    \subfigure{
        \centering
        \hspace{82pt}
        \resizebox{0.35\columnwidth}{!}{\includegraphics{figures/evaluation/simulations/20u_3.6v/legend.pdf}}
    }
    \newline
    \setcounter{subfigure}{0}
    \subfigure[Completion time]{
        \centering
        \resizebox{0.319\columnwidth}{!}{\includegraphics{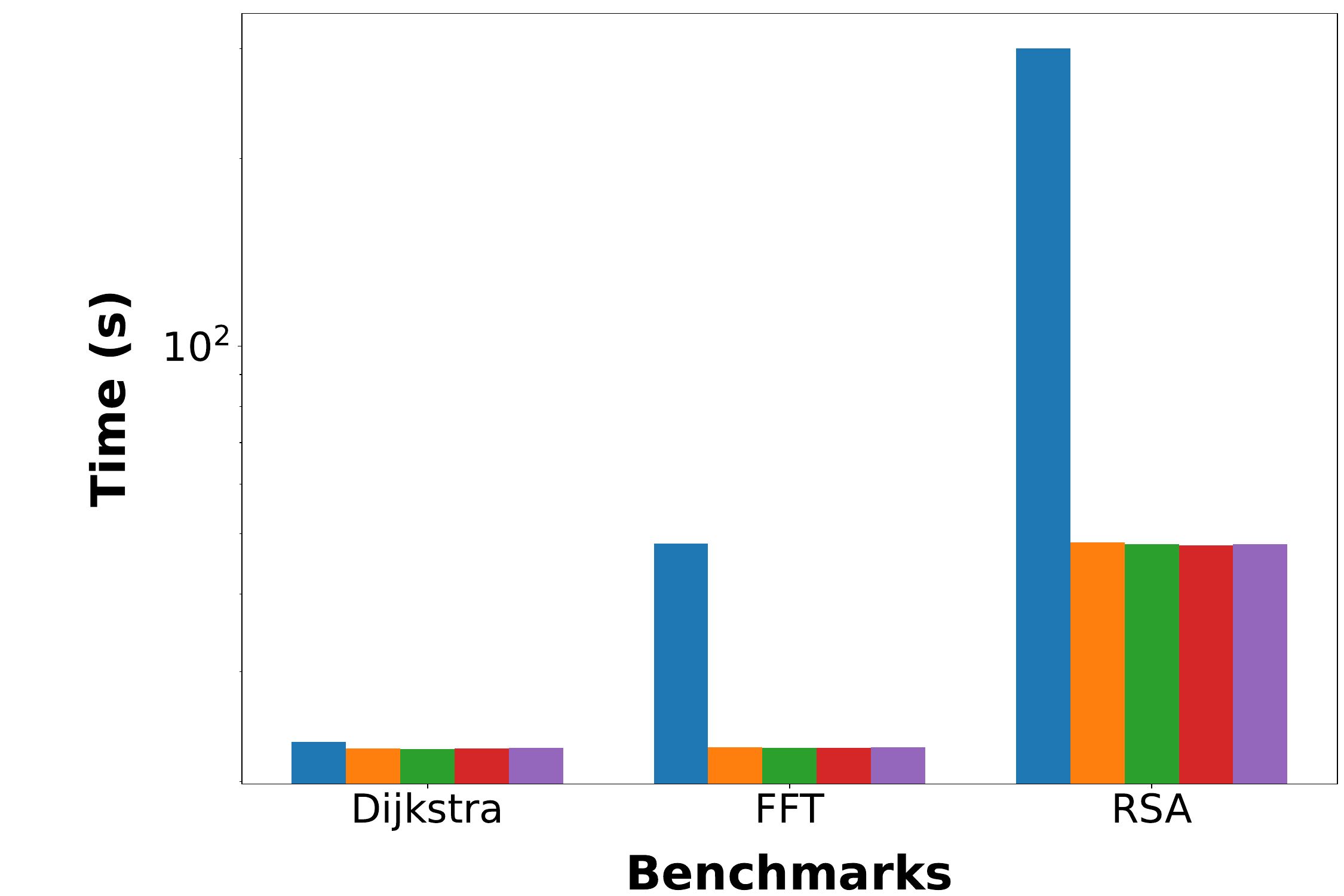}}
        \label{fig:results_rf_20u_36v_mementos_time}
    }
    \subfigure[Recharge time]{
        \centering
        \resizebox{0.319\columnwidth}{!}{\includegraphics{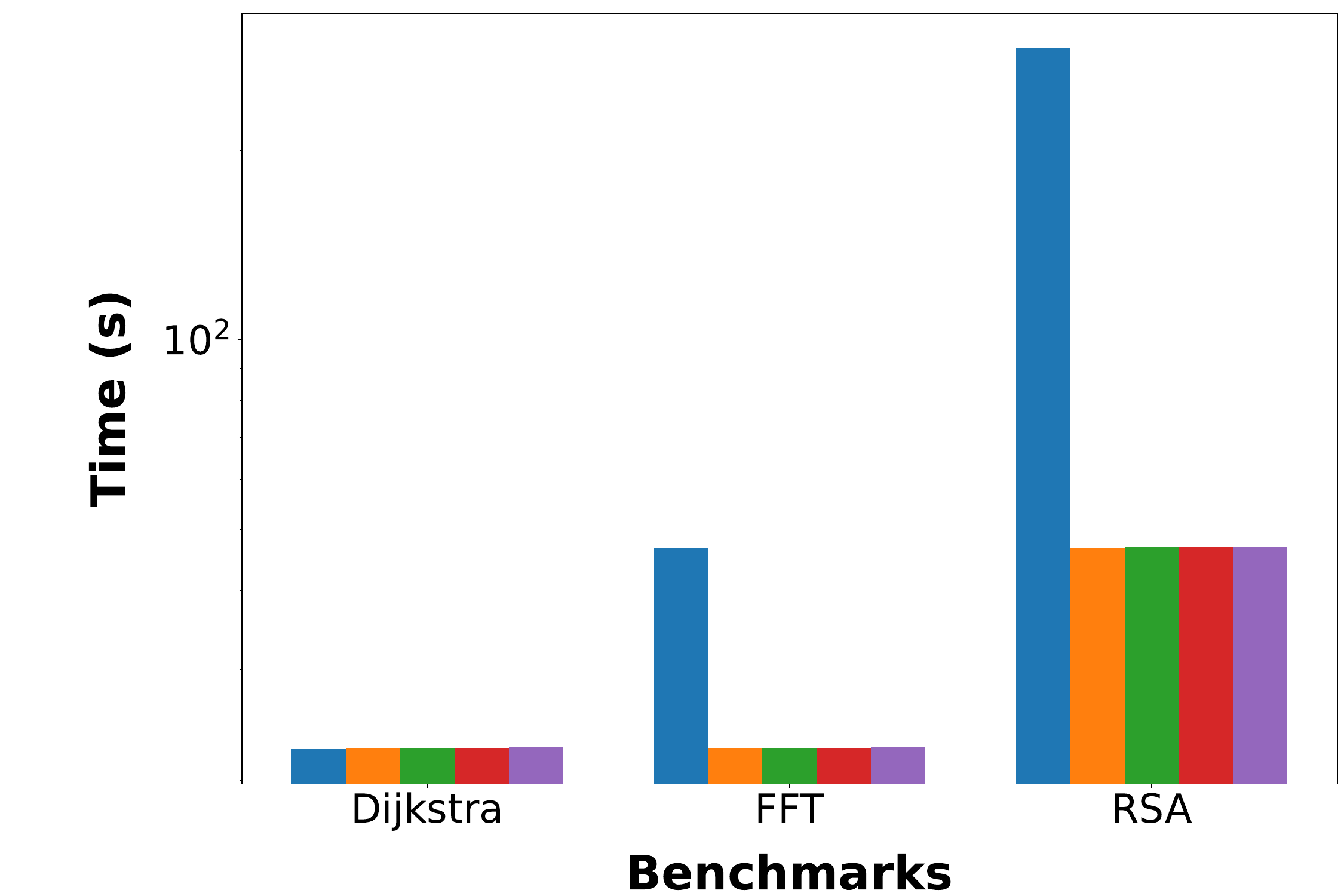}}
        \label{fig:results_rf_20u_36v_mementos_time_recharge}
    }
    \subfigure[Execution time]{
        \centering
        \resizebox{0.319\columnwidth}{!}{\includegraphics{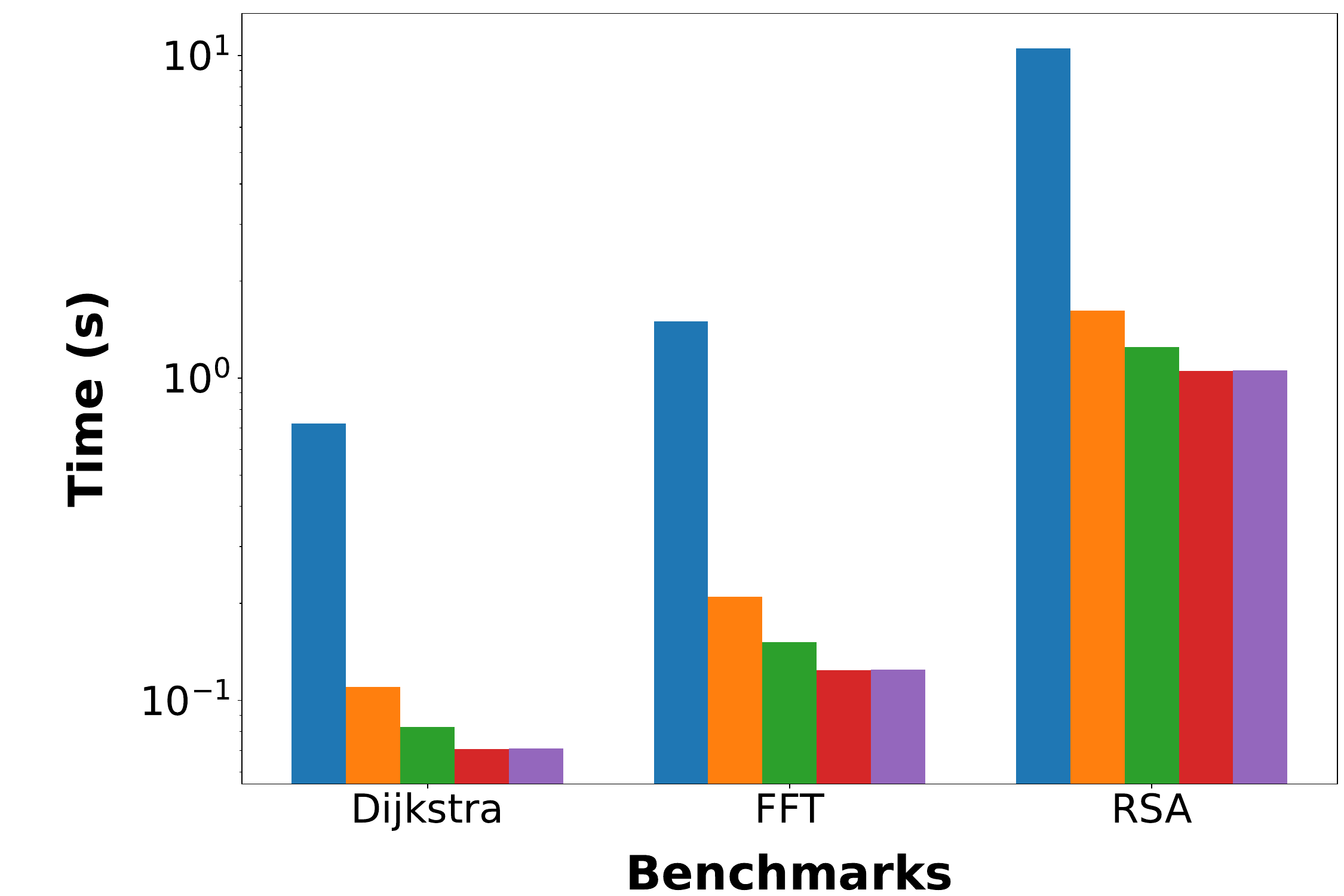}}
        \label{fig:results_rf_20u_36v_mementos_time_execution}
    }
    \subfigure[Energy consumption]{
        \centering
        \resizebox{0.319\columnwidth}{!}{\includegraphics{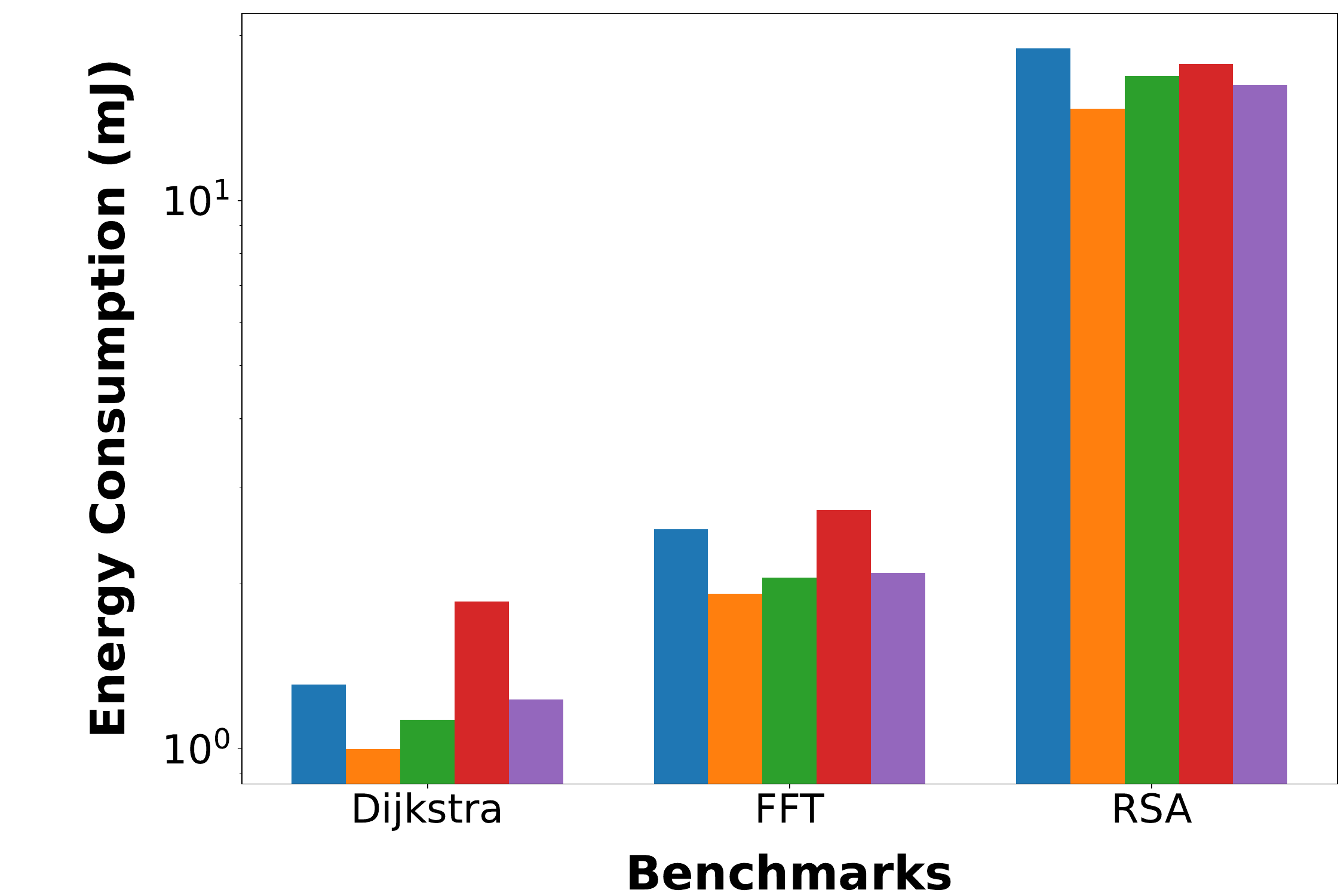}}
        \label{fig:results_rf_20u_36v_mementos_energy}
    }
    \subfigure[Impact of external circuitry]{
        \centering
        \resizebox{0.319\columnwidth}{!}{\includegraphics{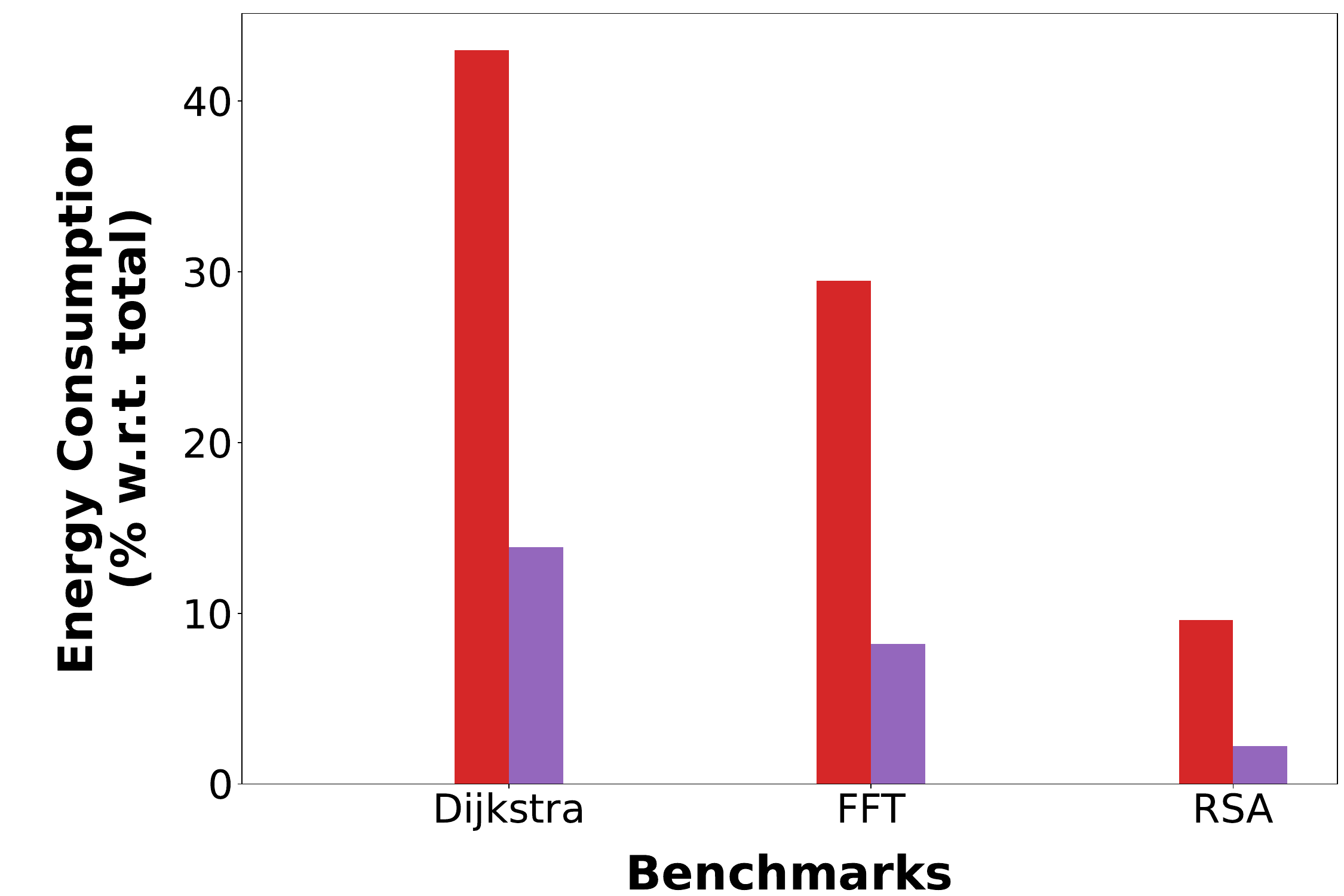}}
        \label{fig:results_rf_20u_36v_mementos_energy_dvfs_ratio}
    }
    \subfigure[Number of energy failures]{
        \centering
        \resizebox{0.319\columnwidth}{!}{\includegraphics{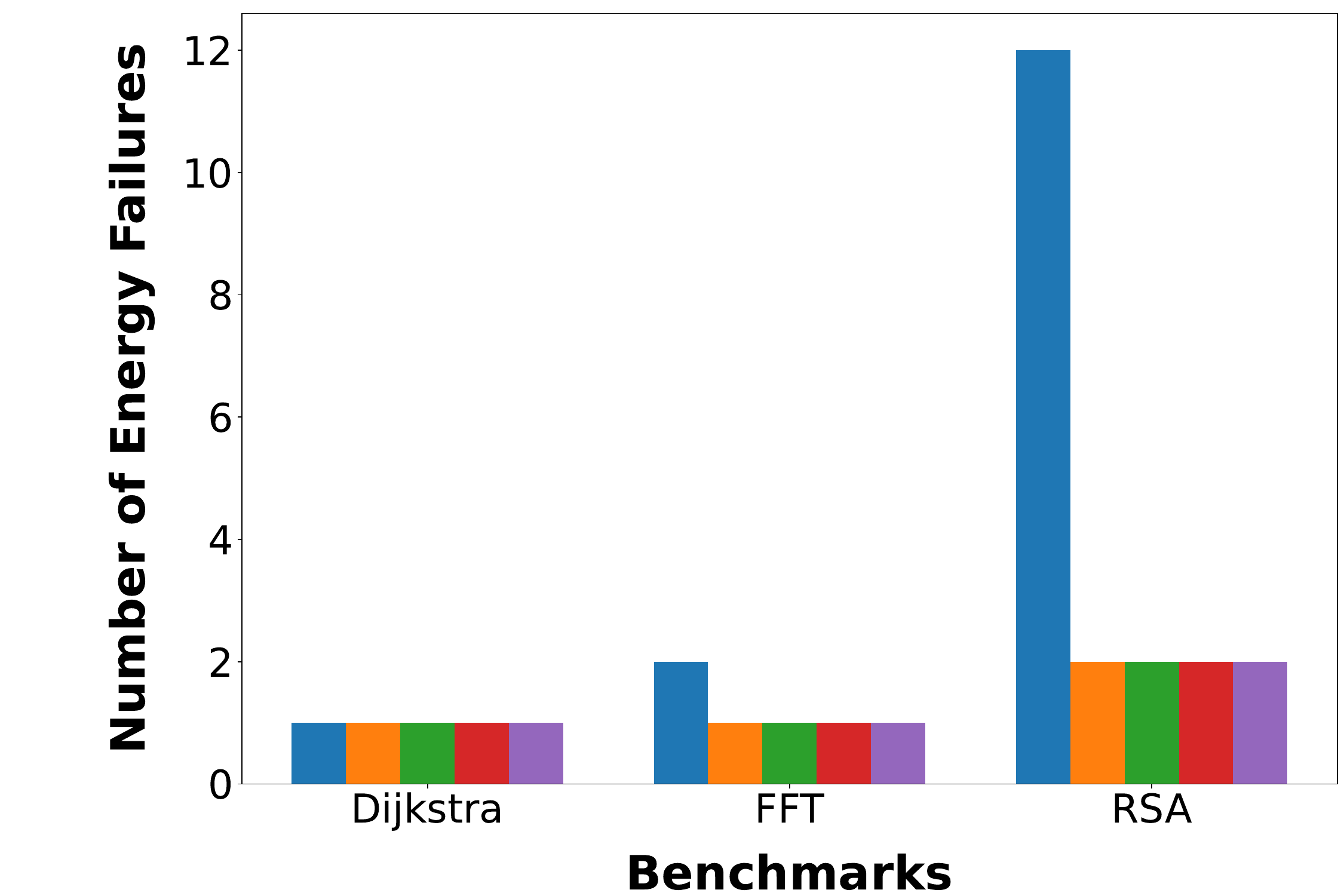}}
        \label{fig:results_rf_20u_36v_mementos_power_failures}
    }
    \caption{Results with the energy-moderate source and Mementos with \twovmin, $\mathbf{C = 20 \mu F}$, and $\mathbf{V_{boot} = 3.6V}$.}
    \label{fig:results_rf_20u_36v_mementos}
\end{figure}

\fakepar{Mementos with $\mathbf{C=80 \mu F}$}
We run the experiments considering the three Mementos configurations, namely \default, \noadcoff, and \twovmin, described in \secref{sec:evaluation-setting}.
The results show no significant change in performance between these configurations.
For these reasons, we report here only the results for \twovmin, as it represents the most reasonable choice for a real-world deployment.

\figref{fig:results_rf_80u_36v_mementos} summarizes the results.
\figref{fig:results_rf_80u_36v_mementos_time} indicates that the static $16Mhz$ configuration has the shortest completion time.
Analogous to the Hibernus experiments, the reduced operating range of this configuration enables the MCU to resume computation sooner after an energy failure, as the capacitor needs less energy to attain the boot voltage $V_{boot}$ again.
With Mementos, the MCU shuts down without entering a hibernation mode, thus avoiding the capacitor discharge caused by the quiescent current consumption of Hibernus's external comparators, since Mementos does not rely on any such components.
The baselines thus recharge back to $V_{boot}$ faster than \dvfs and \fbtc, which may be functioning at a slower, less efficient frequency or are turned off while awaiting the energy buffer to refill to $V_{boot}$.

Similarly to the Hibernus experiments, the completion time is mainly affected by the recharge time, as \figref{fig:results_rf_80u_36v_mementos_time_recharge} and \figref{fig:results_rf_80u_36v_mementos_time_execution} jointly demonstrate.
When running the the Dijkstra and FFT implementations, \dvfs and \fbtc execution times are within the execution time of the $12MHz$ static configuration, whereas in the RSA implementation they match the one of the $16MHz$ static configuration.
Considering that a deployed system runs the same workload indefinitely, in the long run \dvfs and \fbtc may show significantly shorter overall completion times compared to the baselines.

\figref{fig:results_rf_80u_36v_mementos_energy} shows that the static $8MHz$ configuration results in the most efficient energy performance, which however does not translate into the shortest completion time, as seen in \figref{fig:results_rf_80u_36v_mementos_time}.
Among the three benchmarks, \dvfs always shows one of the highest energy consumption, consuming on average $66\%$ more energy than the static $8MHz$ configuration.
Instead, on average, \fbtc consumes $45\%$ less energy than \dvfs and $12\%$ more energy than the static $8MHz$ configuration, always resulting among the most efficient configurations.

Two factors influence \dvfs and \fbtc energy performance in this setting.
As we point out in the experiments with the energy-rich source of \secref{sec:results-rich}, ADC probing introduces a clock cycle penalty that increases with the MCU operating frequency.
Hence, ADC probing makes \dvfs and \fbtc pay a higher penalty than the static $8MHz$ configuration, as the former execute a portion of the program at $16MHz$ and $12MHz$, which incur in a higher penalty than the static $8MHz$ configuration.
Second, \dvfs and \fbtc have a quiescent current draw that does not impact the baselines.

Despite the higher energy consumption, when running the Dijkstra and FFT implementations, \dvfs and \fbtc experience the same number of energy failures of the baselines, as \figref{fig:results_rf_80u_36v_mementos_power_failures} shows.
Instead, with the RSA implementation, \dvfs and \fbtc experience only one energy failure, whereas the baselines experience at least twice that.
This demonstrates that, despite the higher energy consumption, \dvfs and \fbtc can manage energy more efficiently, as they experience fewer energy failures.

The more efficient voltage and frequency scaling circuitry of \fbtc demonstrates, on average, a $45\%$ lower energy consumption and a $3.6\%$ faster completion time than \dvfs.
The higher quiescent current draw of \dvfs components is responsible, on average, for the $33\%$ of the overall energy consumption, whereas \fbtc components bear only a $9\%$ ompact, as shown in \figref{fig:results_rf_80u_36v_mementos_energy_dvfs_ratio}.
This causes \dvfs to consume $4.5x$ more energy than \fbtc when the MCU is powered off and recharges its energy buffer, causing the recharge time of \dvfs to be $4\%$ higher than \fbtc, as seen in \figref{fig:results_rf_80u_36v_mementos_time_recharge}.

\fakepar{Mementos with $\mathbf{C=20 \mu F}$}
As before with Hiubernus, we run experiments with a $20 \mu F$ capacitor, which does not require a voltage doubler with RF energy harvesting.
For the reasons oulined earlier, we discuss only the results for the \twovmin configuration.

\figref{fig:results_rf_20u_36v_mementos} summarizes the results with this configuration.
We note again a different performance compared to the experiments with a $80 \mu F$ capacitor.
\figref{fig:results_rf_20u_36v_mementos_time} shows that there is negligible difference in the completion time between all configurations but the static $1MHz$ one.
With the RSA implementation, that is, the benchmark with the highest number of required clock cycles, \fbtc and \dvfs are $0.11\%$ and $0.37\%$ faster than the static $12MHz$ configuration, respectively, which is the fastest baseline.

The reason for the different performance is the same as with the Hibernus experiments: the capacitor size no longer represents a disadvantage for \dvfs and \fbtc extended voltage range.
\dvfs and \fbtc recharge times are now on par with the baselines, as \figref{fig:results_rf_20u_36v_mementos_time_recharge} shows.
This also demonstrates that the quiescent current of \dvfs and \fbtc external circuitry bears a limited impact on the performance while the MCU is off.
The recharge times of \dvfs and \fbtc are similar to the baselines,
which have no additional hardware and hence no quiescent current draw.

\figref{fig:results_rf_20u_36v_mementos_time_execution} indicates that the execution time of \dvfs and \fbtc is, on average, $3.5x$ shorter than the baselines and at least $16\%$ faster than the best-performing baseline, which is the static $12MHz$ configuration in this case.
The key behind this performance is \dvfs and \fbtc voltage and frequency scaling technique.
Despite the inability of the static $16MHz$ configuration to complete the workload with the $20 \mu F$ capacitor, \dvfs and \fbtc can set the MCU to operate at $16MHz$ \emph{for a portion} of each energy cycle, which is the fastest and most efficient operating frequency.
This makes \dvfs and \fbtc able to extract the most possible performance out of available energy.

Compared to the experiments with a $80 \mu F$ capacitor, there is limited difference among the different system configurations in other performance metrics.
The same performance difference between \dvfs and \fbtc with the $80 \mu F$ capacitor is visible here too, as \fbtc is only $0.28\%$ slower than \dvfs, while demonstrating a $30\%$ lower energy consumption, as shown in \figref{fig:results_rf_20u_36v_mementos_time} and \figref{fig:results_rf_20u_36v_mementos_energy}.
The higher quiescent current draw of \dvfs components is responsible, on average, for $27\%$ of the overall energy consumption, whereas \fbtc components bear only a $8\%$ impact, as shown in \figref{fig:results_rf_20u_36v_mementos_energy_dvfs_ratio}.

%% file: evaluation-results-discharge.tex
\subsection{Results $\rightarrow$ Energy-poor Source}
We discuss here the results of the experiments with the energy-poor source, which only produces short energy cycles and yields a high energy failure rate.
We reproduce this scenario with a synthetic $5V$ energy source that supplies energy only when the device is powered off.
We set $V_{boot}$ to $3.6V$.

In the following and for both Hibernus and Mementos, we only discuss results with a $80 \mu F$ capacitor.
The results with the $80 \mu F$ show almost identical trends, leading to the same conclusions.

\begin{figure}[!t]
    \subfigtopskip = -2pt
    \subfigure{
        \centering
        \hspace{70pt}
        \resizebox{0.35\columnwidth}{!}{\includegraphics{figures/evaluation/simulations/80u_3.6v/legend.pdf}}
    }
    \newline
    \setcounter{subfigure}{0}
    \subfigure[Completion time]{
        \centering
        \resizebox{0.319\columnwidth}{!}{\includegraphics{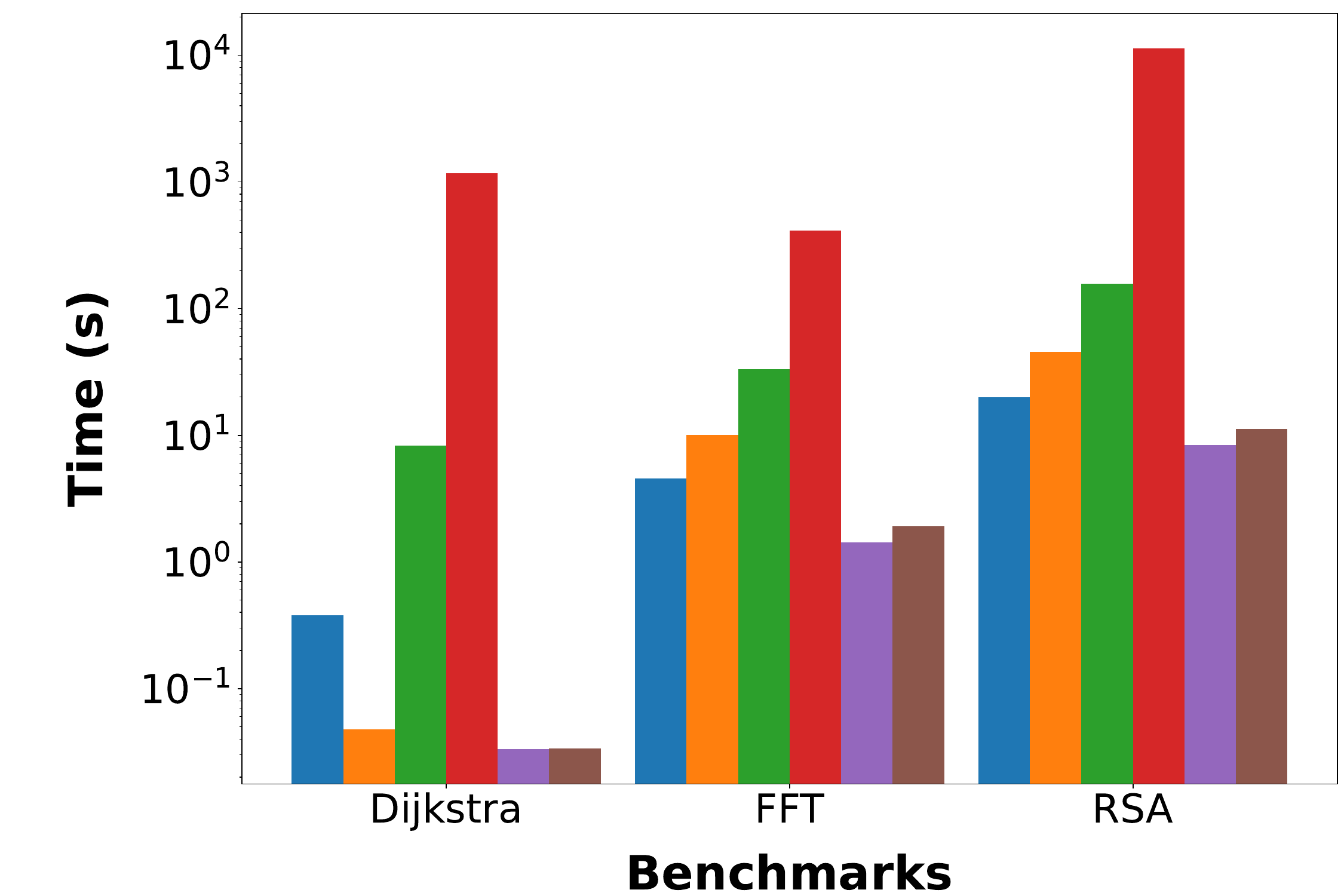}}
        \label{fig:results_discharge_80u_36v_hibernus_time}
    }
    \subfigure[Recharge time]{
        \centering
        \resizebox{0.319\columnwidth}{!}{\includegraphics{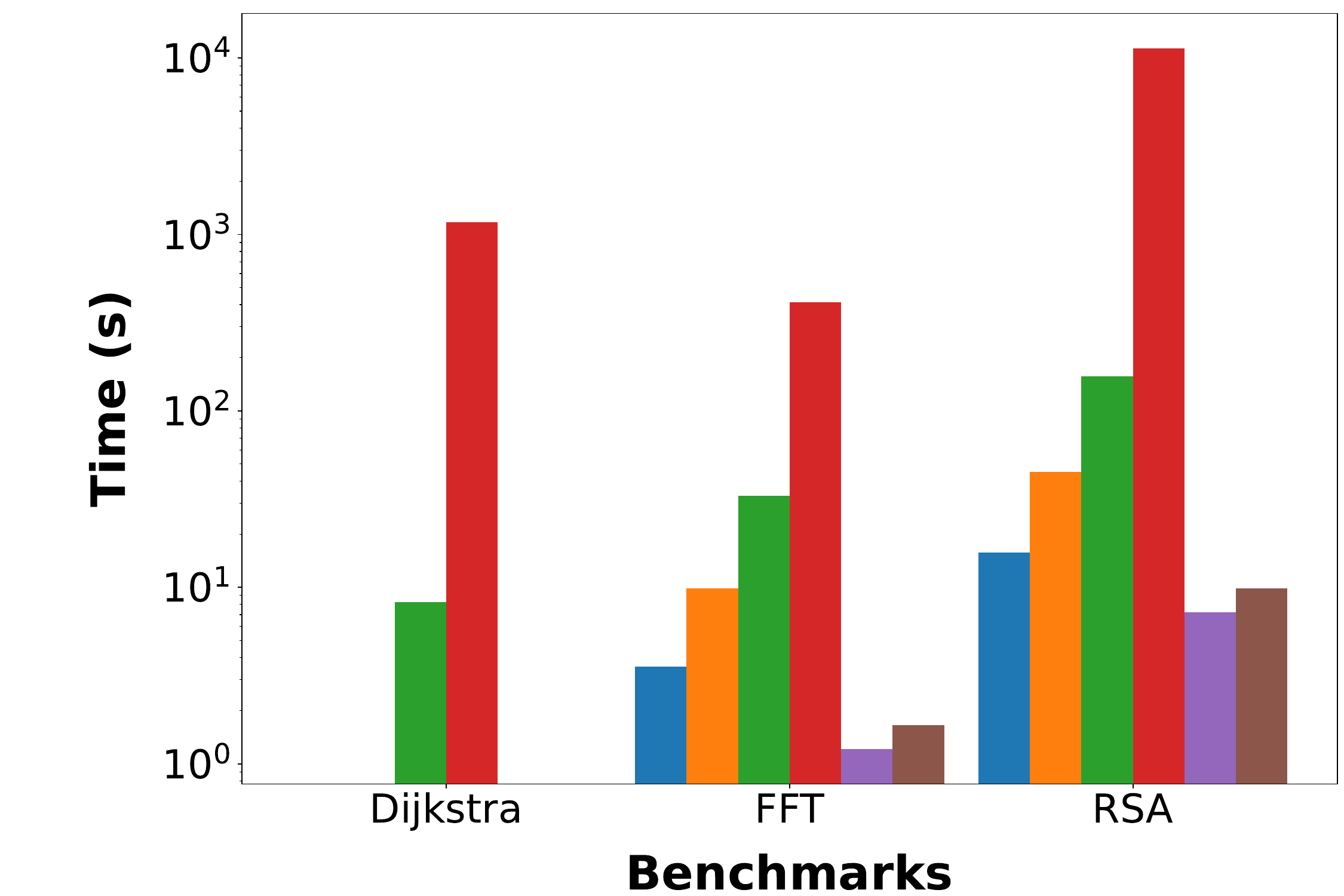}}
        \label{fig:results_discharge_80u_36v_hibernus_time_recharge}
    }
    \subfigure[Execution time]{
        \centering
        \resizebox{0.319\columnwidth}{!}{\includegraphics{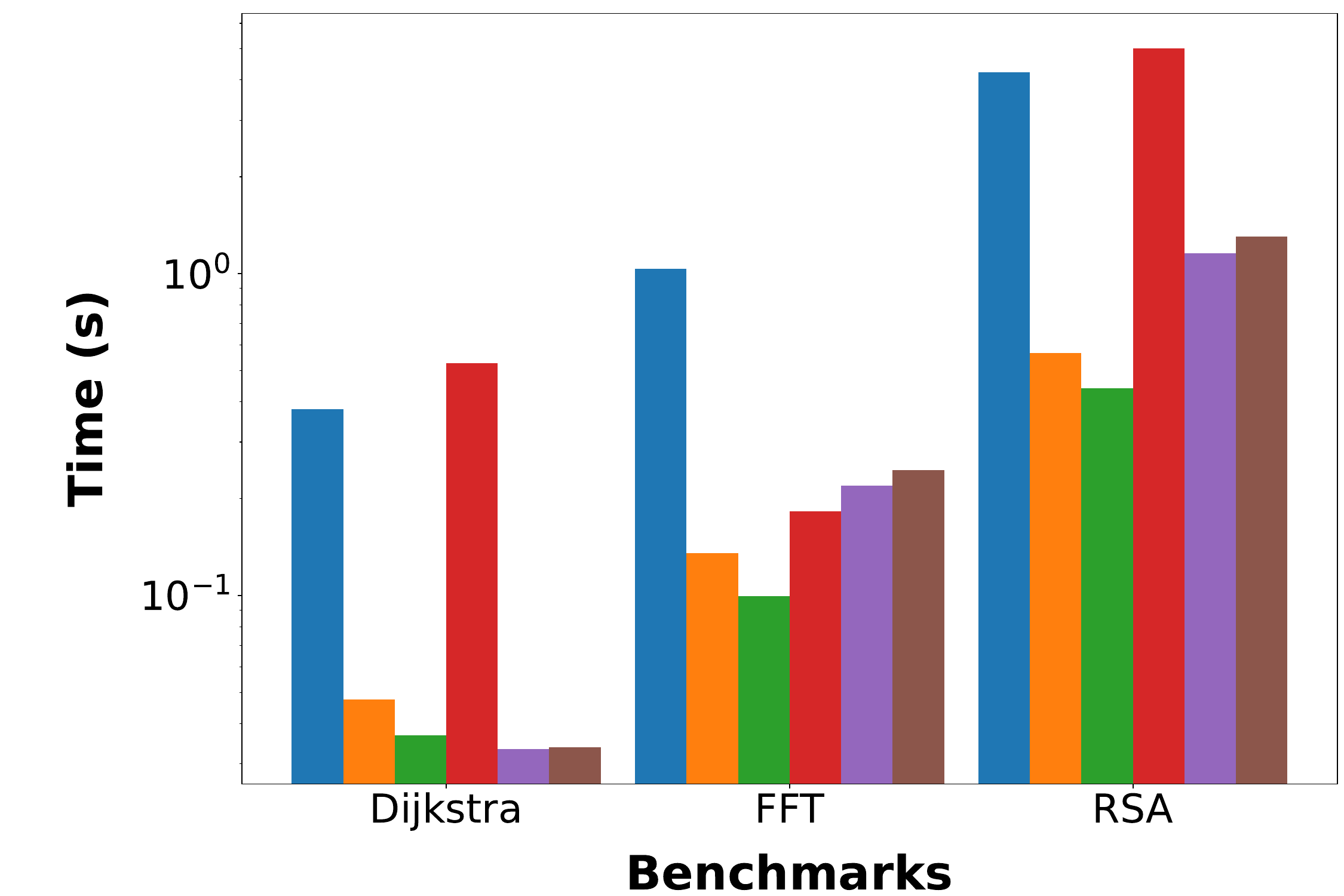}}
        \label{fig:results_discharge_80u_36v_hibernus_time_execution}
    }
    \subfigure[Energy consumption]{
        \centering
        \resizebox{0.319\columnwidth}{!}{\includegraphics{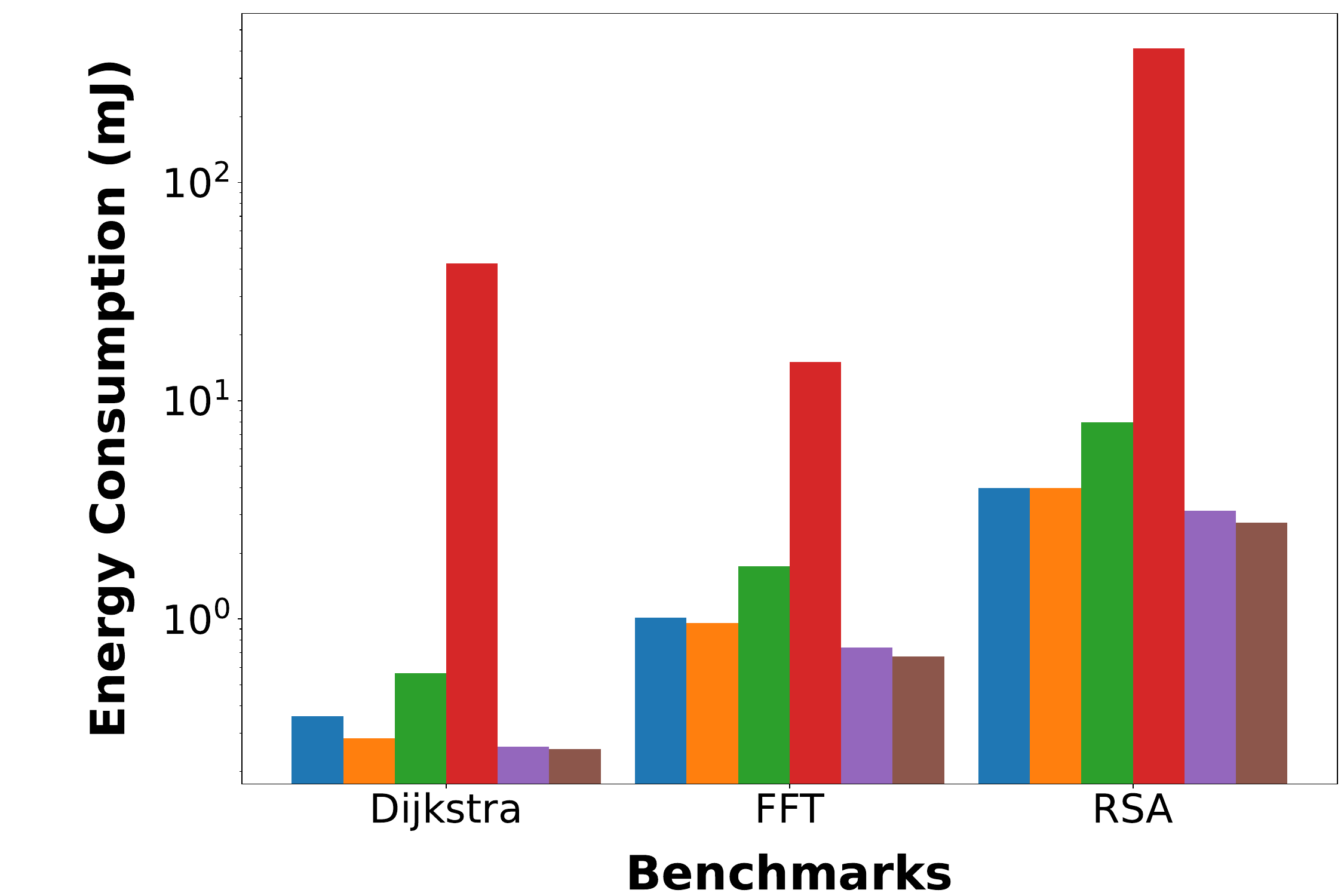}}
        \label{fig:results_discharge_80u_36v_hibernus_energy}
    }
    \subfigure[Impact of external circuitry]{
        \centering
        \resizebox{0.319\columnwidth}{!}{\includegraphics{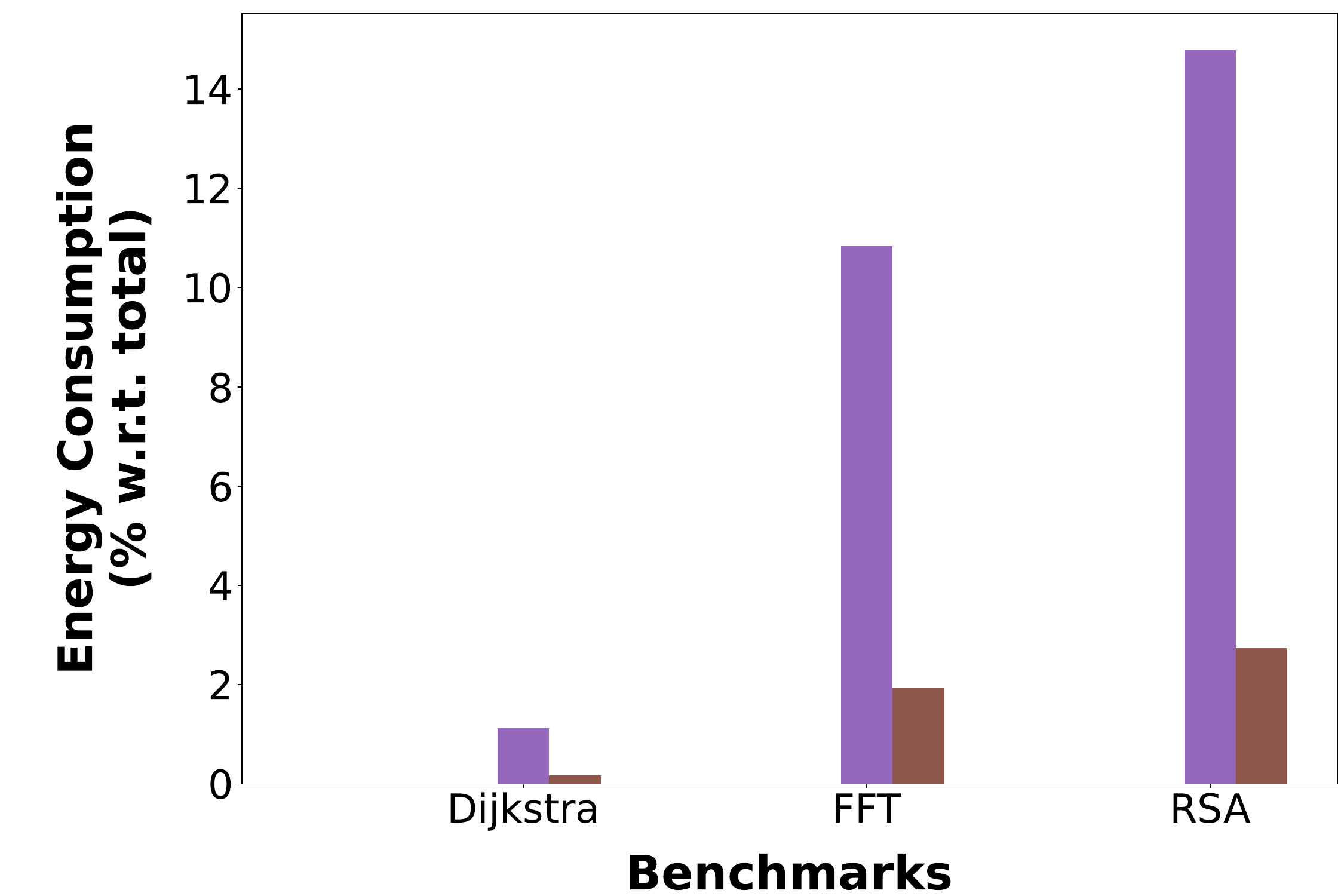}}
        \label{fig:results_discharge_80u_36v_hibernus_energy_dvfs_ratio}
    }
    \subfigure[Number of energy failures]{
        \centering
        \resizebox{0.319\columnwidth}{!}{\includegraphics{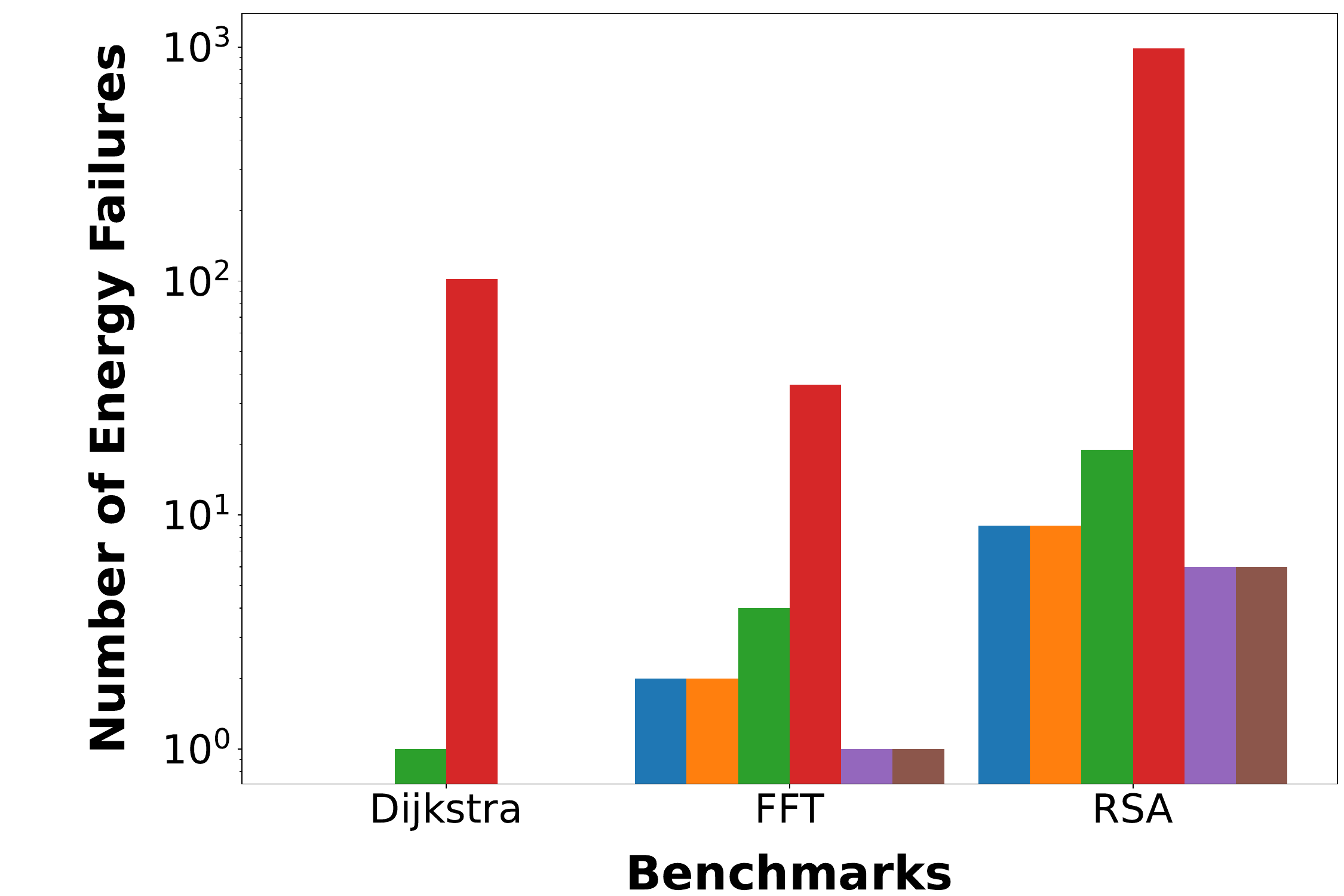}}
        \label{fig:results_discharge_80u_36v_hibernus_power_failures}
    }
    \caption{Results with the energy-poor source and Hibernus, $\mathbf{C = 80 \mu F}$, and $\mathbf{V_{boot} = 3.6V}$.}
    \label{fig:results_discharge_80u_36v_hibernus}
\end{figure}

\fakepar{Hibernus} \figref{fig:results_discharge_80u_36v_hibernus} depicts the results.
\dvfs and \fbtc demonstrate the best overall performance against all baselines.
As the energy-poor source does not supply energy unless the device is off, the duration of an energy cycle only depends on the minimum operating voltage of the selected MCU frequency.
\dvfs and \fbtc ensure that the MCU consistently operates at the maximum possible frequency and minimum possible voltage.
This extends the number of clock cycles executed within a single energy cycle.

\figref{fig:results_discharge_80u_36v_hibernus_time} depicts the completion time of each benchmark.
\dvfs and \fbtc are, on average, three orders of magnitude faster than the baselines.
Extending the energy cycle by lowering the clock frequency also increases, however, the time required to execute, as \figref{fig:results_discharge_80u_36v_hibernus_time_execution} depicts.
\dvfs and \fbtc indeed often show longer execution times than some of the baselines.
For example, when running the FFT and RSA implementations, \dvfs and \fbtc are respectively $91\%$ and $111\%$ slower than the static $12MHz$ configuration, that is, the best-performing baseline with this metric.
The increase in execution time comes in exchange for a higher number of instructions executed within an energy cycle, which significantly lowers the number of energy cycles required to complete the workload.
\dvfs and \fbtc also take less time than the baselines in waiting for new incoming energy, abating recharge times up to two orders of magnitude, as shown in \figref{fig:results_discharge_80u_36v_hibernus_time_recharge}.

Most importantly, \dvfs and \fbtc show a significantly lower energy consumption than all the baselines.
\figref{fig:results_discharge_80u_36v_hibernus_energy} shows that \dvfs and \fbtc consume, on average, $27x$ and $29x$ less energy than the static frequency configurations, respectively.
The voltage and frequency scaling techniques allow them to operate in the most efficient conditions.
Further, \figref{fig:results_discharge_80u_36v_hibernus_power_failures} shows that \dvfs and \fbtc can complete the Dijkstra algorithm implementation within a single energy cycle, whereas with the FFT and RSA implementations, \dvfs and \fbtc experience, on average, $26x$ fewer energy failures than the baselines.
This behavior is a consequence of \dvfs and \fbtc ability to extend the number of instructions executed within an energy cycle, which also results in a reduction of the number of energy cycles required to complete a workload.

The lower quiescent current of \fbtc results, on average, in a $9\%$ lower energy consumption than \dvfs, as shown in \figref{fig:results_discharge_80u_36v_hibernus_energy}.
\dvfs components are responsible for up to $16\%$ of the total energy consumption, wheras \fbtc components do not exceed $3\%$ of it, as shown in \figref{fig:results_discharge_80u_36v_hibernus_energy_dvfs_ratio}.
A higher energy consumption also means a lower equivalent resistance that enables a faster capacitor recharge.
\figref{fig:results_discharge_80u_36v_hibernus_time_recharge} shows that the lower resistance of \dvfs results, on average, in a $37\%$ faster recharge time than \fbtc.
This affects the completion time, as \dvfs shows, on average, a $33\%$ shorter completion time than \fbtc, as \figref{fig:results_discharge_80u_36v_hibernus_time} shows.

\begin{figure}[t]
    \subfigtopskip = -2pt
    \subfigure{
        \centering
        \hspace{70pt}
        \resizebox{0.35\columnwidth}{!}{\includegraphics{figures/evaluation/simulations/80u_3.6v/legend.pdf}}
    }
    \newline
    \setcounter{subfigure}{0}
    \subfigure[Completion time]{
        \centering
        \resizebox{0.319\columnwidth}{!}{\includegraphics{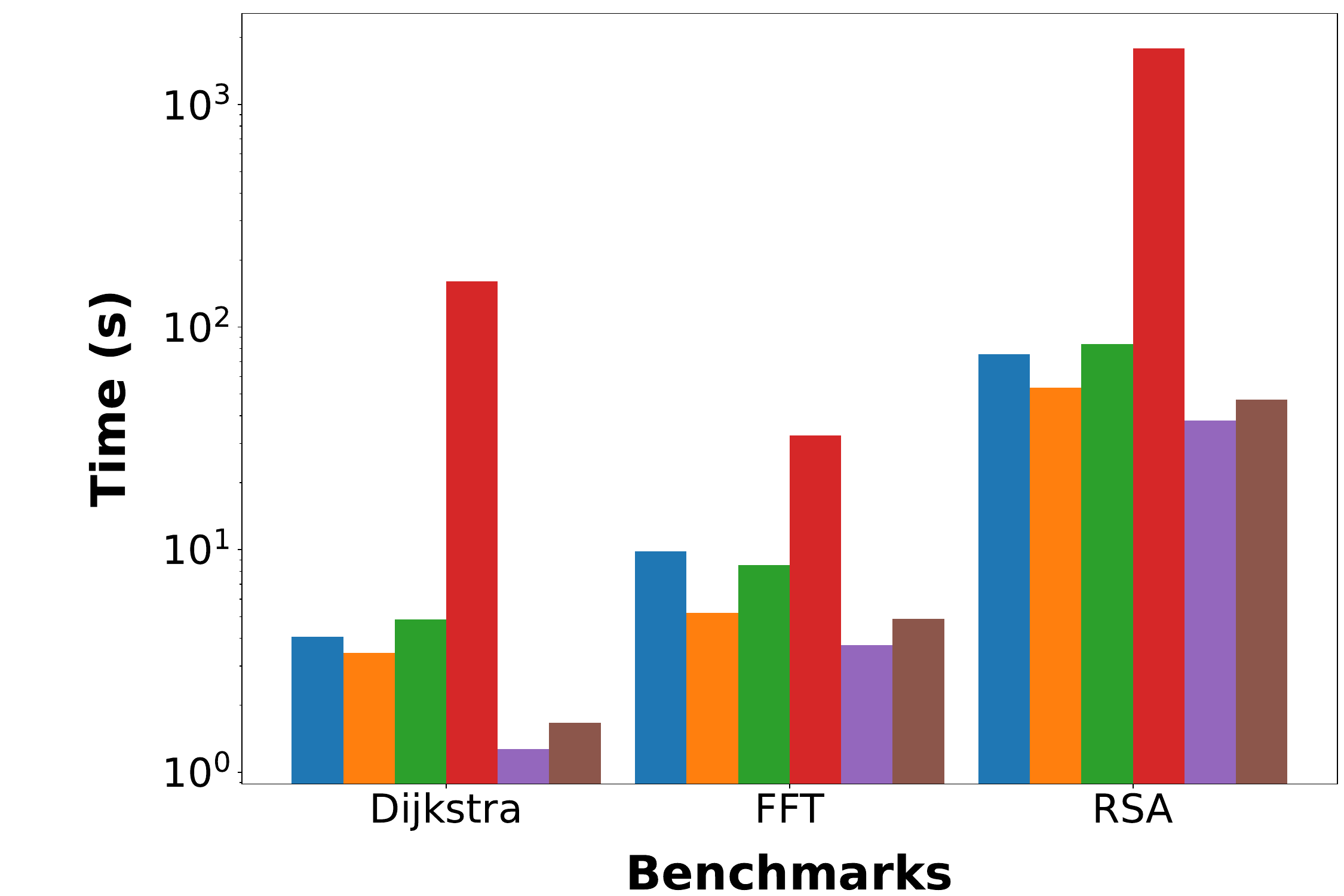}}
        \label{fig:results_discharge_80u_36v_mementos_time}
    }
    \subfigure[Recharge time]{
        \centering
        \resizebox{0.319\columnwidth}{!}{\includegraphics{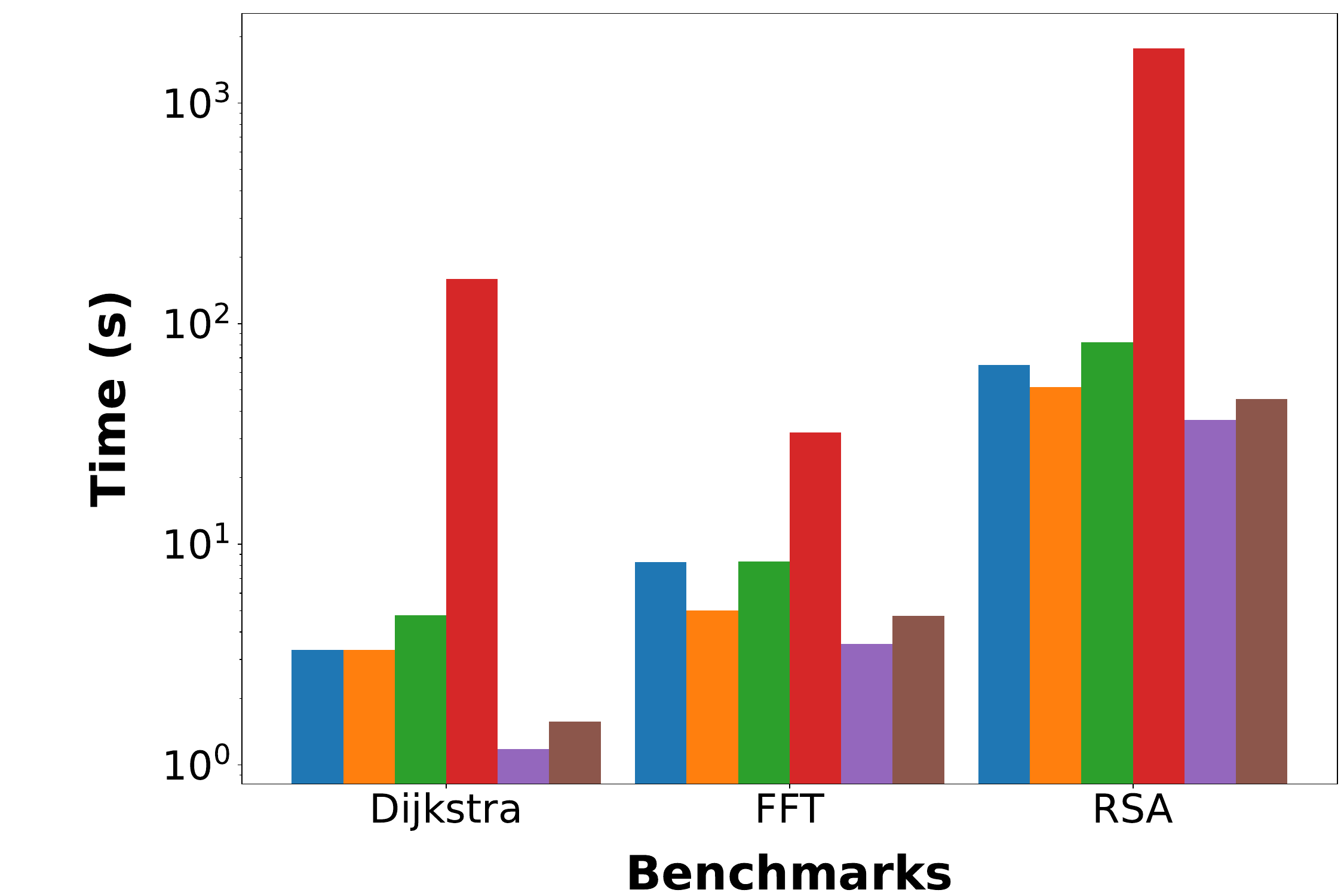}}
        \label{fig:results_discharge_80u_36v_mementos_time_recharge}
    }
    \subfigure[Execution time]{
        \centering
        \resizebox{0.319\columnwidth}{!}{\includegraphics{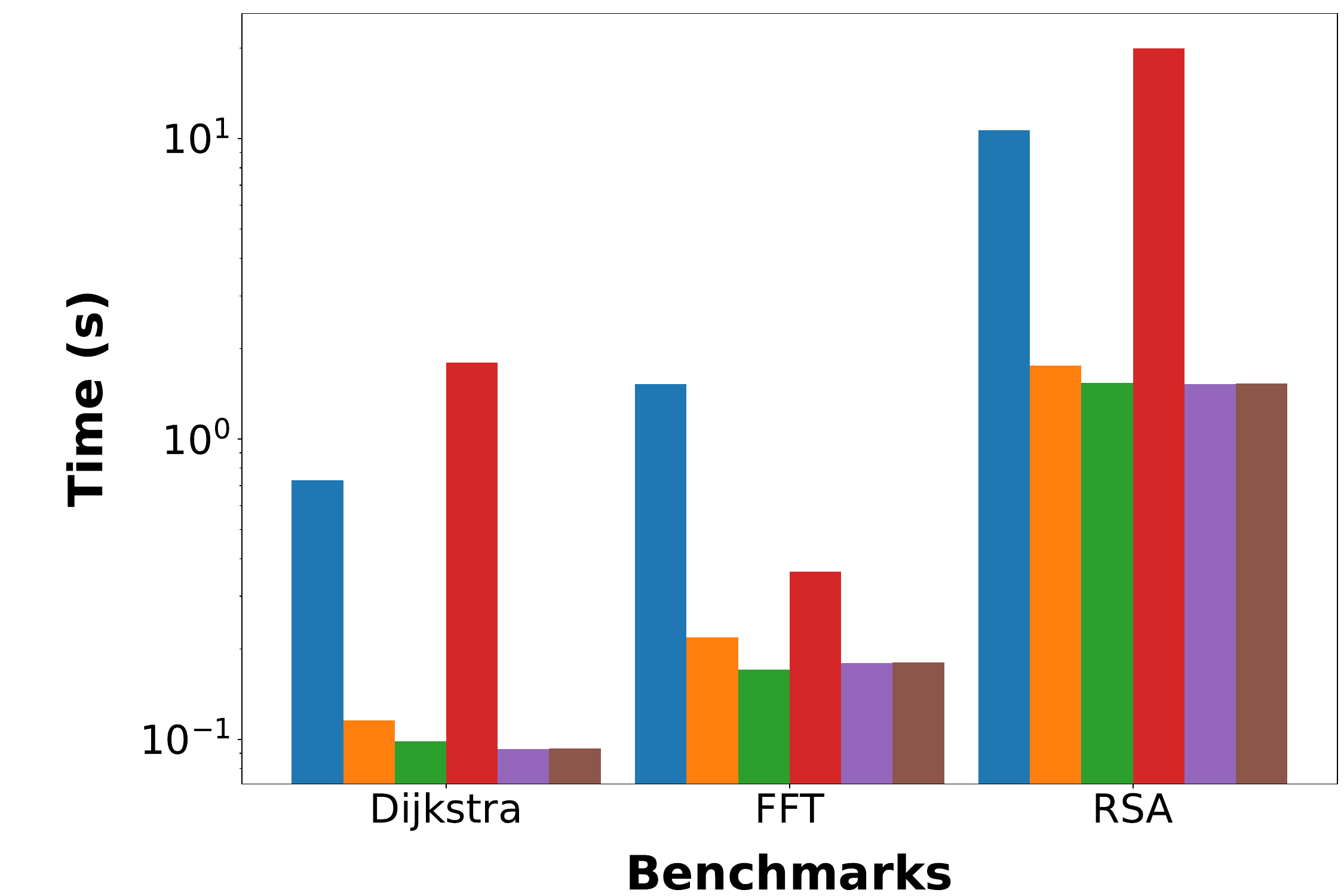}}
        \label{fig:results_discharge_80u_36v_mementos_time_execution}
    }
    \subfigure[Energy consumption]{
        \centering
        \resizebox{0.319\columnwidth}{!}{\includegraphics{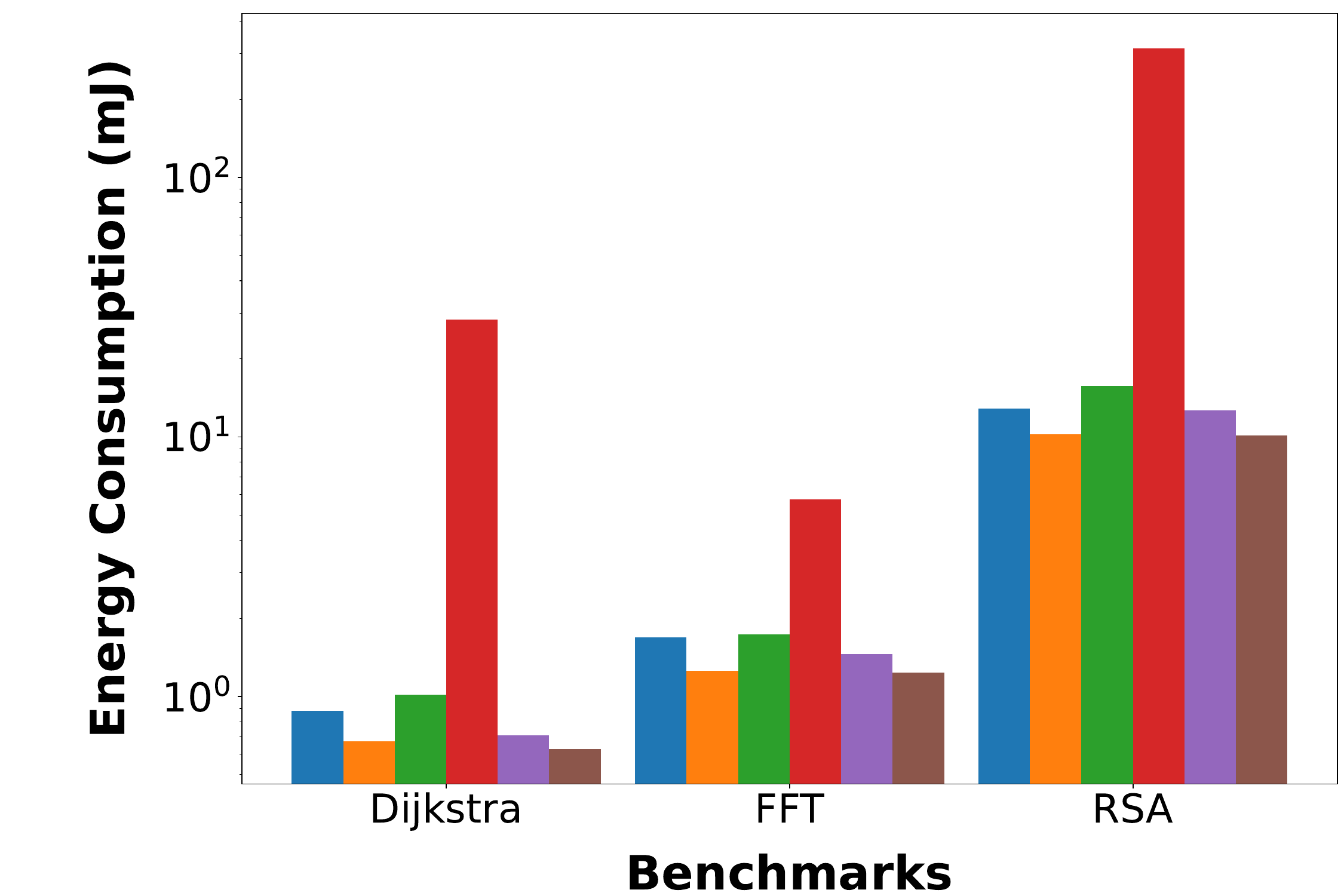}}
        \label{fig:results_discharge_80u_36v_mementos_energy}
    }
    \subfigure[Impact of external circuitry]{
        \centering
        \resizebox{0.319\columnwidth}{!}{\includegraphics{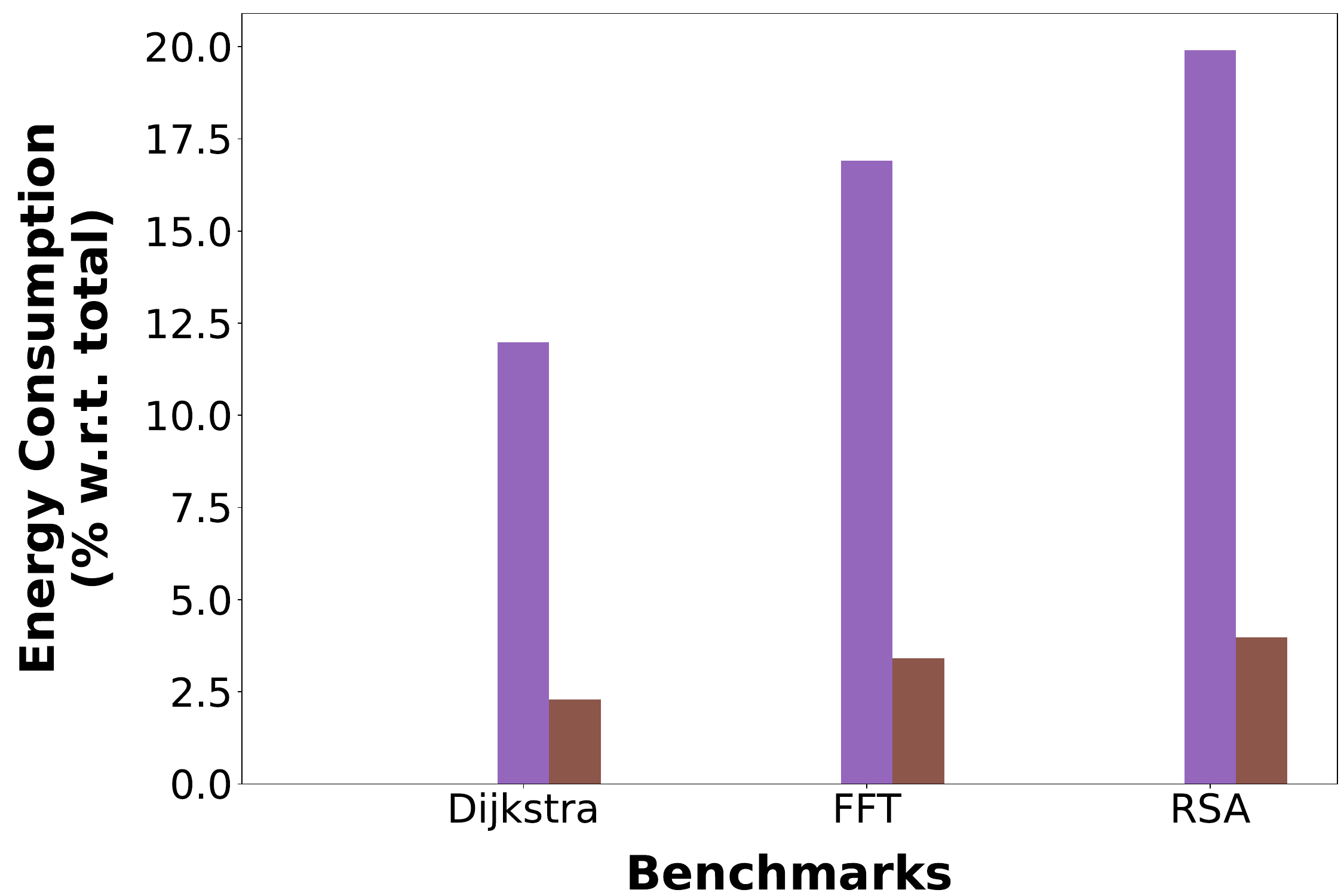}}
        \label{fig:results_discharge_80u_36v_mementos_energy_dvfs_ratio}
    }
    \subfigure[Number of energy failures]{
        \centering
        \resizebox{0.319\columnwidth}{!}{\includegraphics{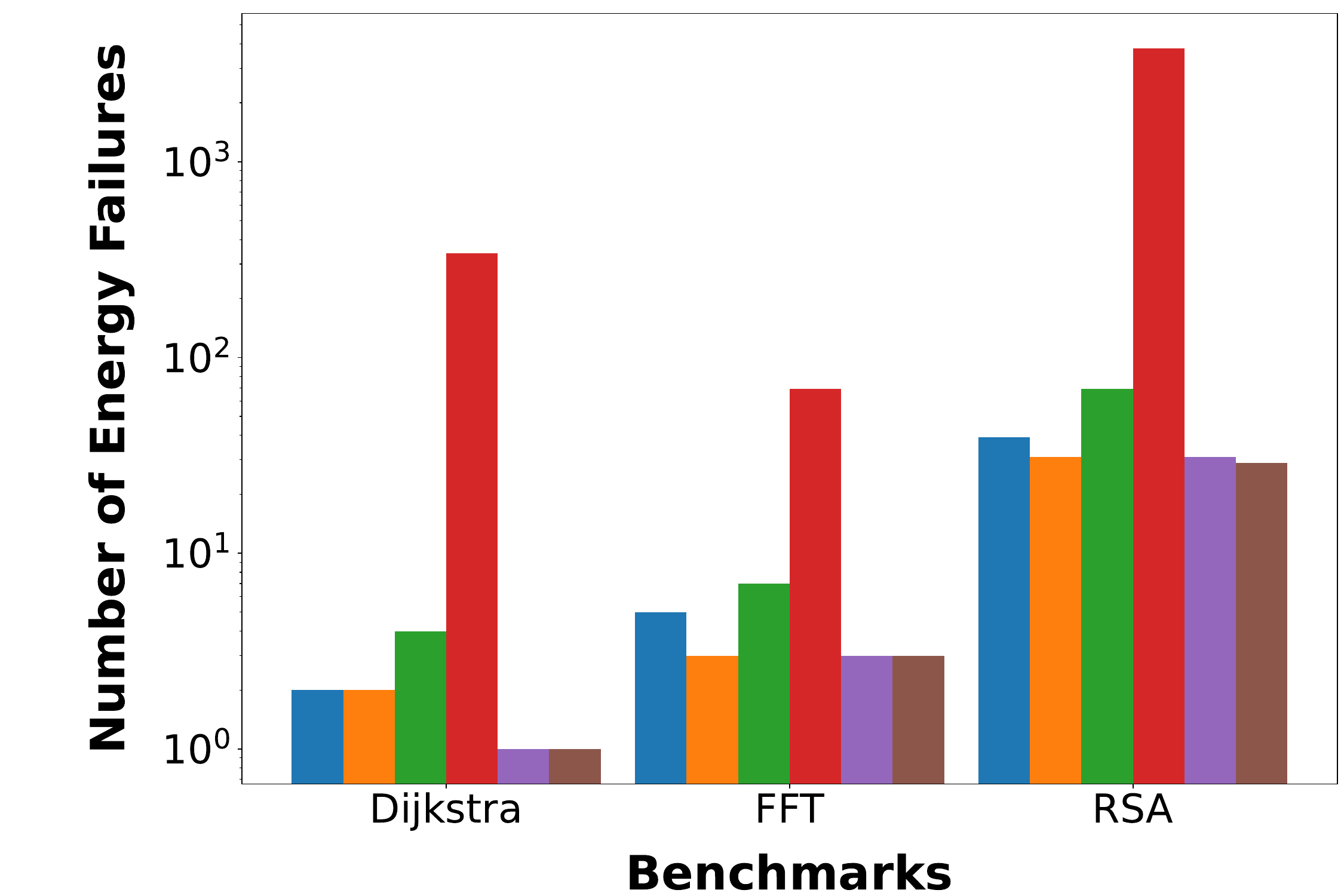}}
        \label{fig:results_discharge_80u_36v_mementos_power_failures}
    }
    \caption{Results with the energy-poor source and Mementos with \twovmin, $\mathbf{C = 80 \mu F}$, and $\mathbf{V_{boot} = 3.6V}$.}
    \label{fig:results_discharge_80u_36v_mementos}
\end{figure}

\fakepar{Mementos}
For reasons similar to \secref{sec:results-rf}, \figref{fig:results_discharge_80u_36v_mementos} only reports the results with the \twovmin ADC configuration.
The performance difference between \dvfs, \fbtc, and the baselines generally shows a trend similar to the Hibernus experiments.
The execution of Mementos' probe function, however, introduces an additional overhead, because each ADC access introduces a latency that increases with the MCU frequency.
Compared with \figref{fig:results_discharge_80u_36v_hibernus}, here the $1MHz$ static configuration pays the highest penalty due to ADC accesses, as the completion times in \figref{fig:results_discharge_80u_36v_mementos_energy} demonstrate.

Both \dvfs and \fbtc outperform the best-performing baselines depending on the metric at hand.
They show, respectively, a $42\%$ and $84\%$ shorter completion times than the static $8MHz$ configuration, as \figref{fig:results_discharge_80u_36v_mementos_time} demonstrates.
\figref{fig:results_discharge_80u_36v_mementos_energy} also indicates that, on average, \fbtc (\dvfs) has a $3.5\%$ ($0.81\%$) lower (higher) energy consumption than the same baseline.
Collectively, the metrics of completion times and energy consumption suggest that \fbtc and \dvfs outperform the static $8MHz$ configuration by finishing tasks more rapidly while consuming comparable amounts of energy. 
Their ability to dynamically scale voltage and frequency enables them to sustain longer energy cycles by operating in the most efficient settings.

Similarly to the Hibernus experiments with the energy-moderate source, \fbtc demonstrates, on average, a $19\%$ lower energy consumption but $29\%$ longer completion times than \dvfs.
The lower quiescent current of \fbtc, as evidenced in \figref{fig:results_discharge_80u_36v_mementos_energy_dvfs_ratio}, accounts for no more than 4\% of the overall energy consumption. In contrast, the components of \dvfs contribute up to 20\% of the energy use.

%% file: conclusion.tex
\section{Conclusion}

In this paper, we delved into the unique challenges faced by intermittently computing devices that harness ambient energy and utilize small capacitors as energy buffers.
Traditional methods of setting clock frequency fall short in addressing the intricate relationship between capacitor voltage, operational frequency's energy efficiency, and the associated operational range.
Existing techniques, designed for conventional devices, prove to be ill-suited due to the extreme energy limitations and distinct hardware attributes of energy-harvesting devices.

Through our exploration, we introduced two innovative hardware/software co-designs that recognize these distinct hardware characteristics.
These designs operate effectively within a constrained energy envelope, each offering its own set of trade-offs and functionalities.
Our experimental assessments, grounded in a mix of real-world and synthetic benchmarks, underscore the potential of these techniques to reshape the landscape of intermittent computing. As ambient energy-harvesting devices continue to gain traction, the strategies presented in this paper lay a foundation for their efficient and sustainable operation.